\documentclass{aa}  

\usepackage{graphicx}
\usepackage{epstopdf}

\usepackage{txfonts}
\usepackage{braket}
\usepackage{subcaption}
\begin{document}

   \title{The atmospheric parameters of FGK stars using wavelet analysis of CORALIE spectra}

   \author{S. Gill
          %\inst{1}
          \and
          P. F. L. Maxted
          \and
          B. Smalley}

   \institute{Astrophysics Group, Keele University, Keele, ST5 5BG, UK\\
              \email{s.gill@keele.ac.uk}
             }

   \date{Accepted January 18, 2018}

% \abstract{}{}{}{}{} 
% 5 {} token are mandatory
 
  \abstract
   { Atmospheric properties of F-,G- and K-type stars can be measured by spectral model fitting or from the analysis of equivalent width measurements. These methods require data with good signal-to-noise ratio and reliable continuum normalisation. This is particularly challenging for the spectra we have obtained with the CORALIE \'{e}chelle spectrograph for FGK stars with transiting M-dwarf companions. The spectra tend to have low signal-to-noise ratios, which makes it difficult to analyse them using existing methods.
    
}      % aims heading (mandatory)
   { Our aim is to create a reliable automated spectral analysis routine to determine $T_{\rm eff}$, [Fe/H], $V \sin i$ from the CORALIE spectra of FGK stars.
 }
   {We use wavelet decomposition to distinguish between noise, continuum trends, and stellar spectral features in the CORALIE spectra. A subset of wavelet coefficients from the target spectrum are compared to those from a grid of models in a Bayesian framework to determine the posterior probability distributions of the atmospheric parameters.}
   {By testing our method using synthetic spectra we found that our method converges on the best fitting atmospheric parameters.  We test the wavelet method on 20 FGK exoplanet host stars for which  higher quality data have been independently analysed using  equivalent width measurements. We find that we can determine $T_{\rm eff}$ to a precision of $85$\,K, [Fe/H] to a precision of 0.06\,dex and $V \sin i$ to a precision of 1.35\,km\,s$^{-1}$ for stars with $V \sin i$ $\geq$ 5\,km\,s$^{-1}$. We find an offset in metallicity $\approx -$0.18 dex relative to the equivelent width fitting method. We can determine $\log g$ to a precision of $0.13$\,dex but find systematic trends with $T_{\rm eff}$. Measurements of $\log g$ are only reliable to confirm dwarf-like surface gravity ($\log g \approx 4.5$).
   }
   {The wavelet method can be used to determine $T_{\rm eff}$, [Fe/H] and $V \sin i$ for FGK stars from CORALIE \'{e}chelle spectra. Measurements of $\log g$ are unreliable but can confirm dwarf-like surface gravity. We find that our method is self consistent, and robust for spectra with SNR$\gtrapprox 40$.}

   \keywords{Techniques: spectroscopic -- binaries: spectroscopic }

   \maketitle
%
%-------------------------------------------------------------------

\section{Introduction}
The WASP project \citep[Wide Angle Search for Planets;][]{PollaccoSkillenCollierCameronEtAl2006} monitors the night sky in search of exoplanet transit signatures around bright and nearby stars. So far, WASP has discovered over 130\footnote{TEPCAT \citep{Southworth2011}} exoplanet systems and continues to be the origin of many exoplanet discoveries. The SuperWASP instruments are sensitive to Jovian planets transiting solar-type stars which periodically attenuates the brightness of the host. Candidate exoplanet systems are subject to radial velocity measurements from the CORALIE \'{e}chelle spectrograph \citep{Queloz2000} to determine whether the transiting companion is indeed of planetary mass. Occasionally, these measurements yield radial velocity differences of 10's of km\,s$^{-1}$ over an orbit, suggesting that the transit is stellar in origin as opposed to Jovian. These are typically low-mass stars ($\leq 0.5\, M_{\sun}$) which have radii comparable to large planets, hence mimicking an exoplanet transit signal very well. These systems are given the flag "EBLM" (eclipsing binary, low mass) in the database used to coordinate observations to exclude them from the planet hunting process. 

% VLMS are most abundant
%       - contraction
%       - ages
%low mass stars in the galaxy 
%       - discrepency in measurements
%       - magentic activity.
%exoplanets around M dwarfs
%       - most likely to transit
%       - examples are trappist 1 and the possible exoplanet around Proxima centuri
%       - must know about host 
%       - Difficult to measure T_eff and [Fe/H]

M-dwarfs (0.08 $\leq$ M $\leq$ 0.6 $\rm M_{\sun}$) are the most abundant stars in the Galaxy \citep{Henry2006}. They slowly descend to the main sequence along a near-vertical Hyashi track where they can remain on the main sequence for billions of years. Those which are small enough to be fully convective restrict the build-up of helium ash in their cores, resulting in main sequence lifetimes greater than the age of the universe \citep{Baraffe1998}. 

% talk about low mass stars and  the deviations observed

Numerous photometric measurements around the infrared region are required to sample the peak of the spectral energy distribution of M-dwarfs. Studies of the secondary eclipses of M-dwarfs around larger FGK stars in the infrared have found temperatures much hotter than evolutionary models predict \citep{Triaud2013,Torres2013}. Empirical measurements of radius share a similar issue, in that they are typically underestimated by at least 3\% \citep{GomezMaqueoChewMoralesFaediEtAl2014,Demory2009,Spada2013,Fernandez2009,Nefs2013}. The favoured explanation is that enhanced magnetic activity in M-dwarf stars inhibits the convective flow of energy to the surface, resulting in an enlarged radius \citep{Morales2010,Fernandez2009,Zhou2014,Kraus}. Many measurements of mass and radius are done from eclipsing binary systems, and some have suggested that tidal interactions (via rapid rotation) is the root of the problem \citep{Kervella2016}. However, there is evidence for radius inflation in single stars \citep{Spada2013} suggesting a real problem with evolutionary models.

%exoplanets around M dwarfs
%       - most likely to transit
%       - examples are trappist 1 and the possible exoplanet around Proxima centuri
%       - must know about host 
%       - Difficult to measure T_eff and [Fe/H]

M-dwarfs are becoming prime targets for transiting exoplanet surveys. With all else constant, the transit of an exoplanet in the habitable zone is more likely to be observed around a star of lower mass to produce deeper eclipses that enable planetary atmospheric studies with transmission spectroscopy \citep{Sedaghati2016,Southworth2016}. Exciting new discoveries such as the TRAPPIST-1 system \citep{Gillon2017} and Proxima Centauri b \citep{Anglada-Escude2016} have attracted attention from the scientific community and public alike, bringing more focus to low-mass stars in exoplanet surveys. 

%Paragraph here explaining the lack of measurements in of M dwarfs: faint e.t.c
%An such example is SPECULOOS \citep{GillonJehinDelrezEtAl2013}.
\iffalse
Using spectroscopy to measure atmospheric parameters of VLMS is non-trivial. Low surface temperatures permit diatomic and triatomic molecules to remain in the photosphere. As a consequence, there are almost no un-blended weak lines in the optical which is detrimental to identifying continuum regions and abundance measurements \citep{Lindgren2015}. Fortunately, there are fewer molecular lines in certain bands of the infrared which can be used to determine [Fe/H] to a precision of 0.07 dex \citep[e.g.][]{Newton2014,Rojas-Ayala2010}. 
\fi

We are presented with an opportunity to use EBLMs discovered by the WASP survey to measure the properties of M-dwarfs. Measurements of $T_{\rm eff}$ and [Fe/H] for the FGK star can be measured accurately from a well understood spectrum and can be combined with radial velocity measurements and transit photometry to obtain the mass, radius and temperature of both components in EBLM systems. Measurements of [Fe/H] for the M-dwarf can be adopted from the FGK companion with the assumption that they both formed from the same parent molecular cloud \citep{Triaud2011}. 

The EBLM project \citep{Triaud2013,GomezMaqueoChewMoralesFaediEtAl2014, vonBoetticher2017} is an ongoing effort to characterise transiting M-dwarfs in EBLM systems. Accurate estimates for temperature and  composition are needed to estimate limb-darkening coefficients and make mass estimates for the primary thi star. For EBLM systems discovered by WASP, these parameters are made with CORALIE spectra using measurements of equivalent widths and by fitting individual spectral lines \citep{Gillon2009a, Doyle2013, Doyle2015}. The  individual analysis of each spectrum is acceptable for a small sample size, but we require a reliable automated procedure to measure $T_{\rm eff}$, [Fe/H] and $V \sin i$ for the entirety of the EBLM database to keep pace with future EBLM discoveries. At the same time, such a method needs to allow for noise and systematic errors present in the CORALIE spectra. The sample of EBLM spectra presented in \cite{GomezMaqueoChewMoralesFaediEtAl2014} typicaly have a signal-to-noise ratio per \AA ngstrom (SNR) between $3$ and $7$. The on-going radial velocity campaign to study EBLMs typically yields between 10 and 40 spectra per star. Co-adding spectra can increase the SNR ($\propto \sqrt{N_{obs}}$) to over 40 in some parts of the spectrum, but the regions of the spectrum near the ends of each echelle order suffer from both large photon noise and systematic errors due to inaccurate order-merging.  

Wavelet decomposition has been used previously as part of methods developed for spectral analysis.  \cite{Manteiga2010} used multi-level wavelet decomposition in connectionist systems (artificial neural networks) to derive fundamental stellar parameters in the low SNR domain (5-25) in preparation for spectra from the Gaia radial velocity spectrograph (RVS). This work was extended by \cite{Dafonte2016} by using a generative artificial neural network resulting in predicted uncertainties of 220 K, 0.32 dex and 0.20 dex for $T_{\rm eff}$, $\log g$ and [Fe/H], respectively for stars with a Gaia magnitude $G_{RVS}$ = 13. Using neural networks to  estimate atmospheric properties has well-known problems such as long training times and a strong dependence on the initial training set. \cite{Li2015} use wavelet decomposition in a regression framework to detect representative spectral features from a spectrum and estimate atmospheric parameters with better precision than those from neural networks ($83\, \rm K$, $0.23\, \rm dex$ and $0.16\, \rm  dex$ for $T_{\rm eff}$, $\log g$ and $\rm [Fe/H]$). 

 Our method determines the best-fitting atmospheric parameters ($T_{\rm eff}$, [Fe/H], $V \sin i$ and $\log g$) for FGK stars by comparing a selected subset of coefficients from a wavelet decomposition to those from a grid of stellar models. This reduces systematic errors in the estimated parameters due to poor continuum normalisation and low-quality regions of the spectrum. These measurements can then be combined with photometric follow-up observations to obtain the mass and radius of M-dwarfs in EBLMs to an accuracy of a few percent. These, in turn,  provide calibratable points for empirical mass-radius relations of low-mass stars \citep[e.g.][]{Demory2009,TorresAndersenGimenez2010}. We introduce wavelet decomposition as it applies to a spectrum in Sec. \ref{Wavelet Analysis} before reviewing our Bayesian approach to determine $T_{\rm eff}$, [Fe/H], $\log g$ and $V \sin i$ in Sec. \ref{Method}. We show that our method converges and is self-consistent in Sec. \ref{self_con} and test against a sample of independently analysed FGK stars in Sec. \ref{D15}.

\section{Wavelet analysis} \label{Wavelet Analysis}

\begin{figure*}[ht!]
\centering
\includegraphics[width=0.95\textwidth]{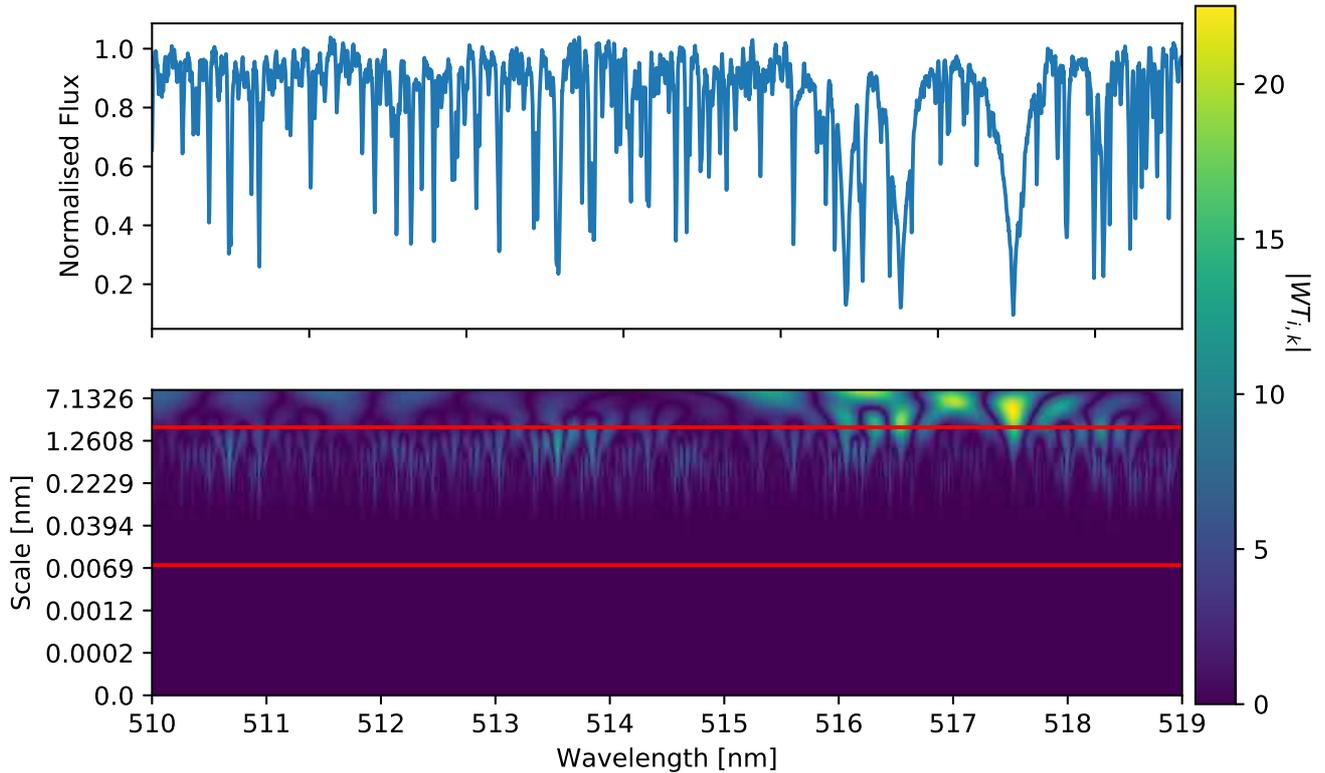}
\caption{The power H\"{o}vmoller of wavelet coefficients (lower panel) for a region around the Mg triplet for WASP-19 (upper panel). There is significant power ($|WT_{i,k}|$ from Eqn. \ref{DWT} ) for scales $\sim$1\,nm in the region of the Mg lines corresponding the wavelets likeness to spectral features. Horizontal red lines represent the scales $0.012-3.125$\,nm.}
\label{wavelet_power}
\end{figure*}

\begin{figure*}
\centering
\begin{subfigure}{.5\textwidth}
  \centering
		\includegraphics[width=0.9\textwidth]{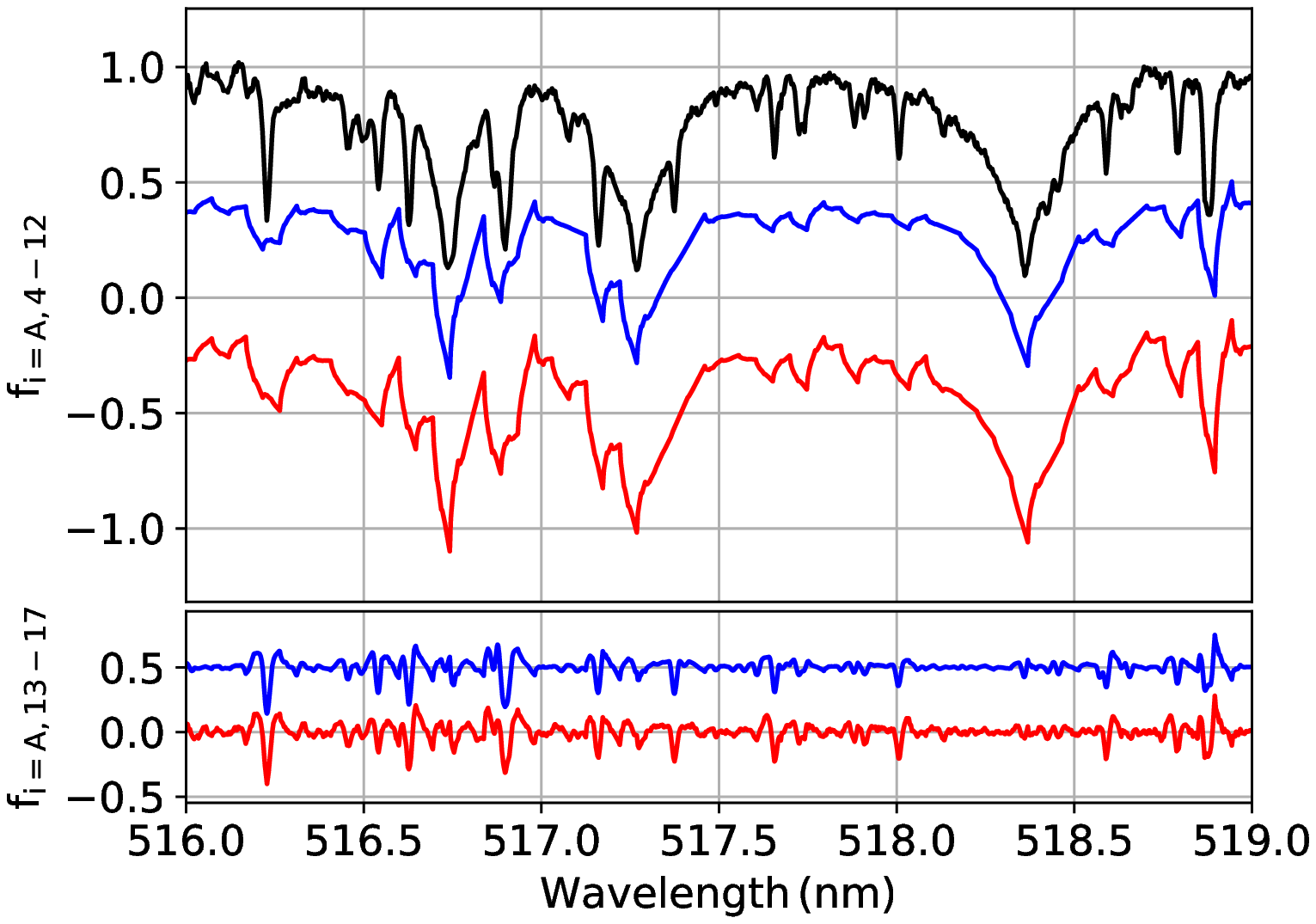}
		\caption{}\label{filt:a}
\end{subfigure}%
\begin{subfigure}{.5\textwidth}
  \centering
		\includegraphics[width=0.9\textwidth]{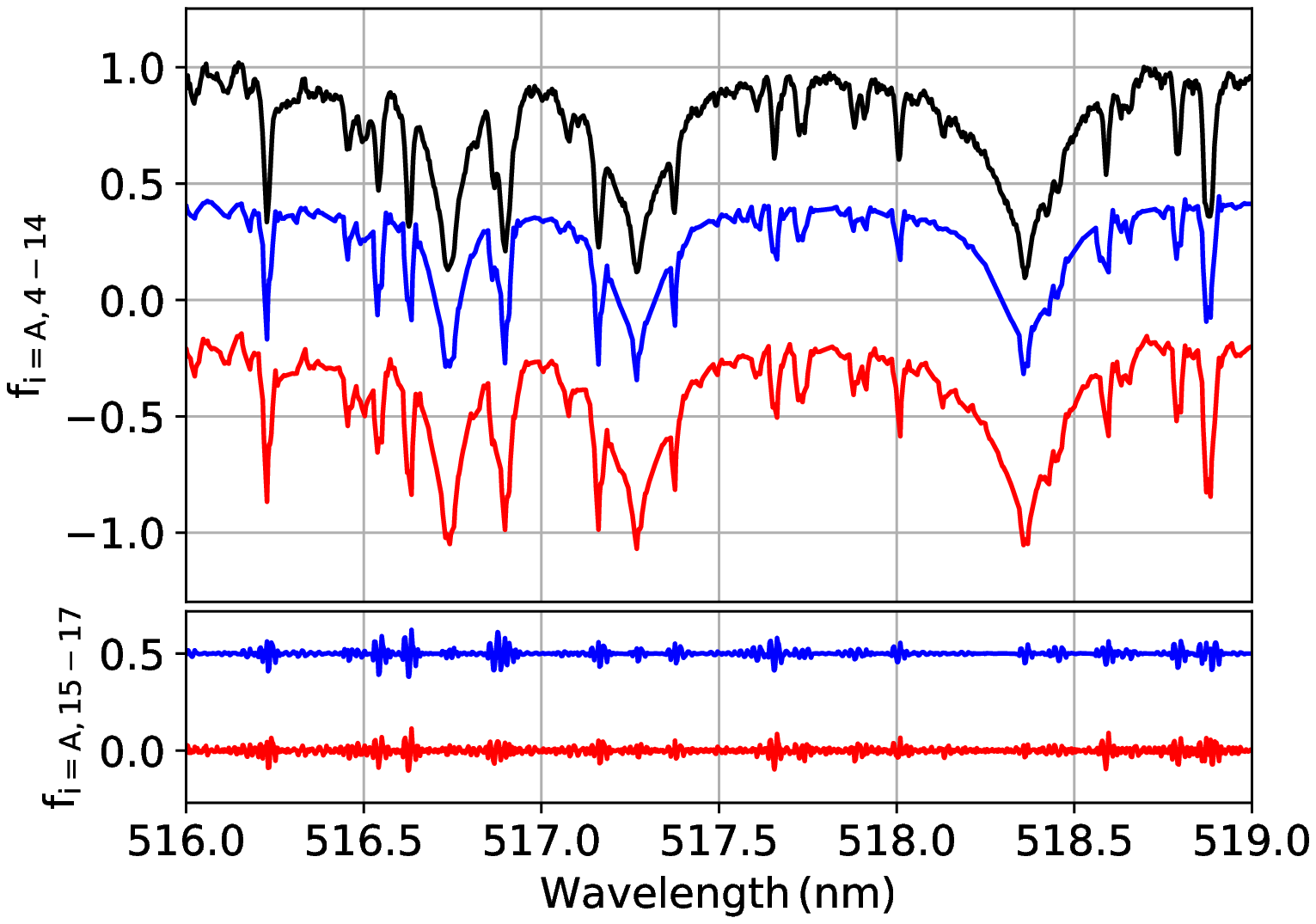}
		\caption{}\label{filt:b}
\end{subfigure}

\caption{The reconstruction of spectra using Eqn. (\ref{IDWT}) for subsets of wavelet coefficients. (Left panel - top) raw spectra for WASP-19 (black) and the flux reconstruction using wavelet coefficients from bands $i=4$-$12$ using the raw spectrum (blue; offset $-0.6$) and the best fitting model for WASP-19 (red; offset $-1.2$). (Left panel - bottom) the reconstruction of the best fitting model for WASP-19 (red) and the raw spectrum (blue; offset $+0.5$) using coefficients $i=13$-$17$. (Right panel) The same as left panel except with reconstructions using coefficients $i=4$-$14$ (top) and coefficients $i=15$-$17$ (bottom).}\label{filt}
\end{figure*}

Analysis of spectral components at different scales can be done using a discrete wavelet transform (DWT). A DWT tiles the wavelength-scale plane by convolving a spectrum, $f(\lambda)$, with variable sized functions \citep{Stumpe2012}. These functions are called daughter wavelets, $\psi_{\rm a,b}(\lambda)$, which are created from a mother wavelet, $\psi(\lambda )$,  using a shift-and-scale operation,

\begin{equation}\label{daugthermother}
\psi_{a,b}(\lambda) = \frac{1}{\sqrt{a}}\psi(\frac{\lambda - b}{a}),\quad a,b \in  \Re, a \neq 0 .
\end{equation}
here,  $a$ is a member of the dyadic sequence,

\begin{equation}\label{dyadic}
a_{i} = 2^{i}, \quad i = 0,1,2,3,...,n
\end{equation}
and $b=kb_{0}$, where $k$ is an integer and $b_0$ is chosen to ensure the recovery of $f(\lambda)$. By employing a DWT, the appropriate values of $b$ are selected to minimise overlap between wavelet convolutions. Following the notation in chapter 8 of \cite{Olkkonen2011}, a discrete wavelet transform can be calculated for each dyadic scale ($i$) and displacement ($k$):

\begin{equation}\label{DWT}
WT_{f(\lambda)}(i,k) = \frac{1}{\sqrt{2^i}} \int f(\lambda)\overline{\psi \left(\frac{\lambda - k2^ib_0}{2^i} \right)} d\lambda = \braket{f(\lambda),\psi_{i,k}(\lambda)}
\end{equation}

The likeness of a wavelet, $\psi_{i,k}$, to a section of the spectrum is given by the wavelet coefficient $WT_{f(\lambda)}(i,k)$ from Eqn. (\ref{DWT}). Performing this calculation over the series of dyadic scales and displacements yields wavelet coefficients which represent different sized structures at different wavelengths. We split coefficients into bands with constant scales, $\lbrace WT_{f(\lambda)}(0,b)\rbrace_k$, which represent the likeness of a single scale across the entire spectrum. The power of each scale, $\lbrace WT_{f(\lambda)}(i,b)\rbrace_k^2$ can be visuallised in a power H\"{o}vmoller (one value of $i$ per row) in Fig. \ref{wavelet_power}. Bands of coefficients which correspond to noise and low order continuum artefacts (such as merged \'{e}chelle orders) can then be excluded. A filtered spectrum may be reconstructed with an inverse DWT (IDWT):

\begin{equation}\label{IDWT}
f(\lambda) = \sum_{i=-\infty }^{\infty} 2^{\frac{-3i}{2}} \int WT_{f(\lambda)}(i,k)\hat{\psi} \left(\frac{\lambda - b}{2^i} \right)db
\end{equation}
where

\begin{equation}
\hat{\psi} \left(\frac{\lambda - b}{2^i} \right) = \frac{\psi \left(\frac{\lambda - b}{2^i} \right)}{\sum_{i=0}^{i=n} \left| \psi \left(\lambda - b \right) \right|^2 }
\end{equation}

The process of reconstructing a spectrum using a subset of wavelet coefficients is called wavelet filtering and is analogous with Fourier filtering. Alternatively, the subset of coefficients may be chosen to meet a threshold criteria (i.e. $\left[ WT_{f(\lambda)}(i,k) \right]^2  \geq \rm 0.01$) which eliminates information that has little contribution to a signal; this is called wavelet compression.

We do not require Eqn. (\ref{IDWT}) to determine atmospheric parameters as we perform a $\chi^2$ fit using a subset of coefficients from Eqn. (\ref{DWT}) to those from a grid of models (see Sect. \ref{Method}). We also do not apply any threshold criterion. The nominal resolving power of the CORALIE spectrograph is R=55\,000, so at least $2^{16}$ values are required to sample a spectrum over the wavelength range 450-650nm.  We decided to use $2^{17}$ values for the wavelet decomposition to ensure no loss of information and to give us more choice in the number of wavelet bands used in our analysis. We use Eqn. (\ref{DWT}) to obtain wavelet coefficients which have information on scales in the range 0.003\,nm--200\,nm. Our wavelet method only uses a subset of $i$ values. To select these, we constructed power H\"{o}vmoller diagrams (similar to Fig. 1) for a variety of regions between 450\,nm and 650\,nm, for different co-added spectra in our sample. We found that power associated with line absorption lies in the range 0.04--4\,nm, with larger scales typically corresponding to systematic trends and shorter scales with noise. This corresponds to values of i=4--12 (0.048--3.125\,nm).  The application of Eqn. (\ref{IDWT}) to the two subsets of coefficients (4--12 and 13--17) is shown in Fig. \ref{filt:a}.  We find that the subset range $i=4-12$ is too restrictive  to reproduce short-scale information (e.g. weak lines) and so we decide to extend this range to $i=4-14$ (0.012--3.125\, nm; Fig. \ref{filt:b}) which better represents the boundary between noise and weak lines. We do not show the reconstruction of subset $i=0-3$ in Fig. \ref{filt:a} and \ref{filt:b} as using only 16 coefficients to reconstruct a spectrum leads to a large Daubechies-4 wavelet with some sub-structure.

\begin{figure*}
\centering
 \begin{subfigure}[b]{0.5\linewidth}
    \centering
    \includegraphics[width=\linewidth]{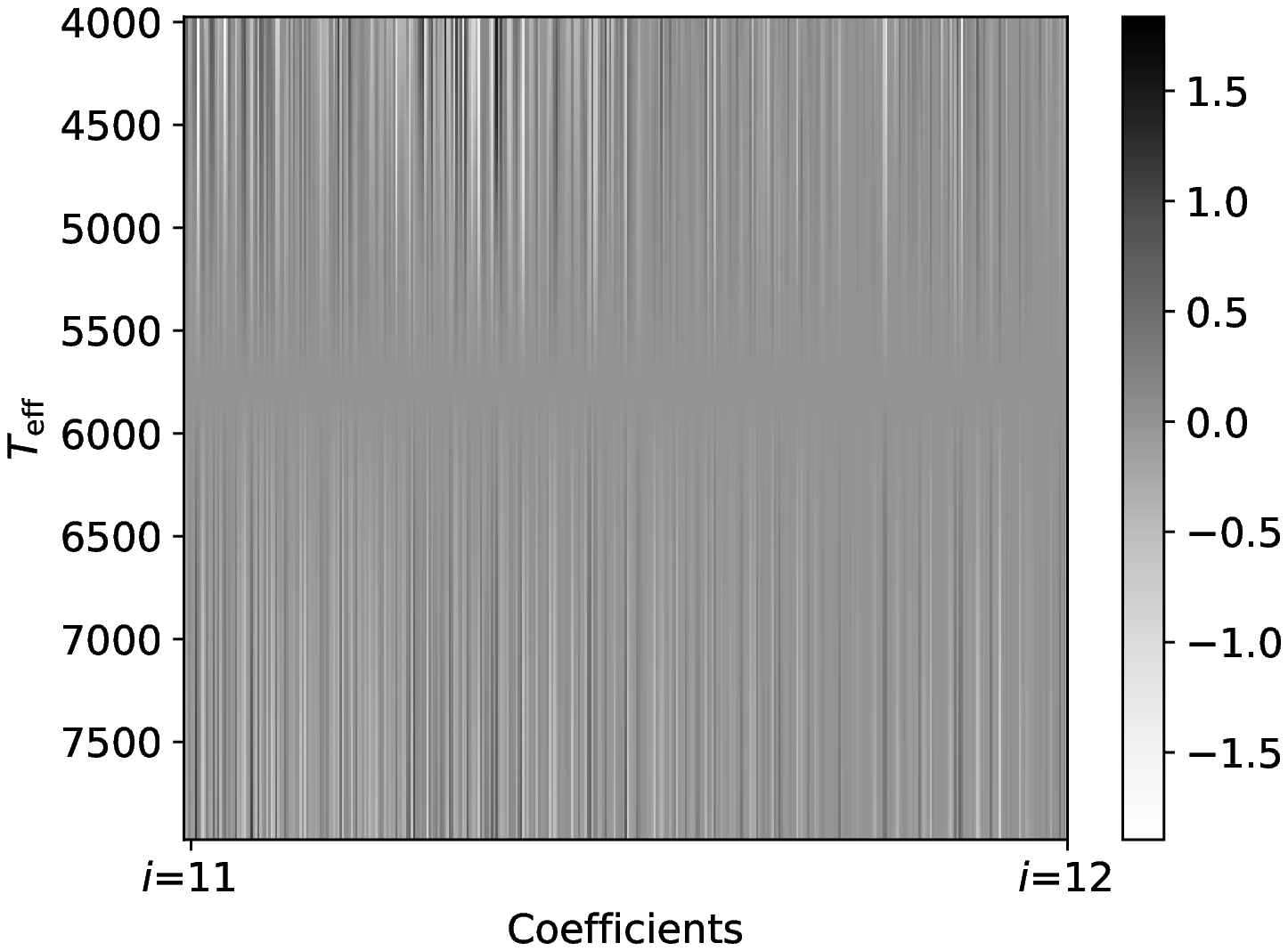} 
    \vspace{4ex}
  \end{subfigure}%% 
  \begin{subfigure}[b]{0.5\linewidth}
    \centering
    \includegraphics[width=\linewidth]{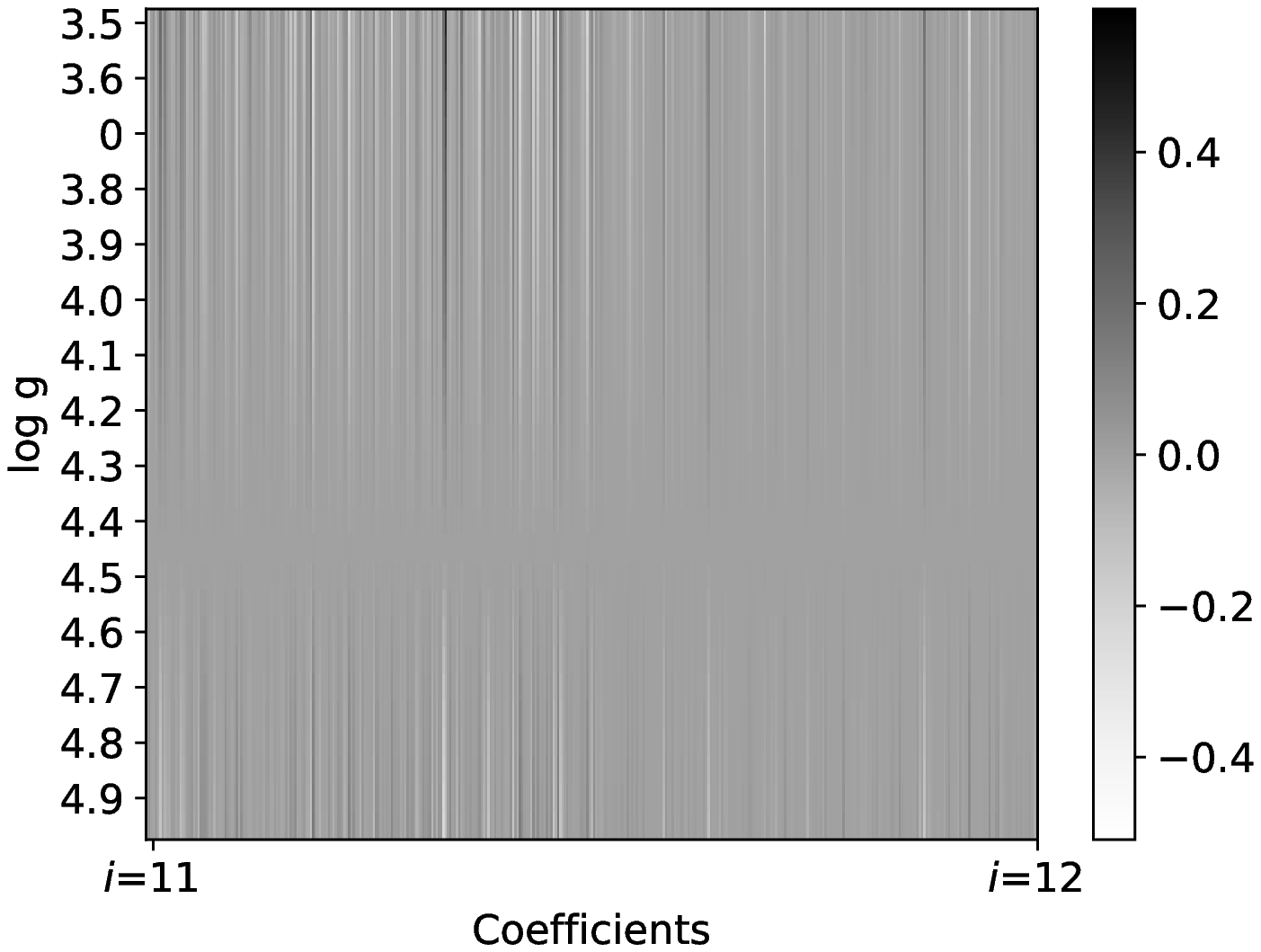} 
    \vspace{4ex}
  \end{subfigure} 
  \begin{subfigure}[b]{0.5\linewidth}
    \centering
    \includegraphics[width=\linewidth]{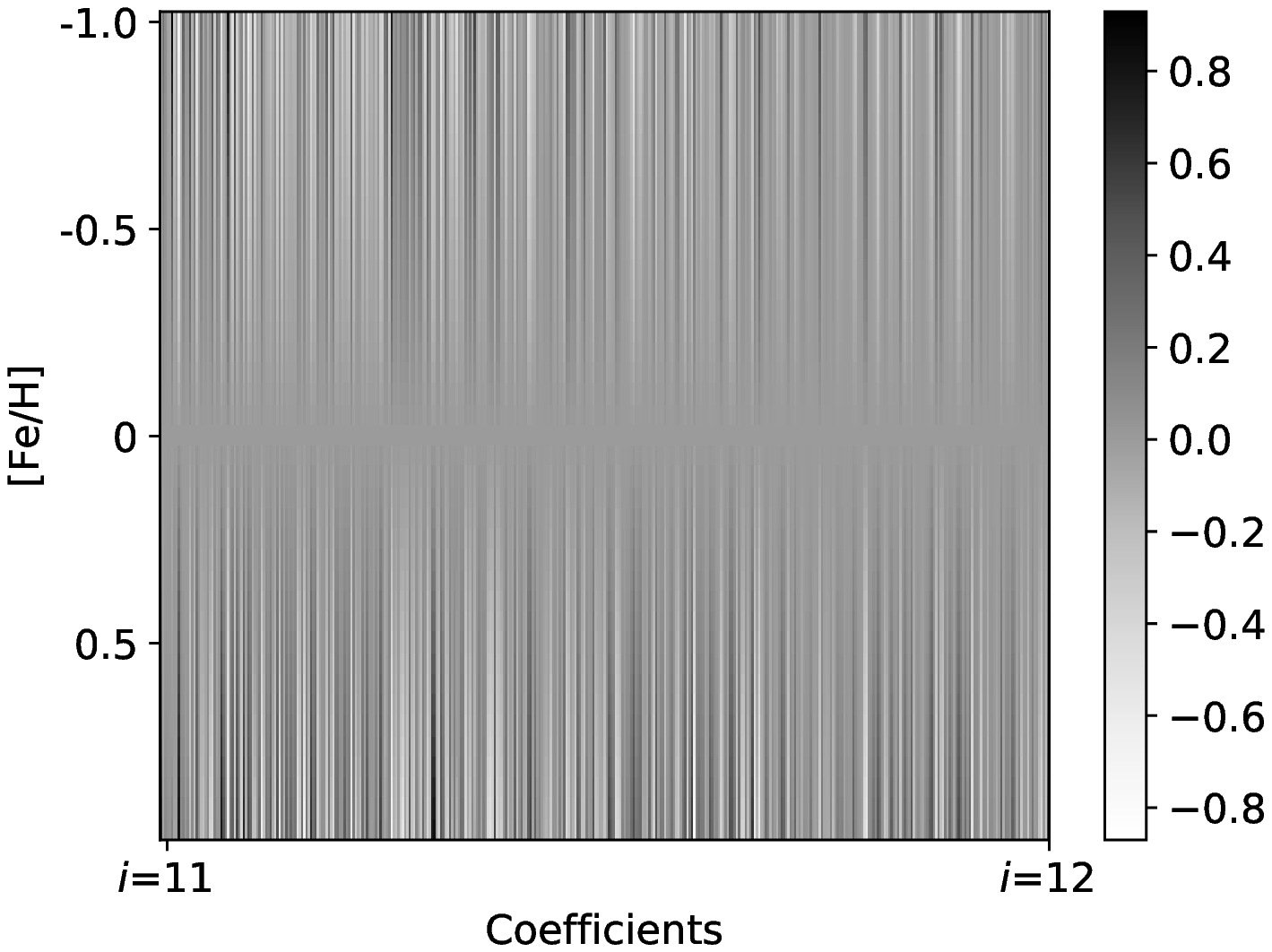} 
  \end{subfigure}%%
  \begin{subfigure}[b]{0.5\linewidth}
    \centering
    \includegraphics[width=\linewidth]{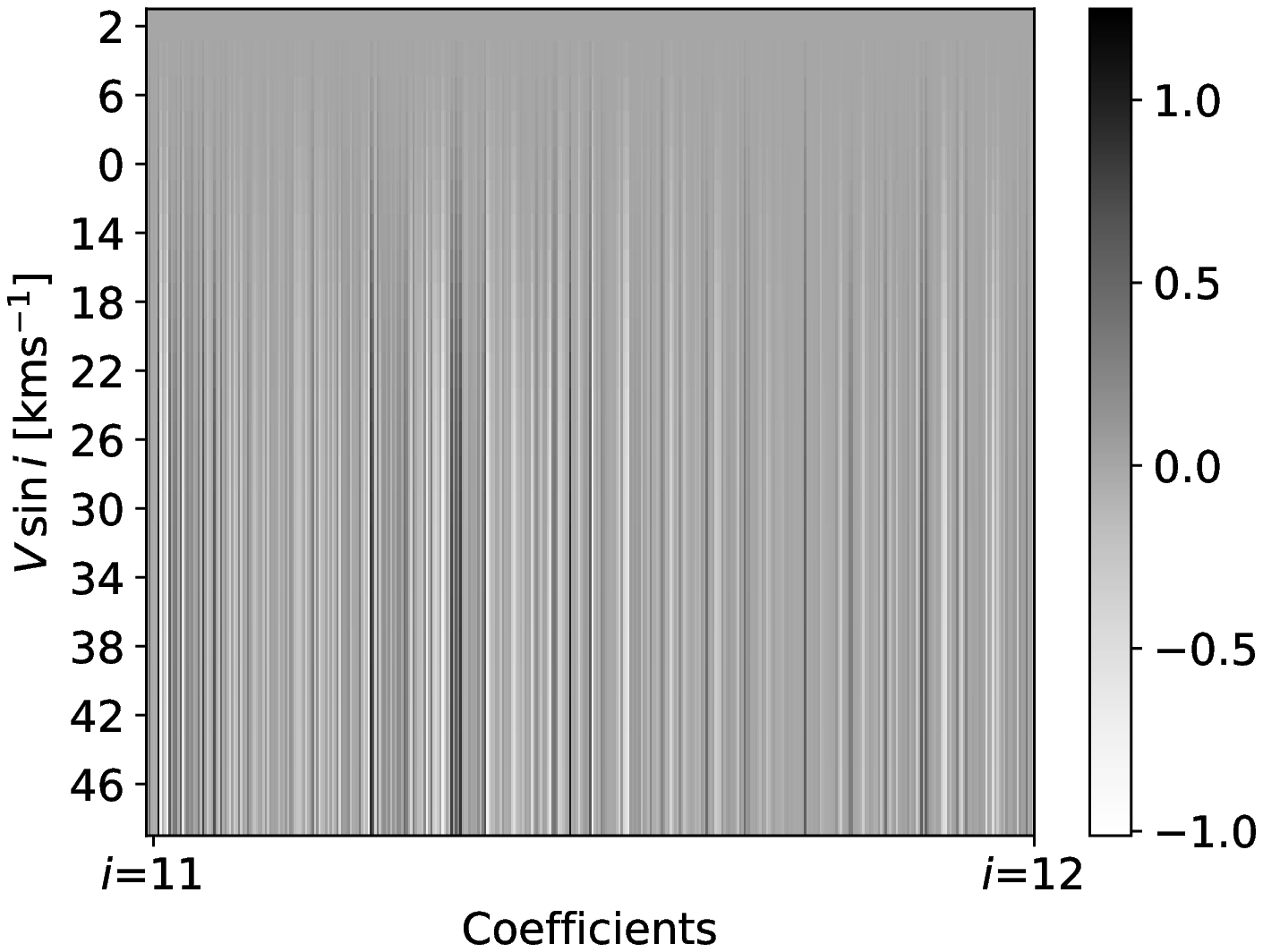} 
  \end{subfigure} 
  \caption{Changes in wavelet coefficients in the range $i=11-12$ for model spectra as a function of atmospheric parameters. The wavelet coefficients from a solar model has been subtracted to emphasise the subtle change in wavelet coefficients for each parameter. A similar result is seen for other values of $i$ between 4 and 14. The colour bar indicates the magnitude of the difference of coefficients. Note that these bars are not on the same scale and highlight the wavelet response to each coefficient.}
  \label{Wavelet_response}
  \end{figure*}

We demonstrate the sensitivity of wavelet coefficients to atmospheric parameters in Fig. \ref{Wavelet_response} for wavelet coefficients in the range $i=11-12$ ($0.04\,-\,0.09\, \rm nm$). We see a slow variation of some wavelet coefficients which correspond to changes in individual spectral line geometries as the each parameter changes.  One thing to note is the sensitivity of each parameter; $T_{\rm eff}$ varies the most, followed by $V \sin i$ and [Fe/H]. Surface gravity is the least varying parameter in wavelet space and is dominated by a few lines sensitive to $\log g$. As $V \sin i$ increases, we see positive and negative structures form and become stronger at higher $V \sin i$ values. This is likely to be a continuum effect as weaker lines are smeared out to average a lower continuum whilst stronger lines persist.
% * <p.maxted@keele.ac.uk> 2017-11-29T14:46:53.002Z:
%
% ^.

The choice of mother wavelet depends on the objective of the work.  A Daubechies wavelet performs well for frequency identification and is widely used in signal processing and data compression \citep[e.g.][]{Belmon2002}. A Haar wavelet, with a more step-like structure, is more suited to identifying discontinuity and is widely used in computer vision projects \citep[e.g.][]{Essaouabi2009}. We investigate the effects of wavelet choice on the determined atmospheric parameters in Sect. \ref{fe_offset}, but proceed with the Daubechies ($k$=4) wavelet for the rest of this paper.

% all of this is from:
%   http://www.intechopen.com/books/discrete-wavelet-transforms-theory-and-applications/discrete-wavelet-transfom-for-nonstationary-signal-processing

\section{Bayesian measurements} \label{Method}

\begin{figure}[ht!]
\centering
\includegraphics[width=0.5\textwidth]{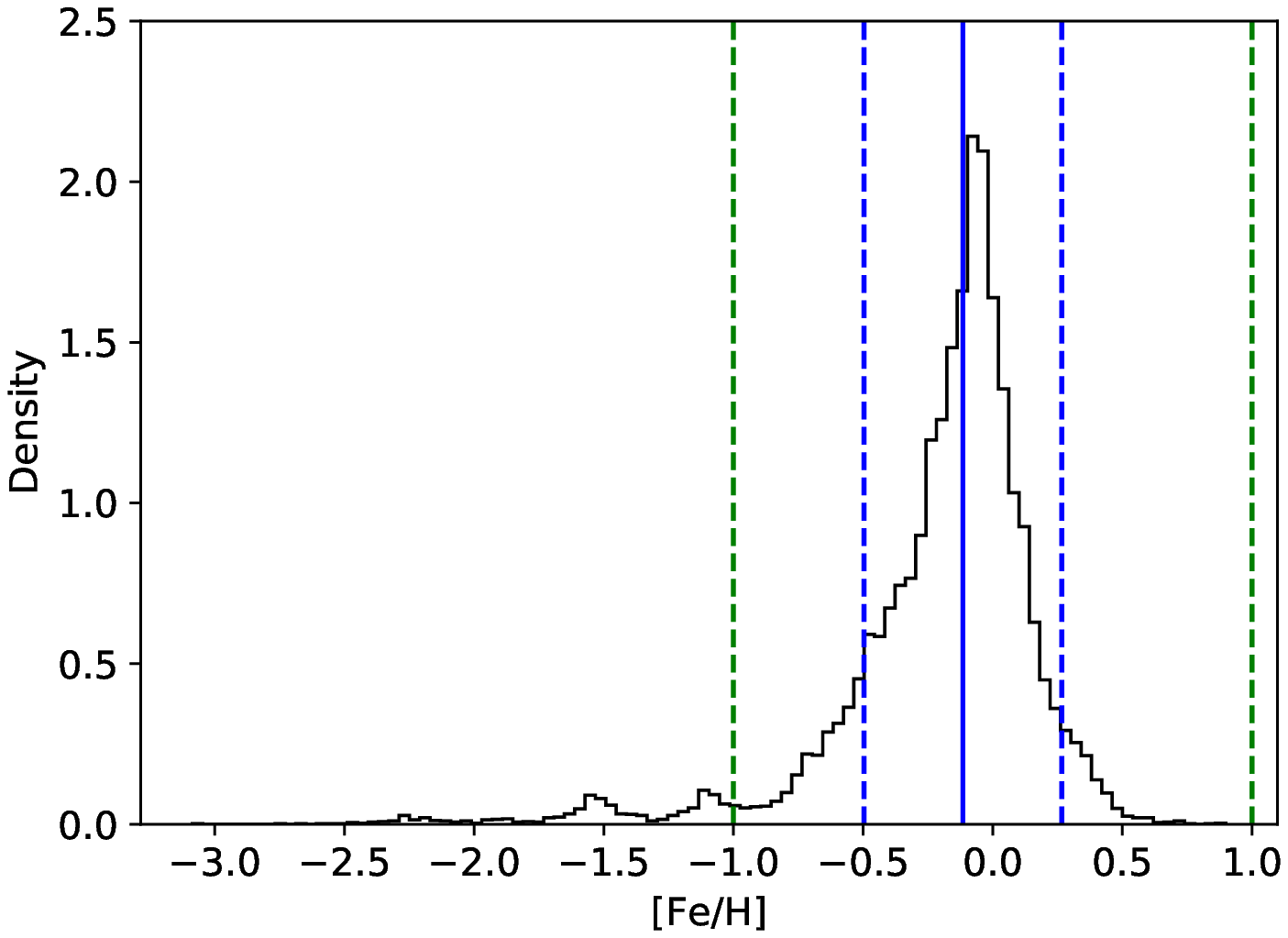}
\caption{Histogram of 14,681 [Fe/H] measurements for  stars from Gaia-ESO data release 3 \protect\citep{SmiljanicKornBergemannEtAl2014}. Plotted is the median value of [Fe/H] (solid blue), with $1 \sigma$ from the median (dashed blue). The grid range used in enclosed by the dashed green lines.}
\label{FE_H}
\end{figure}

We use the Markov chain Monte Carlo method to determine the posterior probability distribution for $T_{\rm eff}$, $\log g$, $V \sin i$ and [Fe/H] given an observed spectrum. Our method is a global $\chi^2$ fitting routine which compares subsets of wavelet coefficients ($i=4-14$) to those from a pre-synthesised grid of spectra. Our grid was synthesised with the radiative transfer code SPECTRUM \citep{Gray1994} using MARCS model atmospheres \citep{Gustafsson2008}, and version 5 of the GES (GAIA ESO survey)  atomic line list provided within iSpec \citep{Blanco-Cuaresma2016} with solar abundances from \cite{Asplund2009}. We computed models spanning 450nm--650nm over a temperature range of 4000 to 8000\,K in steps of 250K, $-$1 to +1\,dex in steps of 0.5 dex for [Fe/H] and 3.5 to 5\,dex in steps of 0.5 for $\log g$. We selected our range of [Fe/H] by looking at composition measurements of over 14,000 FGK stars from Gaia-ESO survey data release 3 \citep[Fig. \ref{FE_H};][]{SmiljanicKornBergemannEtAl2014}. We find that 96\% of stars with measurements of composition had [Fe/H] in the range $-$1 to 1\,dex. This range in [Fe/H] is also much larger than the full  range in [Fe/H] for our benchmark sample described in Sect. \ref{D15}.

Spectra in the grid are calculated with zero instrumental, rotational and macroturbulence broadening. These are accounted for in post-processing by convolving the grid spectra with the appropriate kernels. In this work, we allow $V \sin i$ to have values in the range $0$ - $50 \, \rm kms^{-1}$. The upper limit of $50\,kms^{-1}$ would need to be extended for hotter stars beyond the Kraft break, but it suitable for this work on late-type stars. Macroturbulence are estimated using equation (5.10) from \citet{Doyle2015} and microturbulence was accounted for at the synthesis stage using equation (3.1) from the same source. Spectra in-between grid points are extracted by trilinear interpolation, broadened to the desired value of $V \sin i$ and macroturbulence, and then convolved with a Gaussian to account for  instrumental broadening. For the self-consistency tests in Sect. \ref{self_con} instrumental broadening was ignored, but for the CORALIE spectra in Sect. \ref{D15} we used an instrumental resolving power $R = 55,000$ \citep{Queloz2001,Doyle2015}. We then re-sample between 450\,--\,650\,nm with 2$^{17}$ values (the same as the observed spectrum) and apply Eqn. (\ref{DWT}) to obtain the wavelet coefficients $WT_{f(\lambda)}(4-14,k)$ for the model spectra. 

The subset of wavelet coefficients from the interpolated model, $WT_{\textbf{m}}$, are compared to those from the data, $WT_{\textbf{d}}$, in the following Bayesian framework: the probability of observing a spectrum for a given model is given by $\rm p(\textbf{m}|\textbf{d})\propto \mathcal{L}(\textbf{d}|\textbf{m}) \rm p(\textbf{m})$. The vector of model parameters  is given by $\rm \textbf{m} = \left( \rm T_{\rm eff}, \rm [Fe/H], \log g, \rm V \sin i \right)$ and we assume uniform prior probability for the model parameters within the grid range. We use the likelihood function $\mathcal{L}(\textbf{d}|\textbf{m}) = \exp (-\chi^2/2)$ where

\begin{equation}\label{chi_squared}
\chi^{2} = \frac{(WT_{\textbf{d}} - WT_{\textbf{m}})^{2}}{\sigma_{WT_{\textbf{d}}}^2}
\end{equation}
and 

\begin{equation}\label{sigma}
\sigma_{WT_{\textbf{d}}}^2 = \beta \sigma_{MC}^{2}.
\end{equation}
The term $\sigma_{MC}^{2}$ was calculated by generating 1000 spectra from the co-added spectrum with noise generated from a standard normal distribution centred around $f(\lambda)$ and with $\sigma$ equal to the standard deviation of the spectrum, $\sigma_{f(\lambda)}$ (calculated from the standard deviation in co-added spectra). The blaze function is corrected prior to co-addition of the spectra and so deviations in blaze functions will result in uncertainties propagating through to $\sigma_{f(\lambda)}$, and $\sigma_{MC}^{2}$, effectively down-weighting regions with poor blaze corrections.  The free parameter $\beta$ has been introduced to account for additional noise, incomplete atomic data, deviations from solar metallicity scaling, lines which form under non-local thermodynamic equilibrium, and other unaccounted errors. In principle, we could be used stellar models or empirical relations to set priors on these atmospheric parameters but we have decided not to do this for two reasons. Firstly, allowing the MCMC sampler to explore regions with a-priori low probability gives a better indication of the reliability of our method than using a more constrained solution. Secondly, by imposing a prior from stellar models or empirical relations based on normal stars we may fail to identify interesting examples of anomolous stars in our sample, e.g., helium-rich stars. 
% * <p.maxted@keele.ac.uk> 2017-11-29T15:17:32.295Z:
%
% > A Hertzsprung-Russell diagram could yield prior information about which combinations of atmospheric parameters are more/less probable. Such stars (e.g. He rich stars) are unlikely, but we do not want to coerce  the sampler away from such solutions and since there are no issues with convergence,  we do not include prior information from a Hertzsprung-Russell diagram.
%
% ^.

We sample the model parameter space using the Markov chain Monte Carlo method, implemented by the python package {\sc{emcee}} \citep{Foreman-Mackey2013} . {\sc{emcee}} uses affine-invariant ensemble sampling \citep[parallel stretch move algorithm]{Goodman2010} to split Markov chains into sub-groups and update the position of a chain using the positions of chains in the other subgroups. The algorithms affine-invariance can cope with skewed probability distributions and generally has shorter autocorrelation times than a classic Metropolis-Hastings algorithm.

\begin{figure*}[ht!]
\centering
\includegraphics[width=\textwidth]{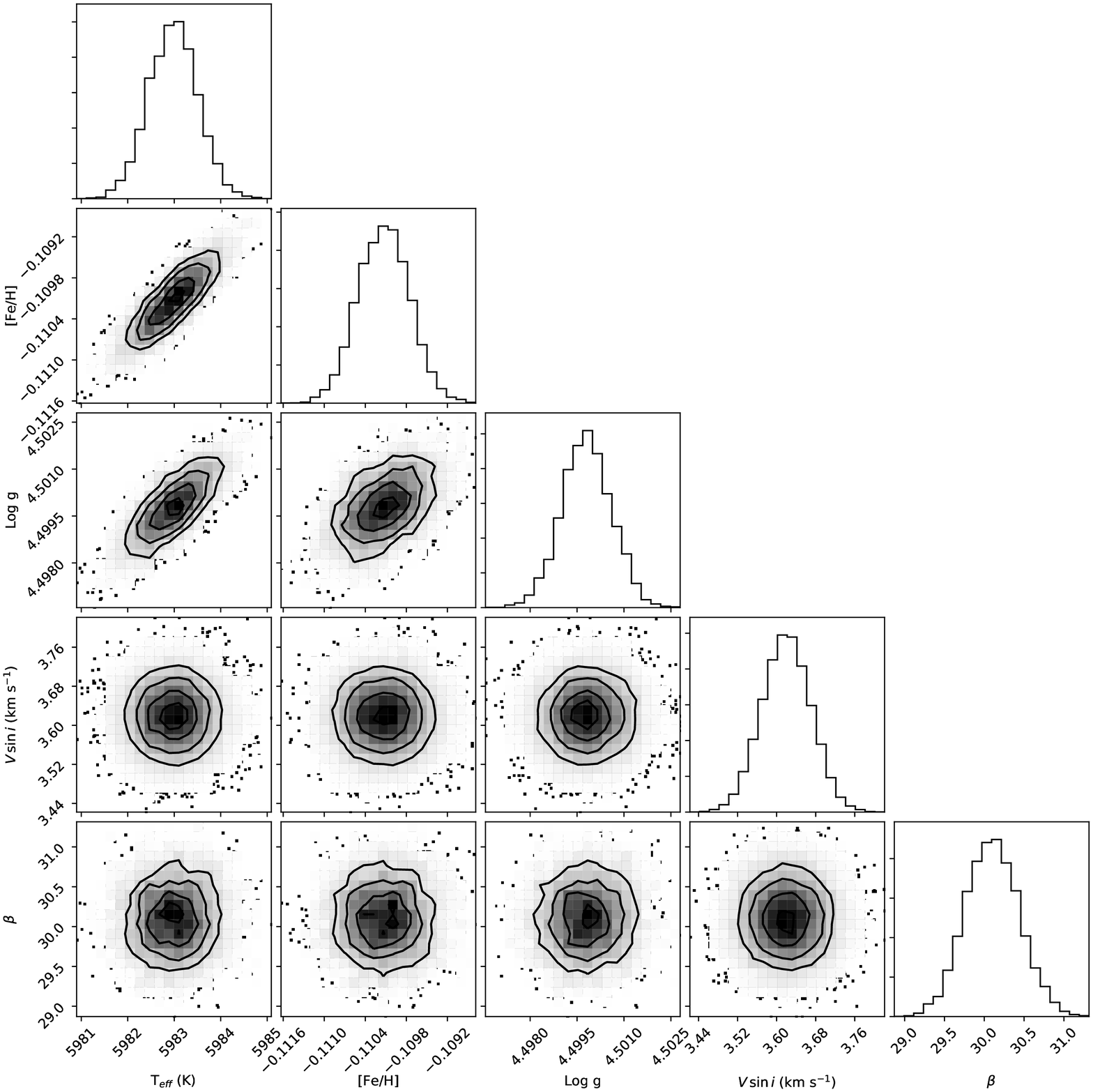}
\caption{Posterior proability distributions for WASP-20.}
\label{wavelet_corner_WASP20}
\end{figure*}

We generate 12 Markov chains of 20,000 draws each to converge on the best atmospheric parameters. We found that the chains converged before the 5000$^{th}$ draw, but as a precaution we discarded the first 10,000 draws. We take the median values of the model parameters in the remaining draws to determine the atmospheric parameters for a spectrum. An example posterior probability distribution for WASP-20 is plotted in Fig. \ref{wavelet_corner_WASP20}. The parameter space is almost symmetric with small degeneracies between $T_{\rm eff}$, [Fe/H] and $\log g.  $Note that the precision of the parameters determined from the standard deviation of each parameter in the Markov Chain is typically an underestimate of the true precision of these parameters because it does not account for systematic errors in the data or the models.

\section{Self consistency}\label{self_con}

 \begin{figure*}
\centering
 \begin{subfigure}[b]{0.5\linewidth}
    \centering
    \includegraphics[width=\linewidth]{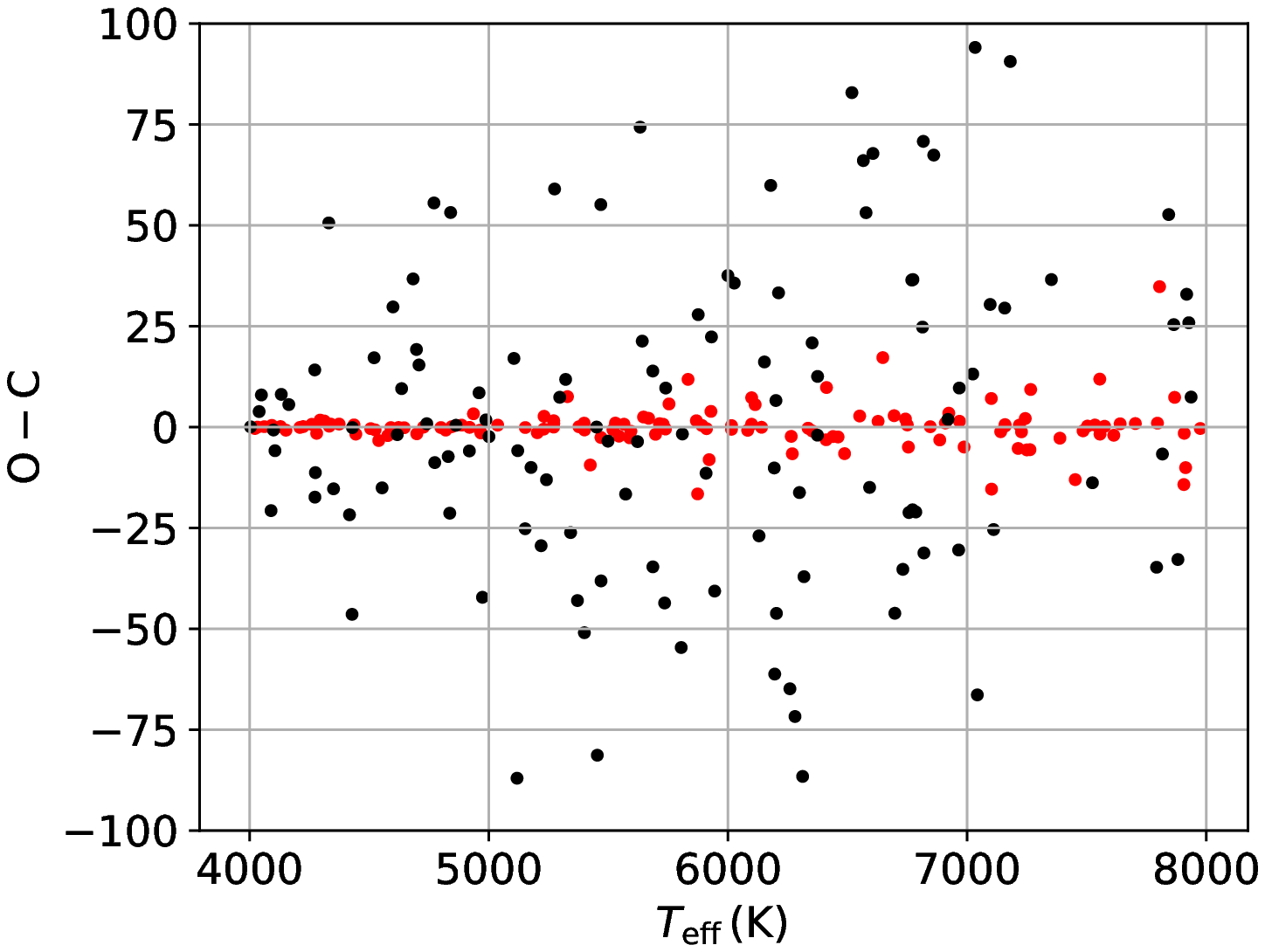} 
    \label{fig7:a} 
    \vspace{4ex}
  \end{subfigure}%% 
  \begin{subfigure}[b]{0.5\linewidth}
    \centering
    \includegraphics[width=\linewidth]{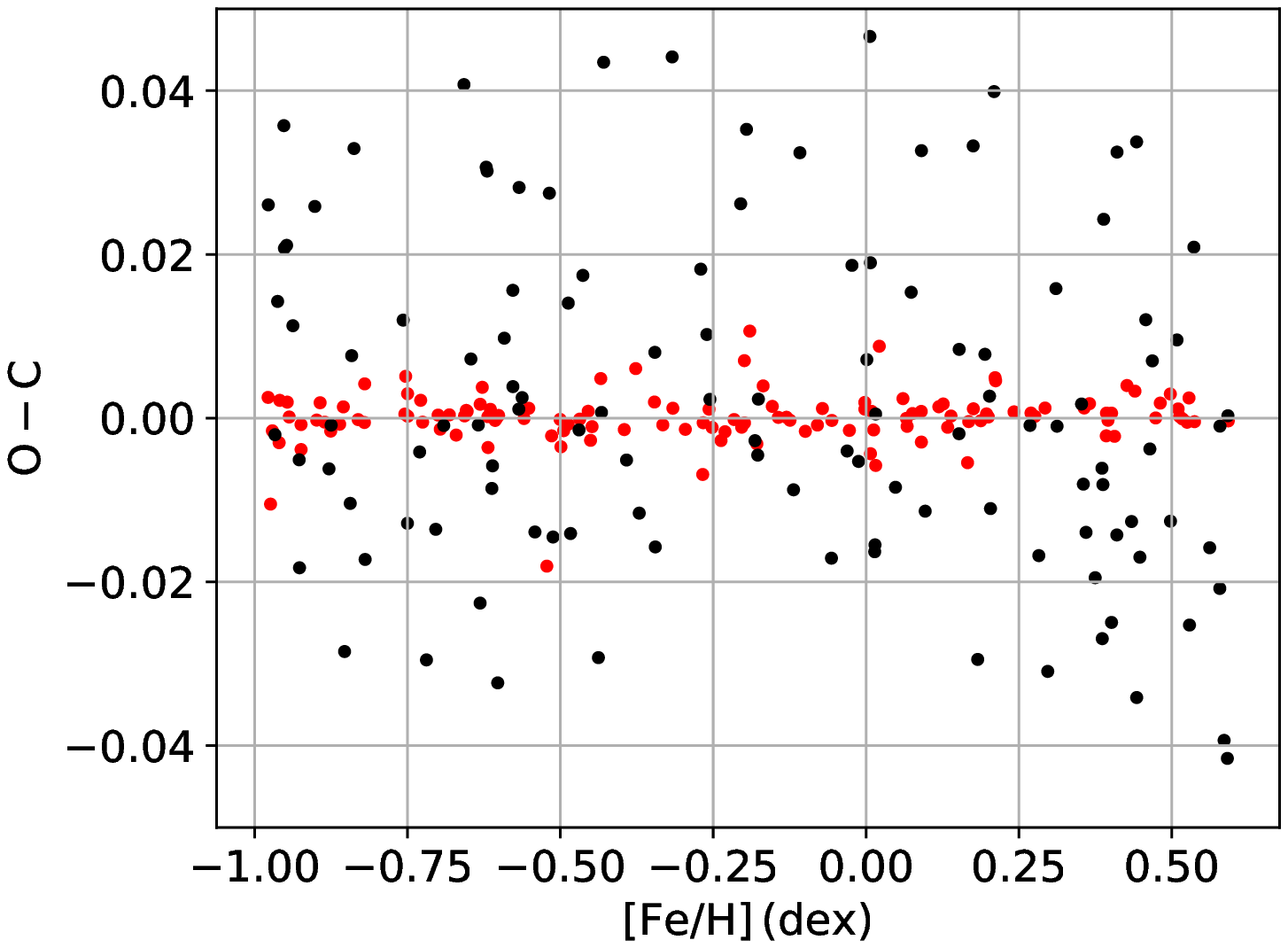} 
    \label{fig7:b} 
    \vspace{4ex}
  \end{subfigure} 
  \begin{subfigure}[b]{0.5\linewidth}
    \centering
    \includegraphics[width=\linewidth]{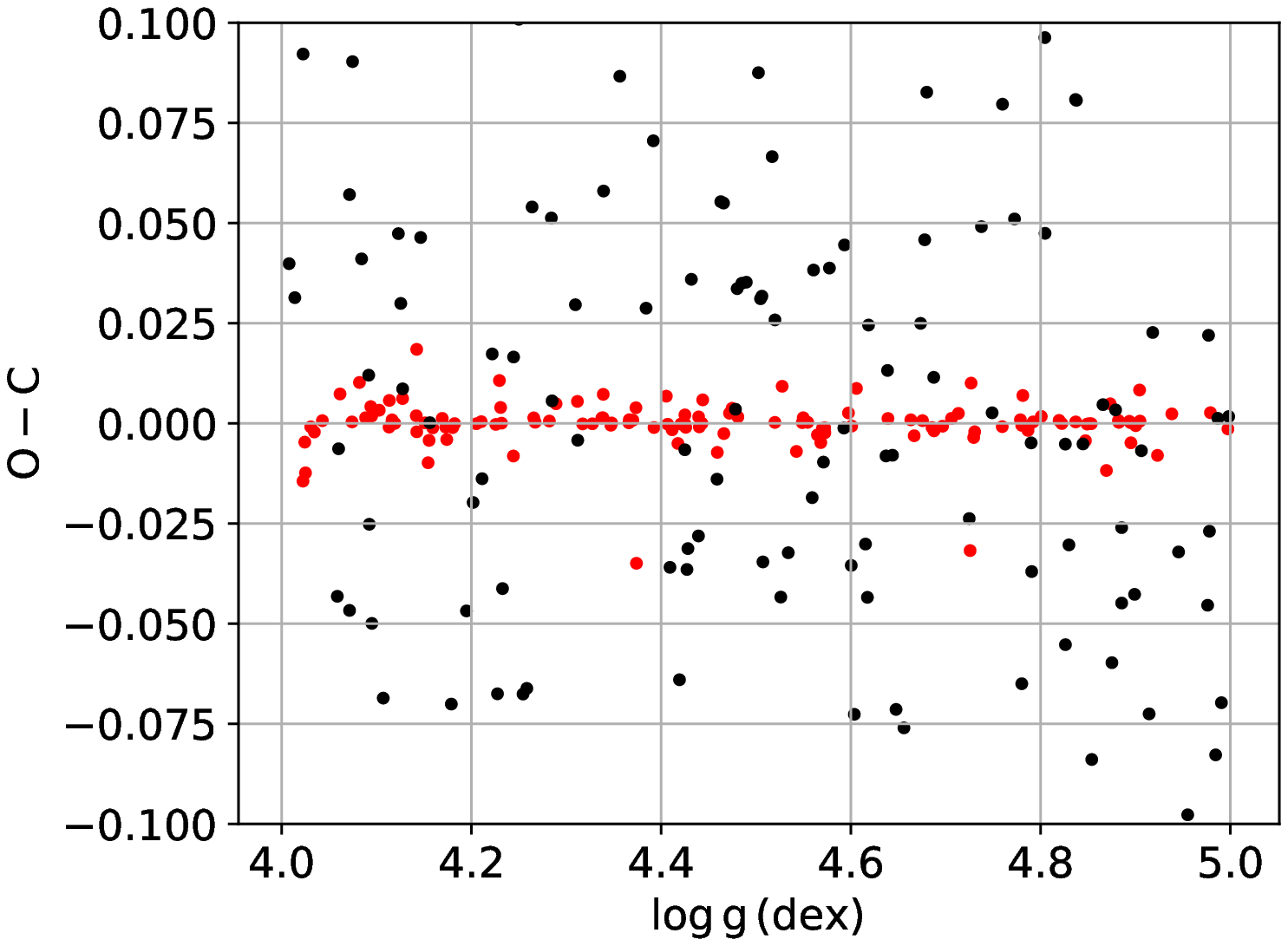} 
    \label{fig7:c} 
  \end{subfigure}%%
  \begin{subfigure}[b]{0.5\linewidth}
    \centering
    \includegraphics[width=\linewidth]{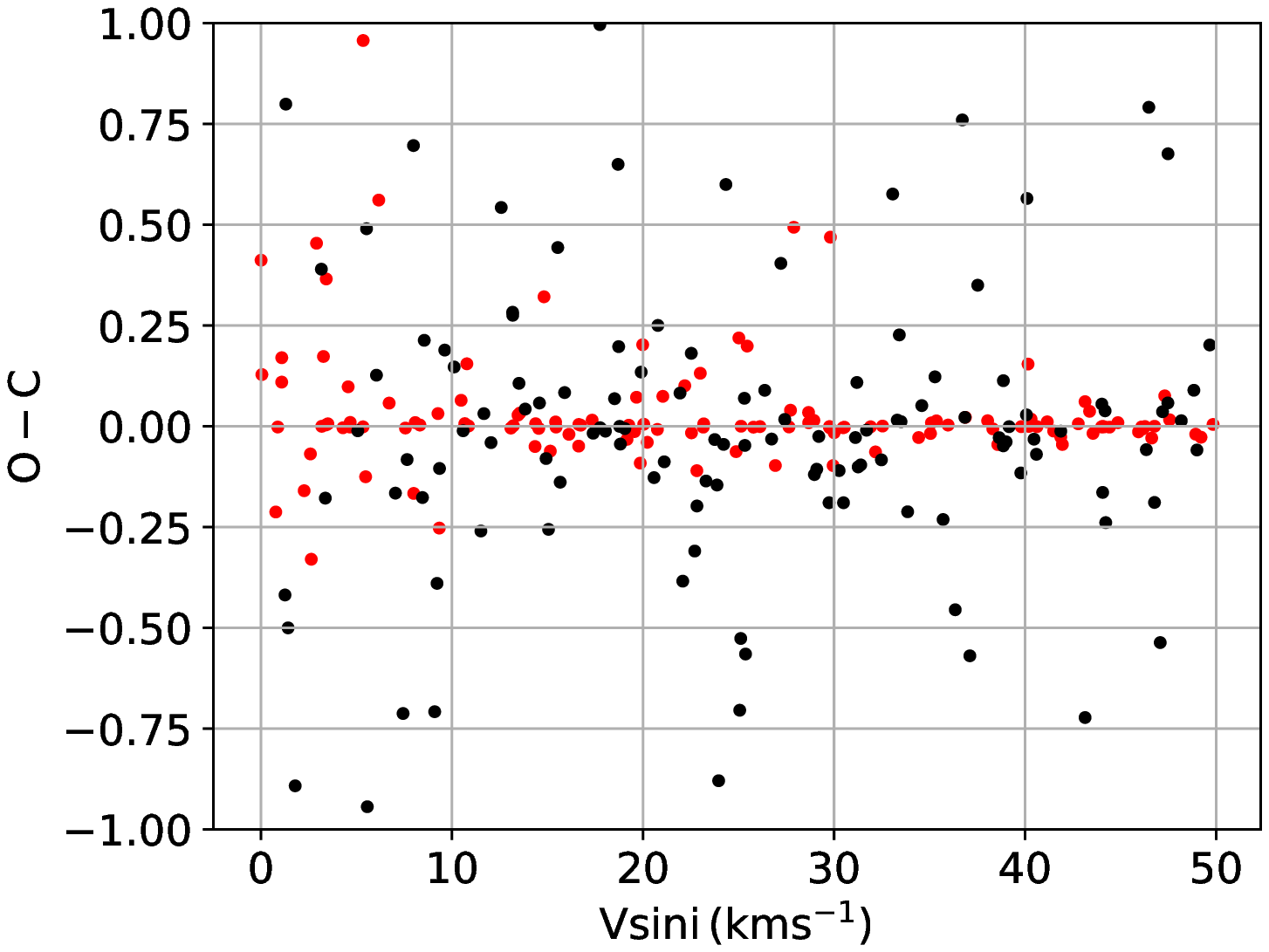} 
    \label{fig7:d} 
  \end{subfigure} 
  \caption{Differences between wavelet determined atmospheric parameters and those used to synthesize spectra with all parameters free (black) and with priors on $\log g$ (red). }

  \label{self}   
\end{figure*}

\begin{table}
\caption{The recovery of atmospheric parameters using the wavelet method for two groups of 256 spectra: one group with no priors on $\log g$ and another with priors imposed from transit photometry. The difference between the value measured by the wavelet method and the input value used to interpolate the spectrum ($\rm x_{\rm out}-\rm x_{\rm in}$) were used to calculate the standard deviation, $\sigma$, and mean offset, $\mu$.}              % title of Table
\label{self_cons_tab}      % is used to refer this table in the text
\centering                                      % used for centering table
\begin{tabular}{l c r r c c c c c c}          % centered columns (4 columns)
\hline\hline                        % inserts double horizontal lines
 & \multicolumn{1}{p{2cm}}{\centering Prior on \\ $\log g$?} & $\sigma$ & $\mu$ & \\
\hline    
$T_{\rm eff}$ (K) & no & 46.0 & -3.2 \\
 & yes & 3.1  &  0.2 \\
$\rm [Fe/H]$ (dex) & no & 0.040 & -0.003 \\
 & yes & 0.020 & -0.001\\
$V \sin i$ (kms$^{-1}$) & no & 0.47 & 0.05 \\
 & yes & 0.17   & -0.06 \\
$\log g$ (dex) & no & 0.060 & -0.002\\
 & yes & 0.020 & 0.001\\

\hline                                             %inserts single line
\end{tabular}
\end{table}

We have assessed the ability of our method to recover atmospheric parameters from synthetic spectra in order to check that our results are self consistent. We interpolated 512 spectra with random values of $T_{\rm eff}$, [Fe/H], $\log g$ and $V \sin i$ selected within the limits of our grid of models. Each spectrum was then re-sampled to have $2^{17}$ values in the range 450--650\,nm to match our choice of coefficients in Sect. \ref{Wavelet Analysis} and the benchmark sample in Sect. \ref{D15}. These spectra were split into two groups and analysed with the aforementioned method. The first group had $\log g$ as a free parameter to probe for any systematics, for the second group we imposed a prior on $\log g$ to simulate the effect of well constrained surface gravity measurement from transit photometry. The  $\log g$ prior probability distribution was assumed to be Gaussian with a mean $\log g$ value equal to the value used to interpolate the spectrum and a dispersion equal to the average uncertainty of transit $\log g$ values from \citet[][hereafter referred to as M13]{Mortier2013} for 44 WASP exoplanet hosts ($\overline{\sigma_{\rm \log g}}$ = 0.02 dex). We decided not to add Gaussian noise to these spectra as noise profiles depend upon stellar parameters and instrumental conditions; this is assessed in Sec. \ref{spec_quality}. We find typical autocorrelation lengths are below 1000 steps for all parameters in the first chain and 12 chains in the second run typically produce an acceptance fraction between $\sim\,0.25\,-\,0.3$.  

The recovery of atmospheric parameters for both groups is shown in Fig.  \ref{self} and summarised in Table \ref{self_cons_tab}. We find that all parameters are recovered well across the range our grid. With no constraints on $\log g$, there were only 2 measurements of $T_{\rm eff}$ that deviated from the input value by more than 150\,K. A prior on $\log g$ significantly decreases the difference between measured and input atmospheric parameters and shows that our method is sensitve to $\log g$. There is a small increase in residual scatter for measurements of $V \sin i$ when the interpolated value of $V \sin i$ below 0.5\,km\,s$^{-1}$; this is seen in both groups and marginally improved with a prior on $\log g$. This is expected as the resolution of the broadening kernel in combination with the edge of parameter space makes it difficult to determine low $V \sin i$ values. The internal precision associated with the wavelet method is remarkably high; by taking $1\,\sigma$ values from the cumulative probability distributions we find precisions around 15\,K, 0.01\,dex, 0.02\,dex, and 0.15\,$\rm km\,s^{-1}$ for $T_{\rm eff}$, [Fe/H], $\log g$, and $V \sin i$ respectively. More realistic uncertainties are determined in the following sections. 

\begin{figure*}
\centering
 \begin{subfigure}[b]{0.5\linewidth}
    \centering
    \includegraphics[width=\linewidth]{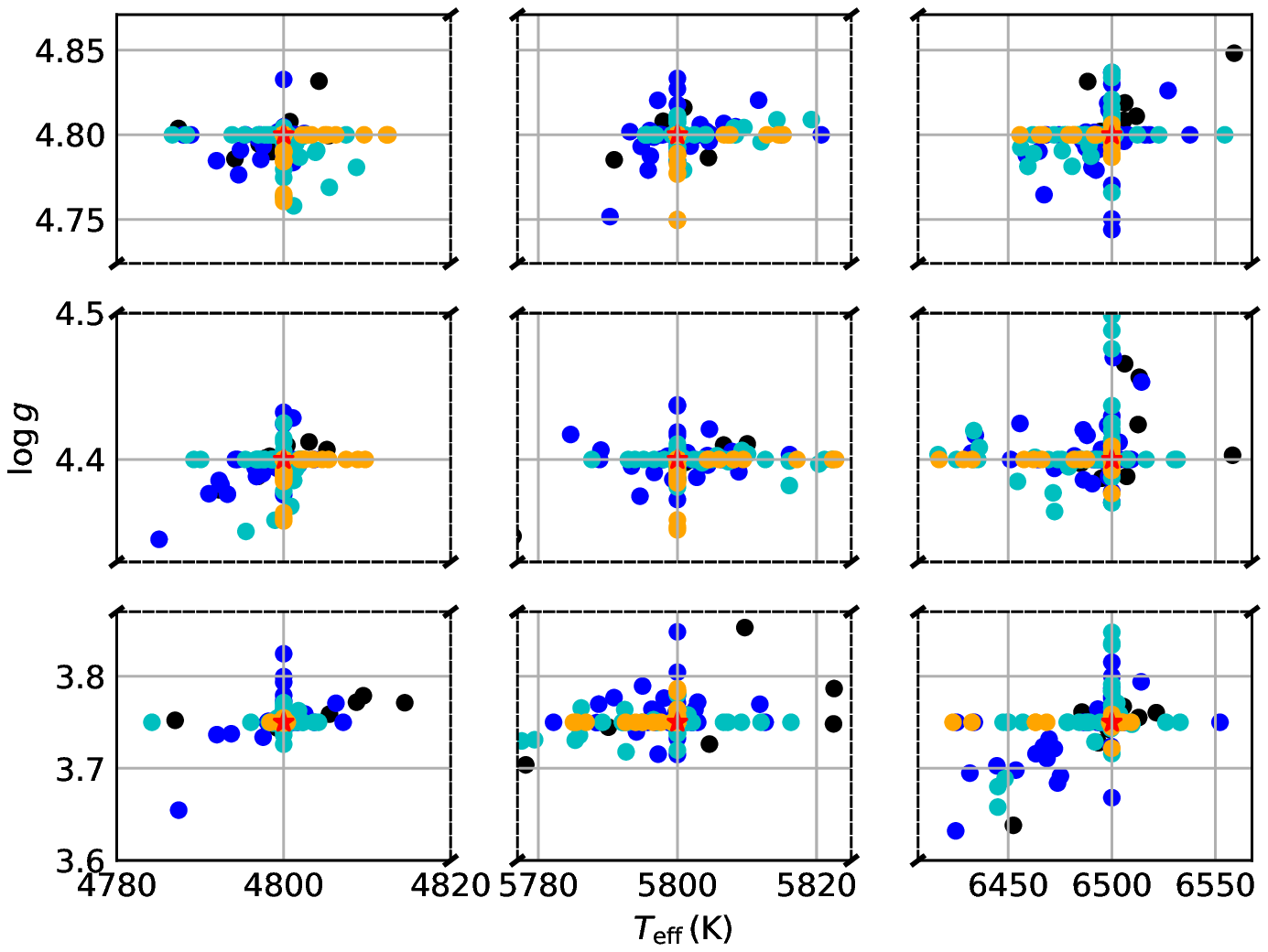} 
    \vspace{4ex}
  \end{subfigure}%% 
  \begin{subfigure}[b]{0.5\linewidth}
    \centering
    \includegraphics[width=\linewidth]{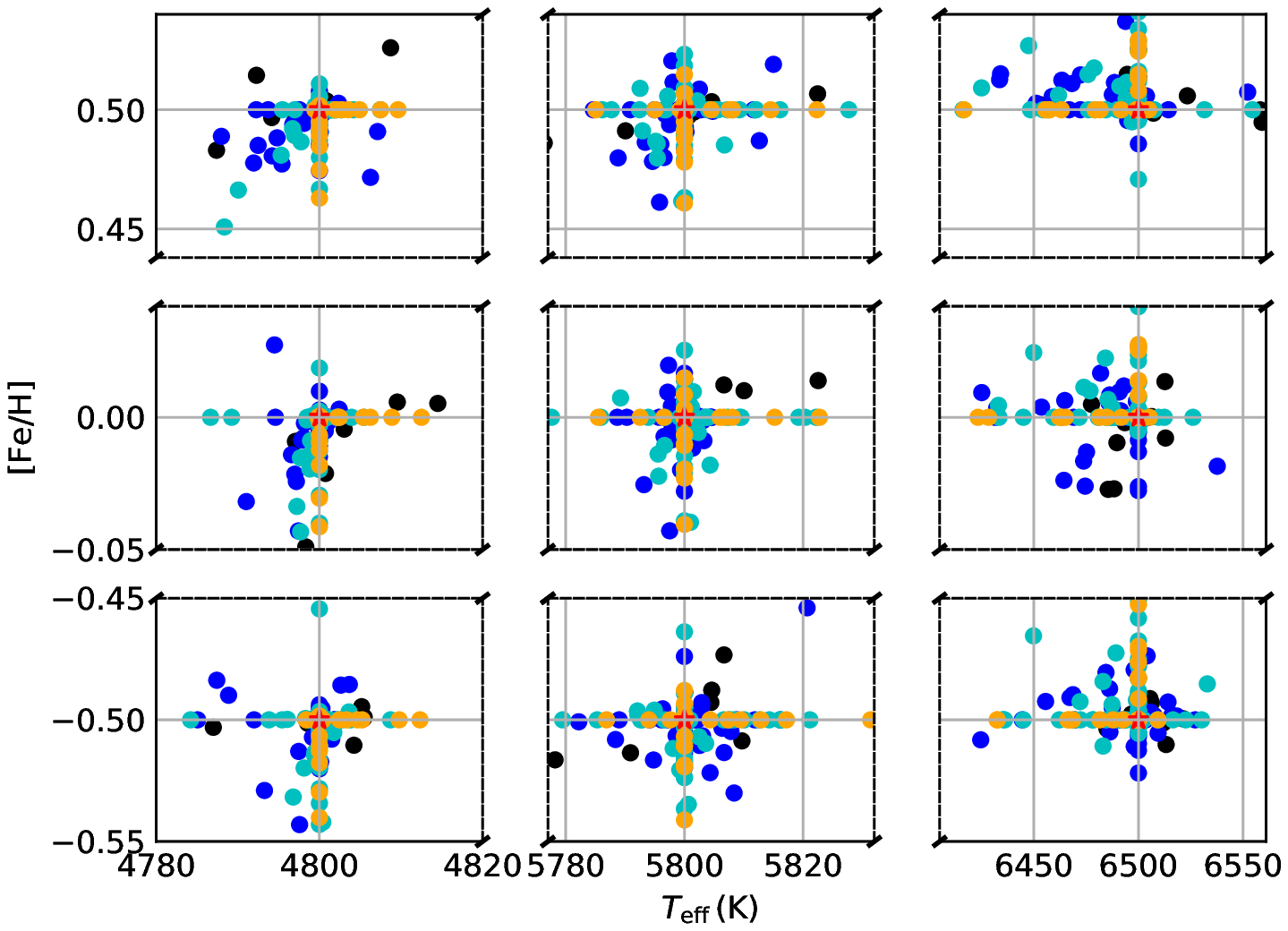} 
    \vspace{4ex}
  \end{subfigure} 
  \caption{Difference between atmospheric parameters determined by the wavelet method for 9 synthetic spectra when some parameters are fixed. We plot atmospheric parameters determined with with all parameters free ($T_{\rm eff}$, [Fe/H], $\log g$, $V \sin i$) in black; with one parameter fixed in blue; with two parameters fixed in cyan; three parameters fixed in orange.  }
  \label{self_2}   
\end{figure*}

We also assess the sensitivity of our method by determining the atmospheric parameters of 9 spectra from a discrete set of grid points with different combinations of fixed parameters. We interpolate 96 spectra from the following grid points -- $4800$ K, $5800$ K and $6500$ K for $T_{\rm eff}$; $-0.5$, $0.0$ and $0.5$ $dex$ for [Fe/H]; $3.75$, $4.40$ and $4.80$ $dex$ for $\log g$; $5$, $10$ and $15\,kms^{-1}$ for $V \sin i$. In total, we determined the atmospheric parameters  for each spectrum 15 times with the wavelet method using every combination of free and fixed parameters (see Fig. \ref{self_2}). We find that constraining one or more parameters increases the internal precision of the wavelet method significantly. A slight degeneracy exists between $T_{\rm eff}$ and $\log g$ resulting in a modest scatter when both parameters are left free. This also highlights the numerical noise introduced by starting walkers at different positions, since walkers explore parameter space by random jumps which may never reach the correct solution, despite prior knowledge that an exact solution lies somewhere within the grid.

\section{Benchmark sample}\label{D15}

Any spectral analysis technique must be tested against stars with high-quality measurements. For this  we use stars from  \citet{Doyle2013} and \citet[][; D15, hereafter]{Doyle2015}. The D15 sample consists of 24 stars analysed by measurements of equivalent width and spectral fitting of high signal-to-noise and high-resolution ($\rm R\,=\,112,000$) data from the HARPS spectrograph \citep{Queloz2001}. We used lower-quality observations from the CORALIE spectrograph to determine $T_{\rm eff}$, [Fe/H], $\log g$ and $V \sin i$ of the same stars with the wavelet method. Only 22 stars in the D15 sample have CORALIE spectra available to use and we further exclude WASP-77A and the close (3") B-component as both component as they are un-resolved in the CORALIE fibre. This leaves a sample of 20 stars for use to calibrate our method (see Table \ref{WASP_STARS}).

\subsection{CORALIE spectra} \label{data}

%converges and detail sample
% HARPS
% dont have wasp 4 - SOPHIE
% dont have WASP 21 - 
% exclude && A and B

Each spectrum was processed with the CORALIE standard data reduction pipeline \citep{Baranne1996}. The radial velocity shift was measured relative to a solar template\footnote{From The Gaia Benchmark Stars Library pipeline which is the result of co-adding asteroid observations by NARVAL.} and corrected into a laboratory frame of reference. The spectra were then median-combined onto an identically sampled wavelength grid. Continuum regions were identified by applying maximum and median filters \citep{Blanco-CuaresmaSoubiranHeiterEtAl2014} and fitted with spline functions (1 every 10nm) for normalisation. The wavelet method was then applied to each spectrum twice: once with no priors on $\log g$ and a second time with priors given by transit photometry. The priors on $\log g$ were set to those from M13 if quoted, or the relevant discovery papers otherwise (see Table \ref{waspstars}).  

\begin{table}[ht!]
\caption{Our benchmark sample of FGK stars from D15. We include the V magnitude, the number of spectra and the signal-to-noise ratio (SNR) of the coadded spectra at 500\,nm.}              % title of Table
\label{WASP_STARS}      % is used to refer this table in the text
\centering                                      % used for centering table
\begin{tabular}{l r r r r}          % centered columns (4 columns)
\hline\hline                        % inserts double horizontal lines
 Star & V mag & \multicolumn{1}{p{2cm}}{\centering \# of \\ spectra} & \multicolumn{1}{p{2cm}}{\centering SNR\\ ($\sim$500 nm)} \\
\hline    
WASP-4  & 12.50 & 12 & 37 \\
WASP-5  & 12.30 & 11 & 35 \\
WASP-6  & 11.90 & 30 & 63 \\
WASP-7  &  9.50 & 13 & 124 \\
WASP-8  &  9.79 & 21 & 137 \\
WASP-15 & 11.00 & 15 & 83\\
WASP-16 & 11.30 & 19 & 77\\
WASP-17 & 11.60 & 42 & 71\\
WASP-18 &  9.30 &  5 & 119 \\
WASP-19 & 12.59 & 28 & 50\\
WASP-20 & 10.68 & 58 & 153 \\
WASP-22 & 12.00 & 29 & 63\\
WASP-23 & 12.68 & 38 & 53\\
WASP-24 & 11.31 & 18 & 53\\
WASP-29 & 11.30 & 14 & 57\\
WASP-30 & 11.90 & 47 & 27\\
WASP-31 & 11.70 & 35 & 53\\
WASP-53 & 12.19  & 35 & 40\\
WASP-69 &  9.88 & 21 & 136 \\
WASP-80 & 11.90 & 37 & 51\\
\hline                                             %inserts single line
\end{tabular}
\end{table}

\subsection{Results}\label{D15_results}

 \begin{figure*}
\centering
 \begin{subfigure}[b]{0.5\linewidth}
    \centering
    \includegraphics[width=\linewidth]{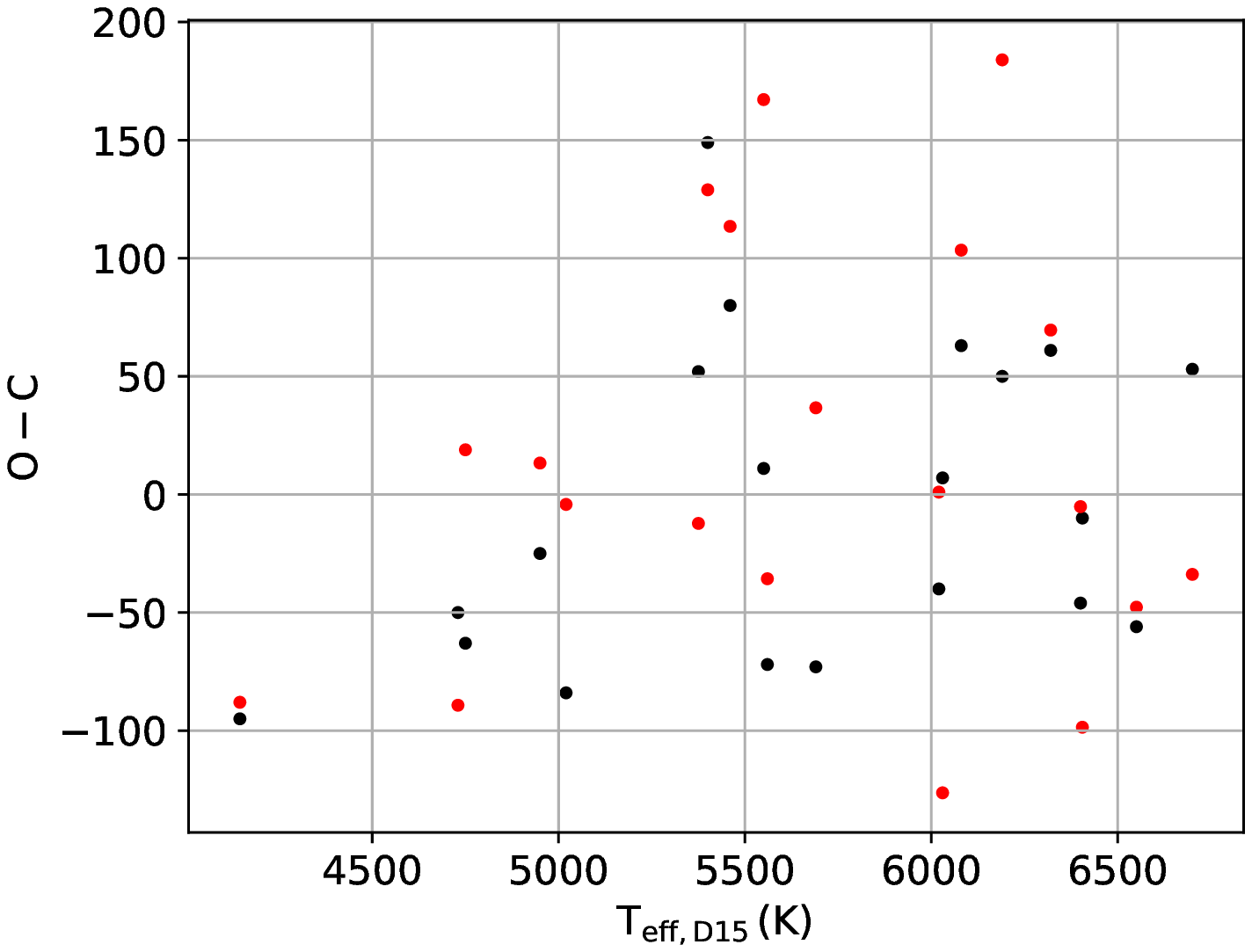} 
    \caption{} 
    \label{doyle:a} 
    \vspace{4ex}
  \end{subfigure}%% 
  \begin{subfigure}[b]{0.5\linewidth}
    \centering
    \includegraphics[width=\linewidth]{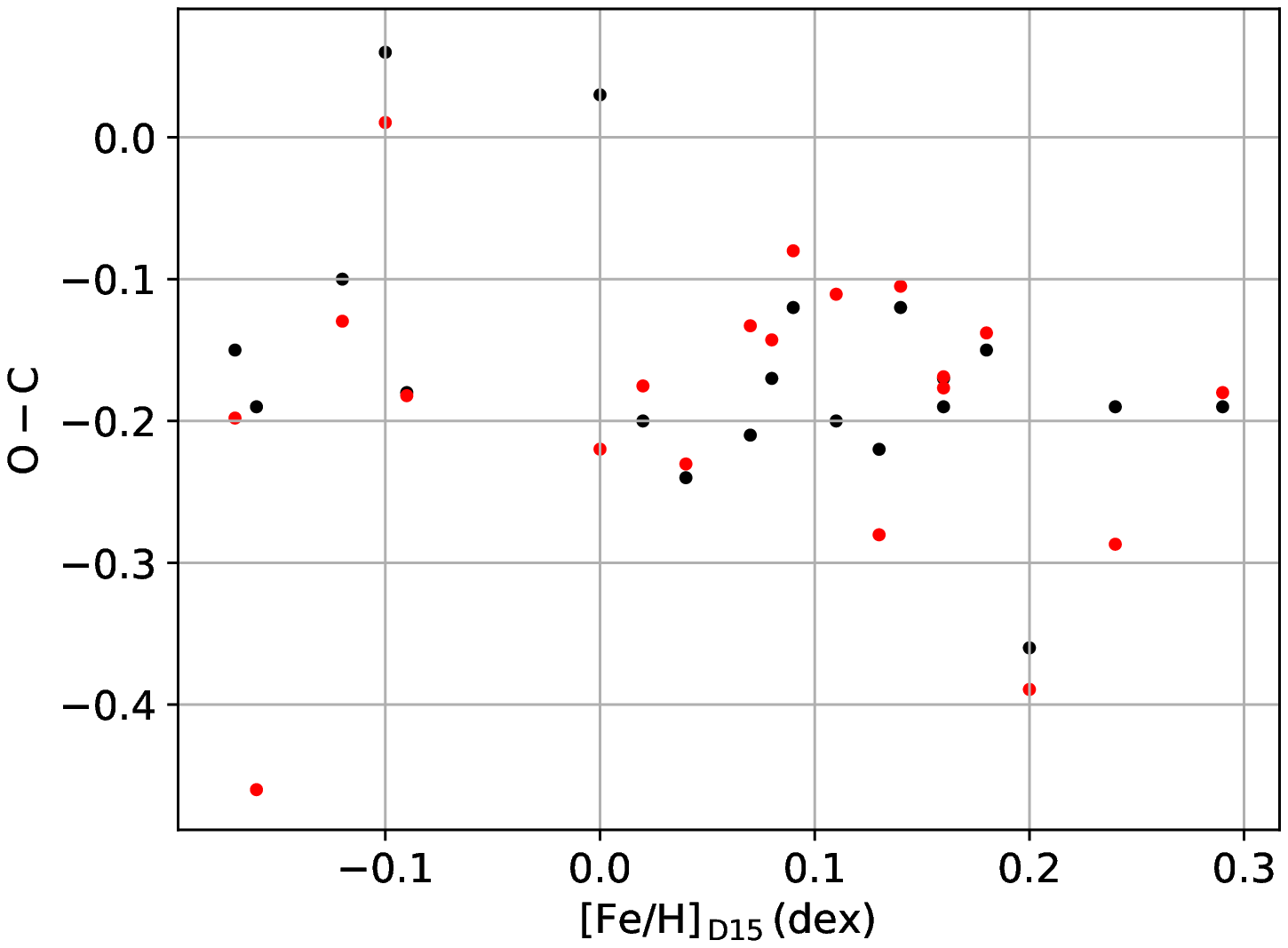} 
    \caption{} 
    \label{doyle:b} 
    \vspace{4ex}
  \end{subfigure} 
  \begin{subfigure}[b]{0.5\linewidth}
    \centering
    \includegraphics[width=\linewidth]{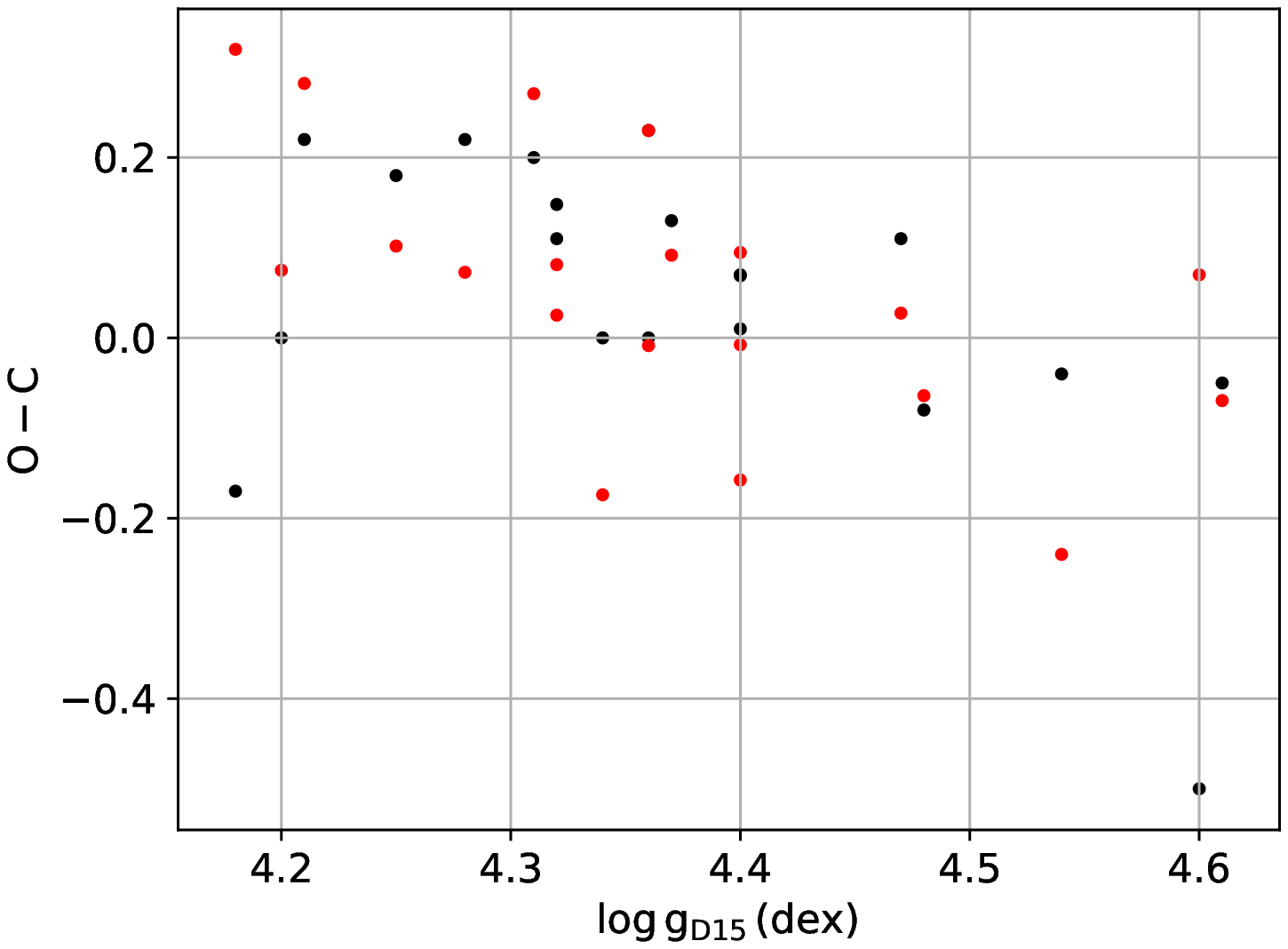} 
    \caption{} 
    \label{doyle:c} 
  \end{subfigure}%%
  \begin{subfigure}[b]{0.5\linewidth}
    \centering
   \includegraphics[width=\linewidth]{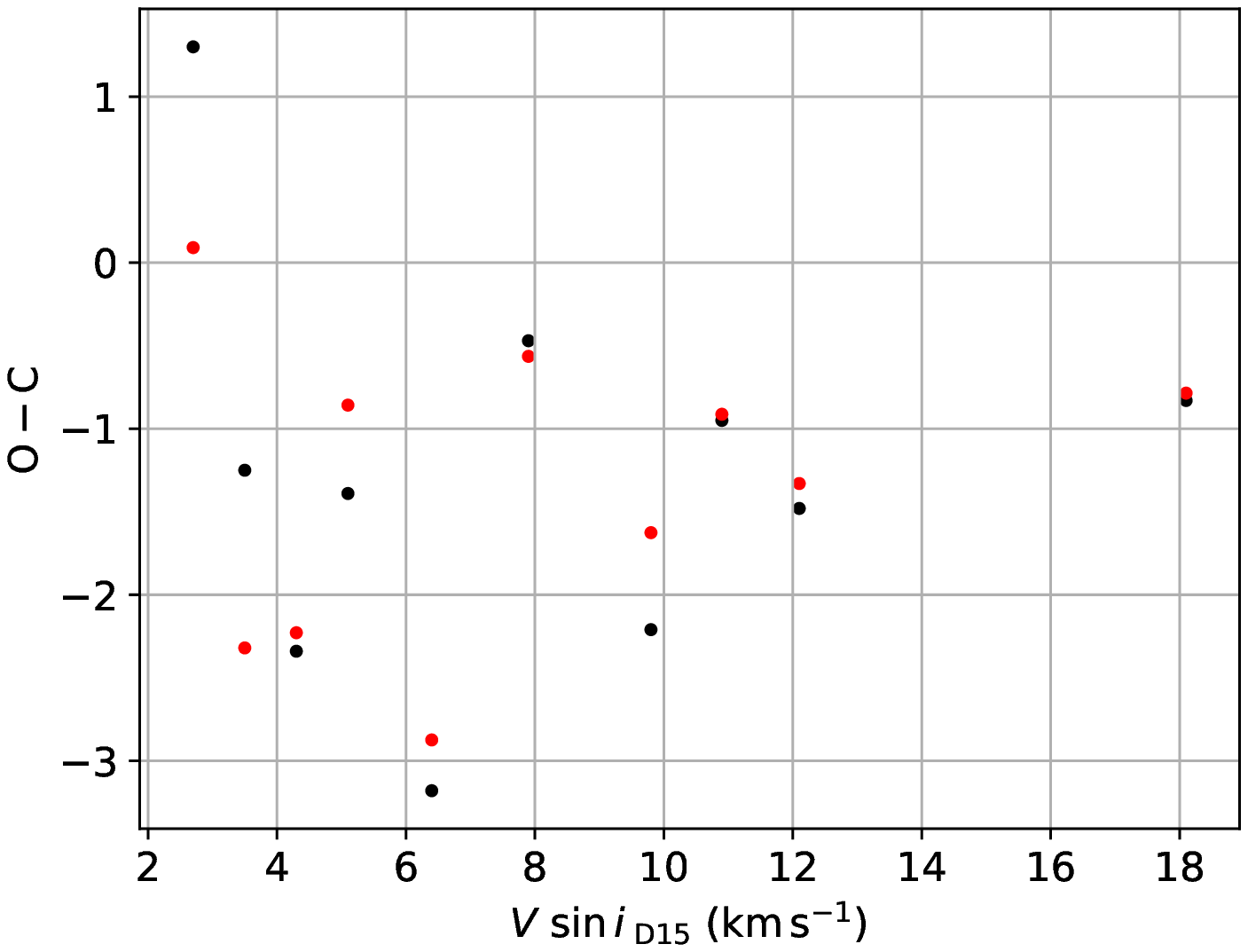} 
    \caption{} 
    \label{doyle:d} 
  \end{subfigure} 
  \caption{Difference between wavelet analysis and D15 (O-C) for each atmospheric parameter in the D15 sample. Each spectrum was measured twice, once with $\log g$ as a free parameter (black) and again with $\log g$ priors imposed from transit photometry (red).  We exclude measurements of $V \sin i$ where macroturbulence, $\xi_t$, was set to $0\,\rm km\,s^{-1}$ to ensure models a best model was converged upon.}
  \label{doyle}   
\end{figure*}

\begin{table}
\caption{Recovery of atmospheric parameters for 20 FGK stars from D15: one group with no priors on $\log g$ and another with priors from transit photometry. The difference between the value measured by the wavelet method and D15 ($x_{wavelet}-x_{D15}$) are used to calculate the mean dispertion, $\sigma$, and mean offset $\mu$.}              % title of Table
\label{doyle_tab}      % is used to refer this table in the text
\centering                                      % used for centering table
\begin{tabular}{l l r r c c }          % centered columns (4 columns)
\hline\hline                        % inserts double horizontal lines
 & \multicolumn{1}{p{2cm}}{\centering Prior on \\ $\log g$?} & $\sigma$ & $\mu$ & \\
\hline    
$T_{\rm eff}$ (K) & no & 85.00 &  31.00 \\
 & yes & 86.00  &  14.00 \\
$\rm [Fe/H]$ (dex) & no & 0.06 & $-$ 0.15 \\
 & yes & 0.10 & $-$ 0.18 \\
$V \sin i$ (kms$^{-1}$) & no & 1.35  & $-$ 0.79 \\
 & yes & 0.62 &  $-$ 1.33 \\
$\log g$ (dex) & no & 0.13 & 0.08 \\
 & yes & 0.14 &  0.05 \\\\\hline                                             %inserts single line
\end{tabular}
\tablefoot{Values of $\sigma$ and $\mu$ for $V \sin i$ excluded stars where macroturbulence, $\xi_t$, was set to $\rm 0\, \rm km\,s^{-1}$.}
\end{table}

The results can be seen in Fig. \ref{doyle} and are summarised in Tables \ref{doyle_tab} and \ref{waspstars}. Our method determines $T_{\rm eff}$ to within 220\,K of the value found by D15. Our measurements of [Fe/H] are systematically offset by approximately\,$-$\,0.18\,dex from those of D15; this is discussed further in Sect. \ref{fe_offset}. It is difficult to constrain $\log g$ spectroscopically and our measurements often differ from those of D15 by up to 0.5 dex. Our measurements of $V \sin i$ converge to 0\,km\,s$^{-1}$ for 7 stars in the sample due to an over estimation of $v_{mac}$ or instrumental resolution. To mitigate this problem we repeated the analysis with $v_{mac}$ = 0\,km\,s$^{-1}$. This allowed these stars to converge on best fitting models without pushing against the edge of parameter space. These stars are marked with (*) in Table \ref{waspstars}. 

\begin{figure}[ht!]
\centering
\includegraphics[width=0.5\textwidth]{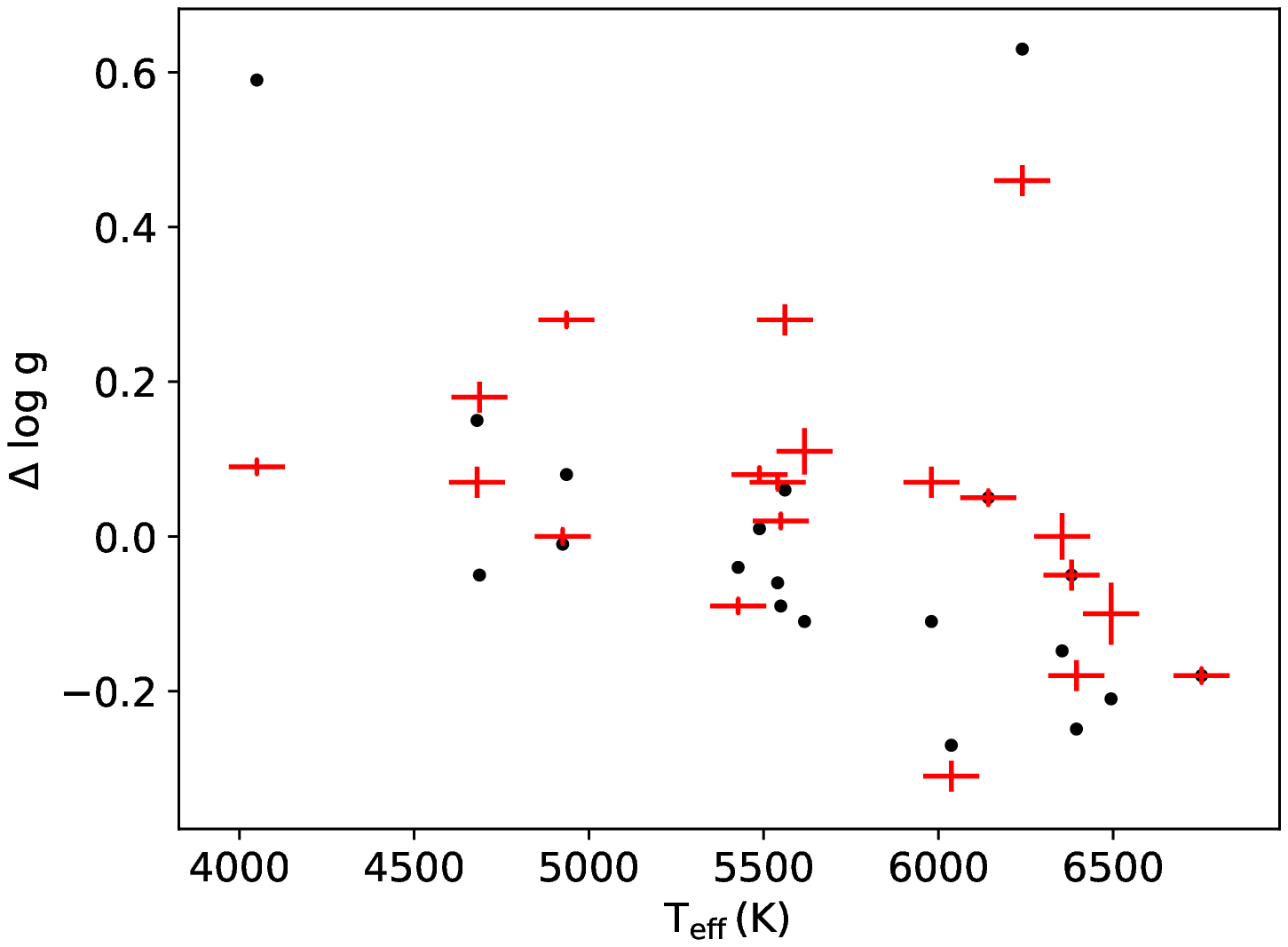}
\caption{The difference between spectroscopic $\log g$ and photometric $\log g$ ($\log g_{ph}$ - $\log g_{sp}$) correlated with $\rm T_{\rm eff, \rm wavelet}$ from this work (black) and from D15 (red).}
\label{mortlogg}
\end{figure}

We find no benefit by using priors on $\log g$. In most cases the use of $\log\,g$ priors increases the standard deviation in differences between atmospheric parameters from our method and published values. We investigated the level of agreement between spectroscopic values of $\log g$ from EW measurements (D15), $\log g$ from the wavelet method and those from transit photometry. Photometric surface gravity is typically measured to better precision than its spectroscopic counterpart, but relies on stellar models and correct limb-darkening parameters which in-turn rely on a constrained effective temperature, composition and surface gravity. Recent work suggests a disagreement between spectroscopic and photometric $\log g$ which is correlated with $T_{\rm eff}$ (see Fig. 4 from \cite{Doyle2017}). We compare the difference between spectroscopic and photometric $\log g$ measurements in Fig. \ref{mortlogg}. We find a statistically significant negative correlation (p-value $\leq$ 10$^{-5}$) between $\Delta \log g$ ($\log g_{\rm photometry} - \log g_{\rm wavelet}$ ) with $T_{\rm eff}$ from our method. The origin of this is unclear, but a similar trend is seen between spectroscopic and asteroseismic measurements \citep[Fig. 6 of][]{Mortier2014} which suggests to us that this is a problem to spectral analysis of late-type stars using plane-parallel non-LTE model atmospheres. For a few stars, we relaxed the $\log g$ prior to have a standard deviation of $0.2$ dex (instead of $0.02$ dex) and found almost no difference between these solutions and those with a uniform prior on $\log g$.

 %when common linelists and atmospheric parameters are used \citep{Hinkel2016}, and we attribute 
 
  %We note that temperature is poorly sampled outside the range between 4750 K - 6750 K and so we limit conclusions outside this range and apply caution when models converge to the edges of our grid of temperatures.
  
  \begin{figure*}[ht!]
\centering
\includegraphics[width=\textwidth, height = 0.4\textwidth]{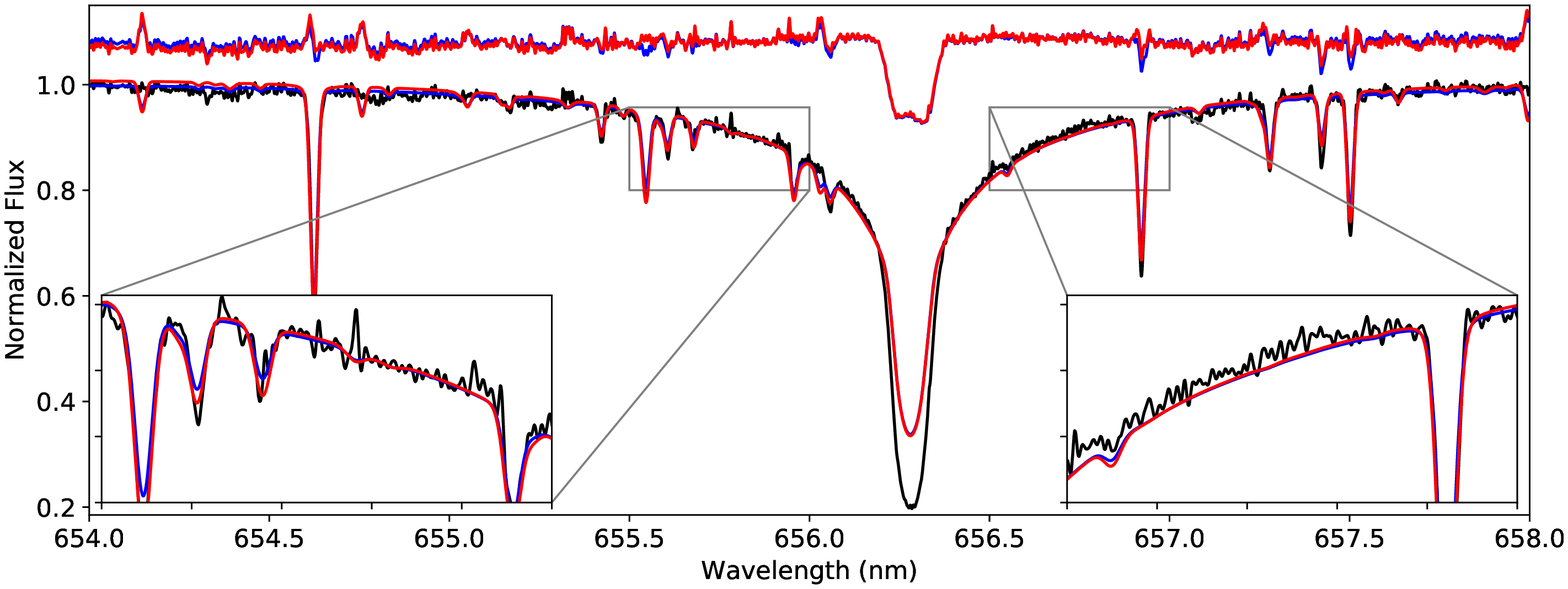}
\caption{The H-$\alpha$ region for WASP-20 (black) fitted with the best fitted model from D15 (red) and the best model from this work (blue). The near horizontal lines at flux ~ 1.2 are the residuals between the D15 model (red) or wavelet model (blue) and the spectrum of WASP-20.}
\label{WASP-20Halpha}
\end{figure*}

\iffalse
  \begin{figure}[ht!]
\centering
%\includegraphics[width=0.5\textwidth]{Felines.eps}
\caption{Fe lines for WASP-20 alongside the best fitting model from D15 (red) and that from this work (blue). We enlarge one of the cores of an Fe line to highlight that the D15 line-depths are a better match than those found by the wavelet method. }
\label{FElines}
\end{figure}
\fi

In Fig. \ref{WASP-20Halpha} we asses the  H$\alpha$ region for the model predicted from D15 and this work for the highest quality spectrum in our sample, WASP-20, with SNR $= 150$. The results from D15 were obtained using a custom line list, whereas we use version 5 of the GES atomic line list provided with iSpec to synthesise the D15 model of WASP-20 using atmospheric parameters, $\nu_{\rm mac}$ and $\nu_{mic}$ from D15. We find both models agree well with the data, with the left wing fitting best and a underestimation in the right wing. The discrepancies between the two wings of the  H$\alpha$ line seen here are the result of the difficulty in calibrating the blaze function in this region of the spectrum.  We find that the majority of Fe lines depths are under-predicted with the wavelet method, with the D15 model better matching individual line profiles. This test demonstrates the need to benchmark against well studied stars and visually inspect the best models against the data.

\subsubsection{Systematic offset in [Fe/H]}\label{fe_offset}

\begin{table}
\caption{The performance of the wavelet method using different mother wavelets. Each analysis was performed on WASP-7 using the same method used in Sect. \ref{D15}. }              % title of Table
\label{wavelet_tab}      % is used to refer this table in the text
\centering                                      % used for centering table
\begin{tabular}{l r r r r c c c c c}          % centered columns (4 columns)
\hline\hline                        % inserts double horizontal lines
Wavelet
& \multicolumn{1}{p{1cm}}{\centering $T_{\rm eff}$  \\ (K)}
& \multicolumn{1}{p{1cm}}{\centering [Fe/H]  \\ ($dex$)}
& \multicolumn{1}{p{1cm}}{\centering $\log g$ \\ ($dex$)}
& \multicolumn{1}{p{1cm}}{\centering $V \sin i$  \\ (km\,s$^{-1}$)} \\
\hline
Daubechies k=4 & 5983 & 0.11 & 4.50 & 17.54\\
Daubechies k=20 & 5975 & 0.11 & 4.36 & 17.55 \\
Harr k=2  & 5962 & 0.11 & 4.34 & 17.52\\
bspline k=20 & 5961 & 0.10 & 4.36 & 17.77\\
\hline                                             %inserts single line
\end{tabular}
\end{table}

\begin{table}
\caption{The regional performance of the wavelet method on WASP-20 using a variety of wavelength ranges. No priors for $\log g$ were used.}              % title of Table
\label{regional_tab}      % is used to refer this table in the text
\centering                                      % used for centering table
\begin{tabular}{l r r r r c c c c c}          % centered columns (4 columns)
\hline\hline                        % inserts double horizontal lines
Range 
& \multicolumn{1}{p{1cm}}{\centering $T_{\rm eff}$  \\ (K)}
& \multicolumn{1}{p{1cm}}{\centering [Fe/H]  \\ ($dex$)}
& \multicolumn{1}{p{1cm}}{\centering $\log g$ \\ ($dex$)}
& \multicolumn{1}{p{1cm}}{\centering $V \sin i$  \\ (km\,s$^{-1}$)} \\

\hline
450\,--\,500\,nm & 5984 & $-$\,0.17 & 4.31 & 3.98  \\
500\,--\,550\,nm & 6076 & $-$\,0.06 & 4.33 & 3.76 \\
550\,--\,600\,nm & 5530 & $-$\,0.34 & 4.00 & 3.60  \\
600\,--\,650\,nm & 6099 & $-$\,0.12 & 4.96 & 3.45\\
400\,--\,600\,nm & 5983 & $-$\,0.11 & 4.50 & 3.63\\
D15 & 6030 & 0.13 & 4.23 & 4.30 \\
 
\hline                                             %inserts single line
\end{tabular}
\tablefoot{50nm windows had 2$^{15}$ samples and the 200nm windows had 2$^{17}$ samples. All were subject to the same analysis in Sect. \ref{D15} with no priors on $\log g$.}
\end{table}

There are many reasons why our method may be produce composition offsets compared with other established techniques. The interested reader should see \cite{Jofre2016} for an excellent review on how the specifics of spectroscopic analysis routines affect abundance measurements. One interesting result from \cite{Jofre2016} is the effect of continuum normalisation which increased the method-to-method scatter in abundance measurements by up to 0.3\,dex (see their Fig.\,5). Wavelet filtering in our method is an alternate approach to normalisation, and so an offset of around 0.18\,dex is not entirely unexpected. We assess if there is a systematically lower continuum placement by adding an free parameter, $C_0$, which is a constant to add to the normalized flux of the model spectra before a discrete wavelet transform in the calculation of log-likliehood. We found values of $C_0$ converged to values between $-0.05$ to $0.05$ and did not affect measurements of [Fe/H] by more than $0.05$\,dex; $T_{\rm eff}$ remained the same for all stars within 150\,K and $\log g$ changed by as much as 0.2\, dex. 

We also looked at components unique to the wavelet method. For instance, the mother wavelet used (Daubechies, k=4) may not capture the true line depths when  convolved with a spectrum. We again measured WASP-20 with 3 alternative wavelets (Daubechies k=20, Harr k=2 and bspline  k=103) across the range 450\,--\,650\,nm  (see Table \ref{wavelet_tab}). We find the choice of mother wavelet has little influence on the determined composition (and  all other atmospheric parameters) for WASP-20 and we find similar results for the rest of the D15 sample.  It is possible that the resolution of the finest wavelet convolution (2 pixels) is not sufficient to capture iron line depths. To assess this, we convolved a few iron lines with the Daubechies k=4 kernel and assessed whether line depths were underdetermined. We found this not to be the case, suggesting no degradation of line depths owing to the choice in wavelets. 

Finally, we consider the possibility that there may be instrumental effects at play with the CORALIE \'{e}chelle spectrograph. A discrepancy in equivalent width measurements for WASP-69 (see Fig. 3.19 in D15) suggests this instrument is prone to scattered light \citep{Doyle2015}. This may be partly responsible for the systematic error in the iron abundance when combined with a low-quality spectrum.

The zero-point of the metalicity scale is a subject of on-going debate \cite[e.g.,][]{Kraft2004}). However, we can conclude that models using parameters found by D15 (as generated with line lists and atmospheres used in the above work) have better fitting line depths for the majority of Fe lines in the D15 sample than our predicted models. For this reason, we advocate using the following correction for the [Fe/H] values measured with the wavelet method to make them consistent with the metallicity scale of D15:
\begin{equation}\label{composition_correction}
\rm [Fe/H]_{\rm corrected} = \rm [Fe/H]_{\rm measured} + 0.18.
\end{equation}

\subsubsection{Systematic trend in $\log g$}

We also observe a negative correlation between residual $\log g$ measurements (wavelet - D15) and $\log g$ measured with the wavelet method (Fig. \ref{doyle:c}). This trend is observed with and without Gaussian priors on $\log g$ from transit photometry. We calculate a Pearson correlation coefficient of -0.501 for measurements with no $\log g$ prior, suggesting a significant negative correlation. We fit this trend with an $1^{st}$ order polynomial and found a gradient of $-0.692$ and a y-intercept of $3.067$. This correlation evaluates to zero at a wavelet $\log g$ value of 4.44. In principle, the following correction can be used to bring our $\log g$ measurements into line with those from D15,

\begin{equation}\label{logg_corr}
\log g_{\rm corrected} = \log g_{\rm wavelet} - 3.067 + 0.692\times \log g_{\rm wavelet}.
\end{equation}
Without knowing the exact cause of this trend, and given the sensitivity of our $log g$ estimates to the continuum placement,  we are reluctant to advise applying this correction and conclude that the wavelet method cannot reliably estimate $\log g$ beyond confirming a dwarf-like surface gravity. Obtaining log g from a spectrum is typical done through ionization balance (balancing the iron abundance measured from the Fe
I and Fe II lines). It is also possible to measure log g by fitting the wings of gravity sensitive lines (e.g. Mg, Na) using model spectra (the synthesis method). This is essentially how the wavelet method operates (in wavelet space rather than normalised flux space). Accurate determinations of log g from the synthesis method requires detailed element abundance measurements for gravity-sensitive Na and Mg lines. Estimating the abundances of these elements by scaling from the solar abundance values and apply some correction for $\alpha$ element enhancement will lead to a systematic error in $\log g $ that is difficult to quantify in individual cases. To investigate this further requires another set of comparison stars with independent $\log g$ values (preferably from binary systems where $\log g$ can be accurately measured and not planet transiting systems). 

\subsubsection{\textbf{Precision of atmospheric parameters}}\label{precision_w}

The high precision of the parameters in Table~\ref{waspstars} shows that the wavelet method can reliably converge to a well-determined set of atmospheric parameters, but to make use of these parameters we also require a reliable estimate of their true precision that accounts for additional uncertainties due to systematic errors in the data and the models. To obtain a realistic estimate of true precision of the parameters from the wavelet method, $\sigma_{\rm wavelet}$, we compare the results from our method with the correction to [Fe/H] described earlier to those from D15. The standard deviation of the residuals between the measured atmospheric parameters made by D15 and from the wavelet method, $\sigma_{\rm D15 - \rm wavelet}$, is a combination of the uncertainties from methods added in quadrature:
\begin{equation}
\sigma_{\rm D15 - \rm wavelet}^2 = \sigma_{\rm D15}^2 + \sigma_{\rm wavelet}^2.
\end{equation}
where $\sigma_{\rm D15}$ is the quoted error on the atmospheric parameters from D15. There are two extreme cases: the first is that the uncertainty from D15 is negligible (or at least much better than what we can achieve) giving $\sigma_{\rm D15 - \rm wavelet}^2 \approx \sigma_{\rm wavelet}^2$; and the second is that the inter-method discrepancy, $\sigma_{\rm D15 - \rm wavelet}^2$, is negligible leaving uncertainties similar to those quoted by D15. In reality, the absolute uncertainty for the wavelet method is somewhere between these two extremes. We adopt a true precision of each parameter from Table \ref{doyle_tab} using a uniform prior on $\log g$ which is to assume that $\sigma_{D15} \langle \langle \sigma_{wavelet}$. We suggest applying a correction of $+0.18\,dex$ to [Fe/H] and not to apply a correction to $\log g$. This means precision of $85\, \rm K$ for $T_{\rm eff}$, $0.06\, dex$ for [Fe/H] and $1.35\,kms^{-1}$ for $V \sin i$. The resulting value of $\log g$ is not likely to be reliable but it good enough to confirm dwarf-like gravity around $\log g = $ 4--5 $dex$. We note that these values are comparable to other methods (e.g., \cite{Bruntt2010}).

\subsubsection{Spectrum quality}\label{spec_quality}

\begin{figure}[ht!]
\centering
\includegraphics[width=0.5\textwidth, height = 0.5\textwidth]{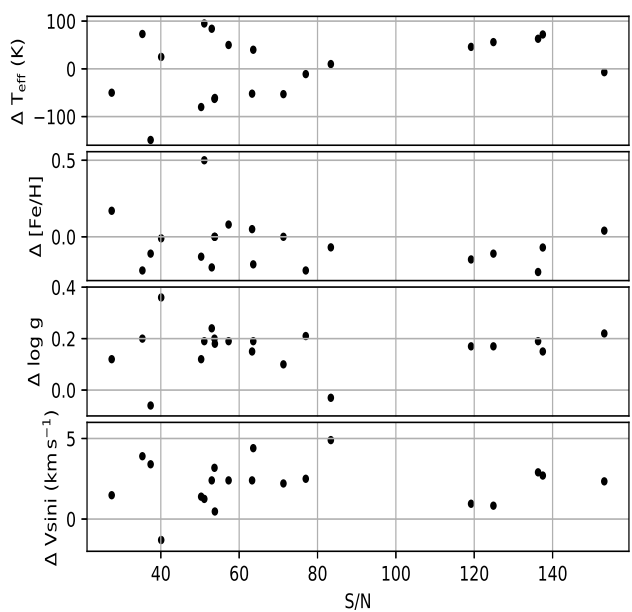}
\caption{Atmospheric parameters from the wavelet method, with no prior on $\log g$, compared to those from D15 ($x_{\rm D15}-x_{\rm wavelet}$) as a function of signal-to-noise.}
\label{snr}
\end{figure}

\begin{figure}[ht!]
\centering
\includegraphics[width=0.5\textwidth]{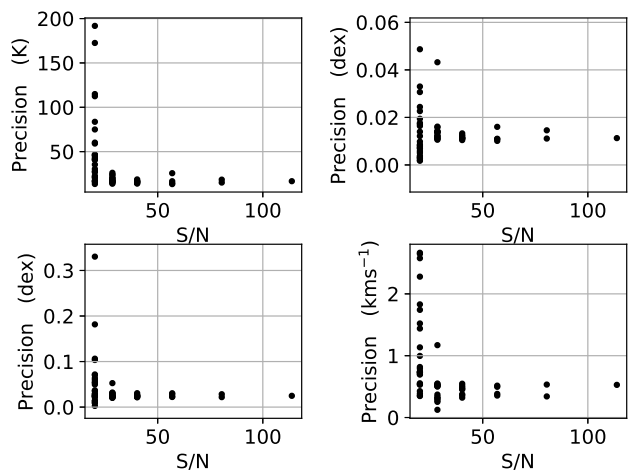}
\caption{Precision of the wavelet method versus signal-to-noise for $T_{\rm eff}$ (top left), [Fe/H] (top right), $\log g$ (bottom left), and $V \sin i$ (bottom right) for WASP-20. }
\label{precision}
\end{figure}

\begin{figure}[ht!]
\centering
\includegraphics[width=0.5\textwidth]{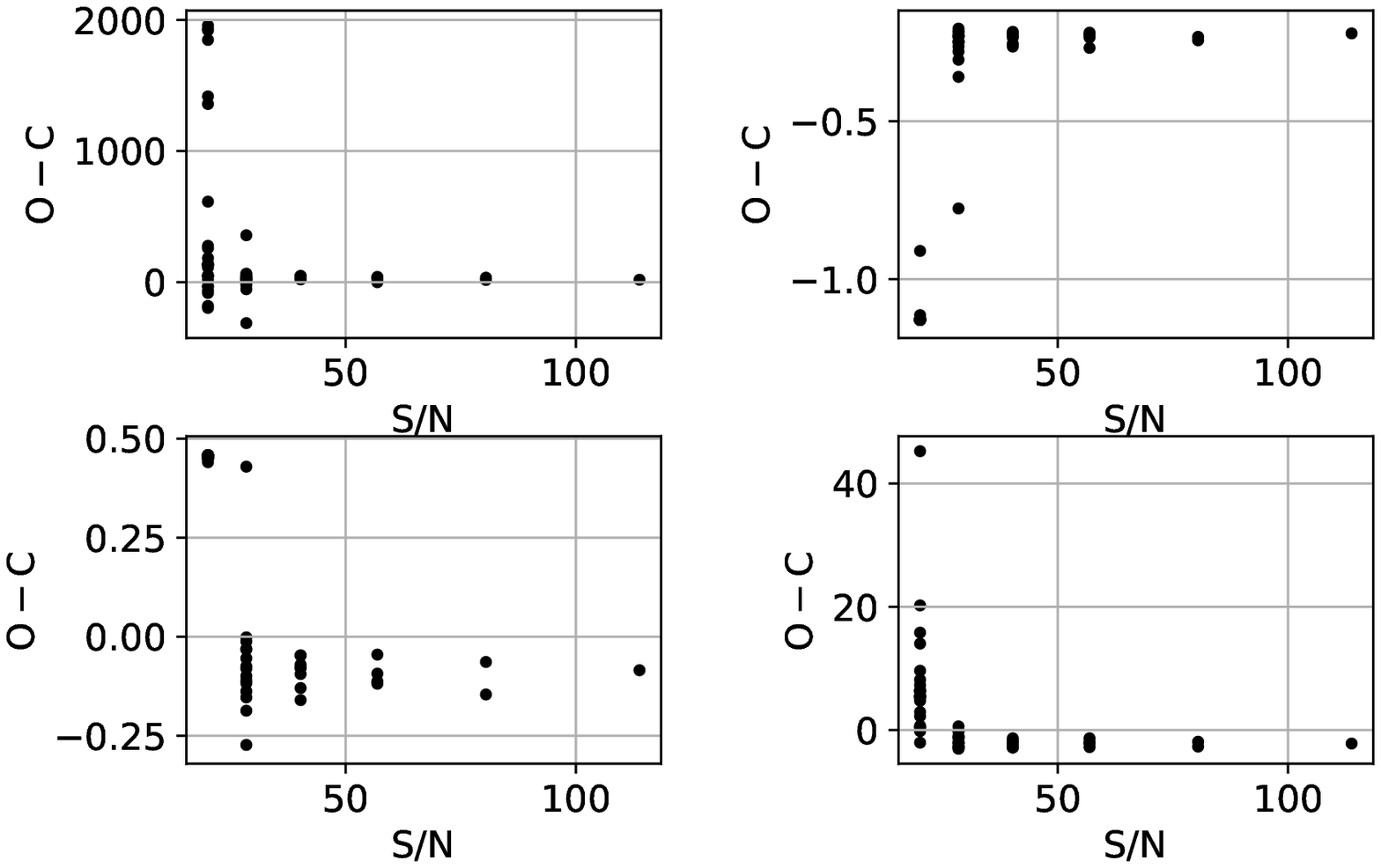}
\caption{Accuracy of the wavelet method versus signal-to-noise for $T_{\rm eff}$ (top left), [Fe/H] (top right), $\log g$ (bottom left), and $V \sin i$ (bottom right) for WASP-20 using the results from D15 as a zero-point.}
\label{accuracy}
\end{figure}

In Fig. \ref{snr} we plot the difference between atmospheric parameters obtained with the wavelet method (with no priors for $\log g$) to those from D15 as a function of singal-to-noise rario (SNR). The sample falls into two categories of quality (those with SNR $\leq$ 90 and those with SNR $\geq$ 120). There is noticeably more scatter in the lower quality group and suggests that the uncertainty of our atmospheric parameters decreases with a better quality spectrum. The noise profile of a spectrum depends on observing conditions, properties of the star and the instrument used to make the observations. This is why adding Gaussian noise to a synthetic spectrum until the atmospheric parameters are no longer recoverable does not give a true reflection of a methods robustness to noise. Instead, we use 32 (out of 58) observations of the star with the highest SNR in the D15 sample - WASP-20. We dyadically split up these spectra and median combine them into different sets. The sets of splits used were 1 spectrum (1 set of 32 spectra), 2 spectra (2 sets of 16 spectra), 4  spectra (4 sets of 8 spectra), ..., 32 spectra (32 sets each containing just 1 spectra). We scale SNR from the coaddition of all 58 spectra:
\begin{equation}
\rm SNR = \rm SNR_{58 \, \rm spectra} \times \sqrt{\frac{N_{\rm set}}{58}}.
\end{equation}
Each set was measured with the afforemention wavelet technique with no prior probability function for $\log g$, and best fitting parameters adopted. The precision and accuracy as a function of SNR are shown in Fig. \ref{precision} and Fig \ref{accuracy}, respectively. We find that systematic errors dominate for a SNR below 40. A similar result is found by \cite{SmiljanicKornBergemannEtAl2014} who measured UVES-FLAMES spectra for FGK stars from the GAIA-ESO survey and found a systematics threshold of SNR$\, \approx 50$.

\section{Conclusion} \label{Conclusion}

% new transfer code to 
% need with surveys
% add to automated methods
% provides another check 

We have shown that our method accurately recovers the atmospheric parameters of synthetic spectra from a grid of models using subsets of wavelet coefficients in a Bayesian framework. The same method was applied to the CORALIE spectra of 20 FGK stars which have been analysed independently by measurements of equivalent widths from higher-quality HARPS spectra. From this we determine a precision for the parameters derived from the wavelet method of $85\, \rm K$ for $T_{\rm eff}$, $0.06\, \rm dex$ for [Fe/H] and $1.35\,  \rm kms^{-1}$ for $V \sin i$. Surface gravity, $\log g$, can also be estimated using our method but it is difficult to assess the precision of this parameter in individual cases. Consequently, we recommend that $\log g$ estimates from our method are only used as an indicator of whether or not a star is a dwarf star ($\log g \approx 4.5$) or not. We find an offset in our metallicity scale compared to the results of \citet{Doyle2013, Doyle2015} in the sense that our values of [Fe/H] are lower by 0.18 dex, despite using a consistent solar abundance, and recommend that this offset is applied at a correction to the [Fe/H] values from our method. We find our method is robust for \'{e}chelle spectra with a SNR above 40. Below this value the uncertainity in the measured atmospheric parameters increases to unusable levels. A further development of this method would include a more sophisticated weighting system for the wavelet coefficients beyond the Monte Carlo approach used here. 
% Did the referee ask for a comment on this? If not I would rather not give them something else to argue about...
%We attempted to develop a wavelet coefficient weighting scheme which takes into account the sensitivity of atmospheric parameters to  certain lines (see Fig. \ref{Wavelet_response}) but were unsuccessful due to the noise-like nature of wavelet coefficients and that each coefficient depends on more than one parameter.  

Our method has already been used to determine the atmospheric parameters of the EBLM~J0555$-$57 \citep{vonBoetticher2017},  which hosts one of the densest main-sequence star currently known. This method is also being used to study other EBLM systems and as part of the on-going exoplanet discovery process with for the WASP survey. {For both exoplanet systems and EBLM binaries, the contribution of the companion star to the optical flux is negligible (they are SB1 binaries) and so our method using models of single stars is appropriate, but it would not be suitable for cases where the companion is detectable in the spectrum (SB2 binaries).  } We have optimised our method for the application to spectra from CORALIE, but the same method should be equally applicable to spectra with moderate SNR from other echelle spectrographs.   We have implemented the wavelet method in a python module called \textit{waveletspec} which is available upon request.

% * <p.maxted@keele.ac.uk> 2018-01-02T16:27:30.481Z:
% 
% You should thank the referee in the acknowledgements. "We thank the anonymous referee for their useful comments on the manuscript that have helped to improve the paper."  - or similar
% 
% ^.
\begin{acknowledgements}
SG acknowledges financial support from the Science and
Technology Facilities Council (STFC). SG thanks A. H. M. J Triaud and S. Udry for data acquisition and support.  We thank the referee for their careful reading of the manuscript and suggestions that have helped to improve this paper.
\end{acknowledgements}


%\documentclass[bibyear]{aa}
\bibliographystyle{aa}
\bibliography{references1.bib}

\begin{thebibliography}{53}
\expandafter\ifx\csname natexlab\endcsname\relax\def\natexlab#1{#1}\fi

\bibitem[{{Anderson} {et~al.}(2014){Anderson}, {Collier Cameron}, {Delrez},
  {Doyle}, {Faedi}, {Fumel}, {Gillon}, {G{\'o}mez Maqueo Chew}, {Hellier},
  {Jehin}, {Lendl}, {Maxted}, {Pepe}, {Pollacco}, {Queloz}, {S{\'e}gransan},
  {Skillen}, {Smalley}, {Smith}, {Southworth}, {Triaud}, {Turner}, {Udry}, \&
  {West}}]{Anderson2014}
{Anderson}, D.~R., {Collier Cameron}, A., {Delrez}, L., {et~al.} 2014, \mnras,
  445, 1114

\bibitem[{{Anderson} {et~al.}(2015){Anderson}, {Collier Cameron}, {Hellier},
  {Lendl}, {Lister}, {Maxted}, {Queloz}, {Smalley}, {Smith}, {Triaud}, {Brown},
  {Gillon}, {Neveu-VanMalle}, {Pepe}, {Pollacco}, {S{\'e}gransan}, {Udry},
  {West}, \& {Wheatley}}]{Anderson2015}
{Anderson}, D.~R., {Collier Cameron}, A., {Hellier}, C., {et~al.} 2015, \aap,
  575, A61

\bibitem[{{Anderson} {et~al.}(2011){Anderson}, {Collier Cameron}, {Hellier},
  {Lendl}, {Maxted}, {Pollacco}, {Queloz}, {Smalley}, {Smith}, {Todd},
  {Triaud}, {West}, {Barros}, {Enoch}, {Gillon}, {Lister}, {Pepe},
  {S{\'e}gransan}, {Street}, \& {Udry}}]{Anderson2011}
{Anderson}, D.~R., {Collier Cameron}, A., {Hellier}, C., {et~al.} 2011, \apjl,
  726, L19

\bibitem[{{Anglada-Escud{\'e}} {et~al.}(2016){Anglada-Escud{\'e}}, {Amado},
  {Barnes}, {Berdi{\~n}as}, {Butler}, {Coleman}, {de La Cueva}, {Dreizler},
  {Endl}, {Giesers}, {Jeffers}, {Jenkins}, {Jones}, {Kiraga}, {K{\"u}rster},
  {L{\'o}pez-Gonz{\'a}lez}, {Marvin}, {Morales}, {Morin}, {Nelson}, {Ortiz},
  {Ofir}, {Paardekooper}, {Reiners}, {Rodr{\'{\i}}guez},
  {Rodr{\'{\i}}guez-L{\'o}pez}, {Sarmiento}, {Strachan}, {Tsapras}, {Tuomi}, \&
  {Zechmeister}}]{Anglada-Escude2016}
{Anglada-Escud{\'e}}, G., {Amado}, P.~J., {Barnes}, J., {et~al.} 2016, \nat,
  536, 437

\bibitem[{{Asplund} {et~al.}(2009){Asplund}, {Grevesse}, {Sauval}, \&
  {Scott}}]{Asplund2009}
{Asplund}, M., {Grevesse}, N., {Sauval}, A.~J., \& {Scott}, P. 2009, \araa, 47,
  481

\bibitem[{{Baraffe} {et~al.}(1998){Baraffe}, {Chabrier}, {Allard}, \&
  {Hauschildt}}]{Baraffe1998}
{Baraffe}, I., {Chabrier}, G., {Allard}, F., \& {Hauschildt}, P.~H. 1998, \aap,
  337, 403

\bibitem[{{Baranne} {et~al.}(1996){Baranne}, {Queloz}, {Mayor}, {Adrianzyk},
  {Knispel}, {Kohler}, {Lacroix}, {Meunier}, {Rimbaud}, \& {Vin}}]{Baranne1996}
{Baranne}, A., {Queloz}, D., {Mayor}, M., {et~al.} 1996, \aaps, 119, 373

\bibitem[{{Belmon} {et~al.}(2002){Belmon}, {Benoit-Cattin}, {Baskurt}, \&
  {Bougeret}}]{Belmon2002}
{Belmon}, L., {Benoit-Cattin}, H., {Baskurt}, A., \& {Bougeret}, J.-L. 2002,
  \aap, 386, 1143

\bibitem[{{Blanco-Cuaresma} {et~al.}(2016){Blanco-Cuaresma}, {Nordlander},
  {Heiter}, {Jofr{\'e}}, {Masseron}, {Casamiquela}, {Tabernero}, {Bhat},
  {Casey}, {Mel{\'e}ndez}, \& {Ram{\'{\i}}rez}}]{Blanco-Cuaresma2016}
{Blanco-Cuaresma}, S., {Nordlander}, T., {Heiter}, U., {et~al.} 2016, ArXiv
  e-prints

\bibitem[{{Blanco-Cuaresma} {et~al.}(2014){Blanco-Cuaresma}, {Soubiran},
  {Heiter}, \& {Jofr{\'e}}}]{Blanco-CuaresmaSoubiranHeiterEtAl2014}
{Blanco-Cuaresma}, S., {Soubiran}, C., {Heiter}, U., \& {Jofr{\'e}}, P. 2014,
  \aap, 569, A111

\bibitem[{{Bruntt} {et~al.}(2010){Bruntt}, {Bedding}, {Quirion}, {Lo Curto},
  {Carrier}, {Smalley}, {Dall}, {Arentoft}, {Bazot}, \& {Butler}}]{Bruntt2010}
{Bruntt}, H., {Bedding}, T.~R., {Quirion}, P.-O., {et~al.} 2010, \mnras, 405,
  1907

\bibitem[{{Dafonte} {et~al.}(2016){Dafonte}, {Fustes}, {Manteiga}, {Garabato},
  {{\'A}lvarez}, {Ulla}, \& {Allende Prieto}}]{Dafonte2016}
{Dafonte}, C., {Fustes}, D., {Manteiga}, M., {et~al.} 2016, \aap, 594, A68

\bibitem[{{Demory} {et~al.}(2009){Demory}, {S{\'e}gransan}, {Forveille},
  {Queloz}, {Beuzit}, {Delfosse}, {di Folco}, {Kervella}, {Le Bouquin},
  {Perrier}, {Benisty}, {Duvert}, {Hofmann}, {Lopez}, \& {Petrov}}]{Demory2009}
{Demory}, B.-O., {S{\'e}gransan}, D., {Forveille}, T., {et~al.} 2009, \aap,
  505, 205

\bibitem[{{Doyle}(2015)}]{Doyle2015}
{Doyle}, A.~P. 2015, PhD thesis, Keele University

\bibitem[{{Doyle} {et~al.}(2017){Doyle}, {Smalley}, {Faedi}, {Pollacco}, \&
  {Gomez Maqueo Chew}}]{Doyle2017}
{Doyle}, A.~P., {Smalley}, B., {Faedi}, F., {Pollacco}, D., \& {Gomez Maqueo
  Chew}, Y. 2017, ArXiv e-prints

\bibitem[{{Doyle} {et~al.}(2013){Doyle}, {Smalley}, {Maxted}, {Anderson},
  {Cameron}, {Gillon}, {Hellier}, {Pollacco}, {Queloz}, {Triaud}, \&
  {West}}]{Doyle2013}
{Doyle}, A.~P., {Smalley}, B., {Maxted}, P.~F.~L., {et~al.} 2013, \mnras, 428,
  3164

\bibitem[{{Essaouabi} {et~al.}(2009){Essaouabi}, {Regragui}, \&
  {Ibnelhaj}}]{Essaouabi2009}
{Essaouabi}, A., {Regragui}, F., \& {Ibnelhaj}, E. 2009, ArXiv e-prints

\bibitem[{{Fernandez} {et~al.}(2009){Fernandez}, {Latham}, {Torres}, {Everett},
  {Mandushev}, {Charbonneau}, {O'Donovan}, {Alonso}, {Esquerdo},
  {Hergenrother}, \& {Stefanik}}]{Fernandez2009}
{Fernandez}, J.~M., {Latham}, D.~W., {Torres}, G., {et~al.} 2009, \apj, 701,
  764

\bibitem[{{Foreman-Mackey} {et~al.}(2013){Foreman-Mackey}, {Hogg}, {Lang}, \&
  {Goodman}}]{Foreman-Mackey2013}
{Foreman-Mackey}, D., {Hogg}, D.~W., {Lang}, D., \& {Goodman}, J. 2013, \pasp,
  125, 306

\bibitem[{{Gillon} {et~al.}(2009){Gillon}, {Smalley}, {Hebb}, {Anderson},
  {Triaud}, {Hellier}, {Maxted}, {Queloz}, \& {Wilson}}]{Gillon2009a}
{Gillon}, M., {Smalley}, B., {Hebb}, L., {et~al.} 2009, \aap, 496, 259

\bibitem[{{Gillon} {et~al.}(2017){Gillon}, {Triaud}, {Demory}, {Jehin}, {Agol},
  {Deck}, {Lederer}, {de Wit}, {Burdanov}, {Ingalls}, {Bolmont}, {Leconte},
  {Raymond}, {Selsis}, {Turbet}, {Barkaoui}, {Burgasser}, {Burleigh}, {Carey},
  {Chaushev}, {Copperwheat}, {Delrez}, {Fernandes}, {Holdsworth}, {Kotze}, {Van
  Grootel}, {Almleaky}, {Benkhaldoun}, {Magain}, \& {Queloz}}]{Gillon2017}
{Gillon}, M., {Triaud}, A.~H.~M.~J., {Demory}, B.-O., {et~al.} 2017, \nat, 542,
  456

\bibitem[{{G{\'o}mez Maqueo Chew} {et~al.}(2014){G{\'o}mez Maqueo Chew},
  {Morales}, {Faedi}, {Garc{\'{\i}}a-Melendo}, {Hebb}, {Rodler}, {Deshpande},
  {Mahadevan}, {McCormac}, {Barnes}, {Triaud}, {Lopez-Morales}, {Skillen},
  {Collier Cameron}, {Joner}, {Laney}, {Stephens}, {Stassun}, {Cargile}, \&
  {Monta{\~n}{\'e}s-Rodr{\'{\i}}guez}}]{GomezMaqueoChewMoralesFaediEtAl2014}
{G{\'o}mez Maqueo Chew}, Y., {Morales}, J.~C., {Faedi}, F., {et~al.} 2014,
  \aap, 572, A50

\bibitem[{Goodman(2010)}]{Goodman2010}
Goodman, J., . W.~J. 2010, Commun. Appl. Math. Comput. Sci., 5

\bibitem[{{Gray} \& {Corbally}(1994)}]{Gray1994}
{Gray}, R.~O. \& {Corbally}, C.~J. 1994, \aj, 107, 742

\bibitem[{{Gustafsson} {et~al.}(2008){Gustafsson}, {Edvardsson}, {Eriksson},
  {J{\o}rgensen}, {Nordlund}, \& {Plez}}]{Gustafsson2008}
{Gustafsson}, B., {Edvardsson}, B., {Eriksson}, K., {et~al.} 2008, \aap, 486,
  951

\bibitem[{Henry {et~al.}(2006)Henry, Jao, Subasavage, Beaulieu, Ianna, Costa,
  \& Méndez}]{Henry2006}
Henry, T.~J., Jao, W.-C., Subasavage, J.~P., {et~al.} 2006, The Astronomical
  Journal, 132, 2360

\bibitem[{{Jofre} {et~al.}(2016){Jofre}, {Heiter}, {Worley}, {Blanco-Cuaresma},
  {Soubiran}, {Masseron}, {Hawkins}, {Adibekyan}, {Buder}, {Casamiquela},
  {Gilmore}, {Hourihane}, \& {Tabernero}}]{Jofre2016}
{Jofre}, P., {Heiter}, U., {Worley}, C.~C., {et~al.} 2016, ArXiv e-prints

\bibitem[{{Kervella} {et~al.}(2016){Kervella}, {M{\'e}rand}, {Ledoux},
  {Demory}, \& {Le Bouquin}}]{Kervella2016}
{Kervella}, P., {M{\'e}rand}, A., {Ledoux}, C., {Demory}, B.-O., \& {Le
  Bouquin}, J.-B. 2016, ArXiv e-prints

\bibitem[{{Kraft} \& {Ivans}(2004)}]{Kraft2004}
{Kraft}, R.~P. \& {Ivans}, I.~I. 2004, Origin and Evolution of the Elements

\bibitem[{{Kraus} {et~al.}(2011){Kraus}, {Tucker}, {Thompson}, {Craine}, \&
  {Hillenbrand}}]{Kraus}
{Kraus}, A., {Tucker}, R., {Thompson}, M., {Craine}, E., \& {Hillenbrand}, L.
  2011, \apj, 728, 48

\bibitem[{{Li} {et~al.}(2015){Li}, {Lu}, {Comte}, {Luo}, {Zhao}, \&
  {Wang}}]{Li2015}
{Li}, X., {Lu}, Y., {Comte}, G., {et~al.} 2015, \apjs, 218, 3

\bibitem[{{Manteiga} {et~al.}(2010){Manteiga}, {Ord{\'o}{\~n}ez}, {Dafonte}, \&
  {Arcay}}]{Manteiga2010}
{Manteiga}, M., {Ord{\'o}{\~n}ez}, D., {Dafonte}, C., \& {Arcay}, B. 2010,
  \pasp, 122, 608

\bibitem[{Morales {et~al.}(2010)Morales, Gallardo, Ribas, Jordi, Baraffe, \&
  Chabrier}]{Morales2010}
Morales, J.~C., Gallardo, J., Ribas, I., {et~al.} 2010, The Astrophysical
  Journal, 718, 502

\bibitem[{{Mortier} {et~al.}(2013){Mortier}, {Santos}, {Sousa}, {Fernandes},
  {Adibekyan}, {Delgado Mena}, {Montalto}, \& {Israelian}}]{Mortier2013}
{Mortier}, A., {Santos}, N.~C., {Sousa}, S.~G., {et~al.} 2013, \aap, 558, A106

\bibitem[{{Mortier} {et~al.}(2014){Mortier}, {Sousa}, {Adibekyan},
  {Brand{\~a}o}, \& {Santos}}]{Mortier2014}
{Mortier}, A., {Sousa}, S.~G., {Adibekyan}, V.~Z., {Brand{\~a}o}, I.~M., \&
  {Santos}, N.~C. 2014, \aap, 572, A95

\bibitem[{{Nefs} {et~al.}(2013){Nefs}, {Birkby}, {Snellen}, {Hodgkin}, {Sip{\H
  o}cz}, {Kov{\'a}cs}, {Mislis}, {Pinfield}, \& {Martin}}]{Nefs2013}
{Nefs}, S.~V., {Birkby}, J.~L., {Snellen}, I.~A.~G., {et~al.} 2013, \mnras,
  431, 3240

\bibitem[{Olkkonen(2011)}]{Olkkonen2011}
Olkkonen, J.~T. 2011, DIscrete Wavelet Transforms, ed. J.~T. Olkkonen (InTech)

\bibitem[{{Pollacco} {et~al.}(2006){Pollacco}, {Skillen}, {Collier Cameron},
  {Christian}, {Hellier}, {Irwin}, {Lister}, {Street}, {West}, {Anderson},
  {Clarkson}, {Deeg}, {Enoch}, {Evans}, {Fitzsimmons}, {Haswell}, {Hodgkin},
  {Horne}, {Kane}, {Keenan}, {Maxted}, {Norton}, {Osborne}, {Parley}, {Ryans},
  {Smalley}, {Wheatley}, \& {Wilson}}]{PollaccoSkillenCollierCameronEtAl2006}
{Pollacco}, D.~L., {Skillen}, I., {Collier Cameron}, A., {et~al.} 2006, \pasp,
  118, 1407

\bibitem[{{Queloz} {et~al.}(2001){Queloz}, {Mayor}, {Udry}, {Burnet},
  {Carrier}, {Eggenberger}, {Naef}, {Santos}, {Pepe}, {Rupprecht}, {Avila},
  {Baeza}, {Benz}, {Bertaux}, {Bouchy}, {Cavadore}, {Delabre}, {Eckert},
  {Fischer}, {Fleury}, {Gilliotte}, {Goyak}, {Guzman}, {Kohler}, {Lacroix},
  {Lizon}, {Megevand}, {Sivan}, {Sosnowska}, \& {Weilenmann}}]{Queloz2001}
{Queloz}, D., {Mayor}, M., {Udry}, S., {et~al.} 2001, The Messenger, 105, 1

\bibitem[{{Queloz} {et~al.}(2000){Queloz}, {Mayor}, {Weber}, {Bl{\'e}cha},
  {Burnet}, {Confino}, {Naef}, {Pepe}, {Santos}, \& {Udry}}]{Queloz2000}
{Queloz}, D., {Mayor}, M., {Weber}, L., {et~al.} 2000, \aap, 354, 99

\bibitem[{{Sedaghati} {et~al.}(2016){Sedaghati}, {Boffin}, {Je{\v
  r}abkov{\'a}}, {Garc{\'{\i}}a Mu{\~n}oz}, {Grenfell}, {Smette}, {Ivanov},
  {Csizmadia}, {Cabrera}, {Kabath}, {Rocchetto}, \& {Rauer}}]{Sedaghati2016}
{Sedaghati}, E., {Boffin}, H.~M., {Je{\v r}abkov{\'a}}, T., {et~al.} 2016,
  ArXiv e-prints

\bibitem[{{Smiljanic} {et~al.}(2014){Smiljanic}, {Korn}, {Bergemann}, {Frasca},
  {Magrini}, {Masseron}, {Pancino}, {Ruchti}, {San Roman}, {Sbordone}, {Sousa},
  {Tabernero}, {Tautvai{\v s}ien{\.e}}, {Valentini}, {Weber}, {Worley},
  {Adibekyan}, {Allende Prieto}, {Barisevi{\v c}ius}, {Biazzo},
  {Blanco-Cuaresma}, {Bonifacio}, {Bragaglia}, {Caffau}, {Cantat-Gaudin},
  {Chorniy}, {de Laverny}, {Delgado-Mena}, {Donati}, {Duffau}, {Franciosini},
  {Friel}, {Geisler}, {Gonz{\'a}lez Hern{\'a}ndez}, {Gruyters}, {Guiglion},
  {Hansen}, {Heiter}, {Hill}, {Jacobson}, {Jofre}, {J{\"o}nsson}, {Lanzafame},
  {Lardo}, {Ludwig}, {Maiorca}, {Mikolaitis}, {Montes}, {Morel}, {Mucciarelli},
  {Mu{\~n}oz}, {Nordlander}, {Pasquini}, {Puzeras}, {Recio-Blanco}, {Ryde},
  {Sacco}, {Santos}, {Serenelli}, {Sordo}, {Soubiran}, {Spina}, {Steffen},
  {Vallenari}, {Van Eck}, {Villanova}, {Gilmore}, {Randich}, {Asplund},
  {Binney}, {Drew}, {Feltzing}, {Ferguson}, {Jeffries}, {Micela}, {Negueruela},
  {Prusti}, {Rix}, {Alfaro}, {Babusiaux}, {Bensby}, {Blomme}, {Flaccomio},
  {Fran{\c c}ois}, {Irwin}, {Koposov}, {Walton}, {Bayo}, {Carraro}, {Costado},
  {Damiani}, {Edvardsson}, {Hourihane}, {Jackson}, {Lewis}, {Lind}, {Marconi},
  {Martayan}, {Monaco}, {Morbidelli}, {Prisinzano}, \&
  {Zaggia}}]{SmiljanicKornBergemannEtAl2014}
{Smiljanic}, R., {Korn}, A.~J., {Bergemann}, M., {et~al.} 2014, \aap, 570, A122

\bibitem[{{Southworth}(2011)}]{Southworth2011}
{Southworth}, J. 2011, \mnras, 417, 2166

\bibitem[{{Southworth} {et~al.}(2016){Southworth}, {Mancini}, {Madhusudhan},
  {Molliere}, {Ciceri}, \& {Henning}}]{Southworth2016}
{Southworth}, J., {Mancini}, L., {Madhusudhan}, N., {et~al.} 2016, ArXiv
  e-prints

\bibitem[{{Spada} {et~al.}(2013){Spada}, {Demarque}, {Kim}, \&
  {Sills}}]{Spada2013}
{Spada}, F., {Demarque}, P., {Kim}, Y.-C., \& {Sills}, A. 2013, \apj, 776, 87

\bibitem[{{Stumpe} {et~al.}(2012){Stumpe}, {Smith}, {Van Cleve}, {Jenkins},
  {Barclay}, {Fanelli}, {Girouard}, {Kolodziejczak}, {McCauliff}, {Morris}, \&
  {Twicken}}]{Stumpe2012}
{Stumpe}, M.~C., {Smith}, J.~C., {Van Cleve}, J., {et~al.} 2012, in American
  Astronomical Society Meeting Abstracts, Vol. 220, American Astronomical
  Society Meeting Abstracts \#220, 330.04

\bibitem[{{Torres}(2013)}]{Torres2013}
{Torres}, G. 2013, Astronomische Nachrichten, 334, 4

\bibitem[{{Torres} {et~al.}(2010){Torres}, {Andersen}, \&
  {Gim{\'e}nez}}]{TorresAndersenGimenez2010}
{Torres}, G., {Andersen}, J., \& {Gim{\'e}nez}, A. 2010, \aapr, 18, 67

\bibitem[{{Triaud} {et~al.}(2013){Triaud}, {Anderson}, {Collier Cameron},
  {Doyle}, {Fumel}, {Gillon}, {Hellier}, {Jehin}, {Lendl}, {Lovis}, {Maxted},
  {Pepe}, {Pollacco}, {Queloz}, {S{\'e}gransan}, {Smalley}, {Smith}, {Udry},
  {West}, \& {Wheatley}}]{Triaud2013}
{Triaud}, A.~H.~M.~J., {Anderson}, D.~R., {Collier Cameron}, A., {et~al.} 2013,
  \aap, 551, A80

\bibitem[{{Triaud} {et~al.}(2016){Triaud}, {Neveu-VanMalle}, {Lendl},
  {Anderson}, {Collier Cameron}, {Delrez}, {Doyle}, {Gillon}, {Hellier},
  {Jehin}, {Maxted}, {S{\'e}gransan}, {Smalley}, {Queloz}, {Pollacco},
  {Southworth}, {Tregloan-Reed}, {Udry}, \& {West}}]{Triaud2016}
{Triaud}, A.~H.~M.~J., {Neveu-VanMalle}, M., {Lendl}, M., {et~al.} 2016, ArXiv
  e-prints

\bibitem[{{Triaud} {et~al.}(2011){Triaud}, {Queloz}, {Hellier}, {Gillon},
  {Smalley}, {Hebb}, {Collier Cameron}, {Anderson}, {Boisse}, {H{\'e}brard},
  {Jehin}, {Lister}, {Lovis}, {Maxted}, {Pepe}, {Pollacco}, {S{\'e}gransan},
  {Simpson}, {Udry}, \& {West}}]{Triaud2011}
{Triaud}, A.~H.~M.~J., {Queloz}, D., {Hellier}, C., {et~al.} 2011, \aap, 531,
  A24

\bibitem[{{von Boetticher} {et~al.}(2017){von Boetticher}, {Triaud}, {Queloz},
  {Gill}, {Lendl}, {Delrez}, {Anderson}, {Collier Cameron}, {Faedi}, {Gillon},
  {G{\'o}mez Maqueo Chew}, {Hebb}, {Hellier}, {Jehin}, {Maxted}, {Martin},
  {Pepe}, {Pollacco}, {S{\'e}gransan}, {Smalley}, {Udry}, \&
  {West}}]{vonBoetticher2017}
{von Boetticher}, A., {Triaud}, A.~H.~M.~J., {Queloz}, D., {et~al.} 2017, ArXiv
  e-prints

\bibitem[{{Zhou} {et~al.}(2014){Zhou}, {Bayliss}, {Hartman}, {Bakos}, {Penev},
  {Csubry}, {Tan}, {Jord{\'a}n}, {Mancini}, {Rabus}, {Brahm}, {Espinoza},
  {Mohler-Fischer}, {Ciceri}, {Suc}, {Cs{\'a}k}, {Henning}, \&
  {Schmidt}}]{Zhou2014}
{Zhou}, G., {Bayliss}, D., {Hartman}, J.~D., {et~al.} 2014, \mnras, 437, 2831

\end{thebibliography}

%%%%%%%%%%%%%%%%%%%%%%%%%%%5
% Table of main results
%%%%%%%%%%%%%%%%%%%%%%%%%%%%% 

\begin{sidewaystable*}[t!]
\caption{Descriptions of 20 WASP targets used for this work.}              
\centering
\resizebox{\linewidth}{!}{%
\begin{tabular}{l r r r r r r r r r r r r r r}     % 7 columns 
\hline\hline   
                      % To combine 4 columns into a single one 
%Star & \multicolumn{1}{p{2cm}}{\centering $SNR$ \\ (at 500$nm$)} & \multicolumn{1}{p{2cm}}{\centering $T_{eff}$ \\ (K) \\ \protect\cite{Doyle2015}} & \multicolumn{1}{p{2cm}}{\centering $T_{eff}$ \\ (K) \\This Work}\\
                  
 %  4  & 34.4 & \multicolumn{1}{p{2cm}}{\centering 5400 \\$\pm$ 90} & \multicolumn{1}{p{2cm}}{\centering 5432 \\ $\pm$ 80}\\ 
%   5  & 32.9 & \multicolumn{1}{p{2cm}}{\centering 5690 \\$\pm$ 80} & \multicolumn{1}{p{2cm}}{\centering 5715 \\ $\pm$ 80}\\ 

Star & 
\multicolumn{1}{p{2cm}}{\centering $T_{eff}$ \\ (K) \\ This Work} & 
\multicolumn{1}{p{2cm}}{\centering $T_{eff}$ \\ (K) \\ $log g$ prior}  & 
\multicolumn{1}{p{2cm}}{\centering $T_{eff}$ \\ (K) \\ D15} & 
\multicolumn{1}{p{2cm}}{\centering $Log$ $g$\\ (c.g.s) \\ This Work}  & 
\multicolumn{1}{p{2cm}}{\centering $Log$ $g$\\ (c.g.s)\\ D15} & 
\multicolumn{1}{p{2cm}}{\centering $Log$ $g$\\ (c.g.s)\\ photometry}  & 
\multicolumn{1}{p{2cm}}{\centering $[Fe/H]$ \\ (dex) \\ This Work}  & 
\multicolumn{1}{p{2cm}}{\centering $[Fe/H]$ \\ (dex) \\ $log g$ prior} & 
\multicolumn{1}{p{2cm}}{\centering $[Fe/H]$ \\ (dex)\\ D15} & 
\multicolumn{1}{p{2cm}}{\centering $V.sini$ \\ $kms^{-1}$ \\ This Work} & 
\multicolumn{1}{p{2cm}}{\centering $V.sini$ \\ $kms^{-1}$ \\ $log g$ prior} & \multicolumn{1}{p{2cm}}{\centering $V.sini$ \\ $kms^{-1}$ \\ D15}  \\
%& \multicolumn{1}{p{2cm}}{\centering SNR \\ at 500$nm$} \\
\hline  
% 4   & 5432 & 80 & 5400 & 90  & 4.58  & 0.1 & 4.47 & 0.11  & -0.18  & 0.07 & -0.1  & 0.1  & 0.92  & 0.8 & 3.4  & 0.3  & 12.5  & 34.4   \\ 
 %done  

 WASP-4 *
& 5524 $\pm$ 11
& 5549 $\pm$ 10 
& 5400 $\pm$ 90
& 4.5  $\pm$ 0.02
& 4.47  $\pm$ 0.11
&  4.49 $\pm$ 0.01 $^1$
& -0.08  $\pm$ 0.01
& -0.04  $\pm$ 0.01 
&-0.1 $\pm$ 0.1 
& 4.62  $\pm$ 0.11 
& 4.92  $\pm$ 0.11 
& 3.40 $\pm$ 0.30 \\
%&  34.4 & 12.5\\
                     
%WASP-4 *
%& \multicolumn{1}{p{2cm}}{\centering 5524 \\ $\pm$ 11} 
%& \multicolumn{1}{p{2cm}}{\centering 5549 \\$\pm$ 10} &
%\multicolumn{1}{p{2cm}}{\centering 5400 \\$\pm$ 90} 
%& \multicolumn{1}{p{2cm}}{\centering 4.5 \\ $\pm$ 0.02}  
%& \multicolumn{1}{p{2cm}}{\centering 4.47 \\ $\pm$ 0.11} 
%& \multicolumn{1}{p{2cm}}{\centering 4.49 \\ $\pm$ 0.01 $^1$} 
%& \multicolumn{1}{p{2cm}}{\centering -0.08 \\ $\pm$ 0.01} &
%multicolumn{1}{p{2cm}}{\centering -0.04 \\ $\pm$ 0.01} 
%&\multicolumn{1}{p{2cm}}{\centering -0.1 \\ $\pm$ 0.1} 
%& \multicolumn{1}{p{2cm}}{\centering 4.62 \\ $\pm$ 0.11} 
%&\multicolumn{1}{p{2cm}}{\centering 4.92 \\ $\pm$ 0.11} 
%& \multicolumn{1}{p{2cm}}{\centering 3.40 \\ $\pm$ 0.30} \\
%&  34.4 & 12.5\\

%done
%5     & 5715 & 80 & 5690 & 80  & 4.50  & 0.1 & 4.28 & 0.9   & 0.0    & 0.07 & 0.11  & 0.1  & 2.62  & 0.8 & 3.9  & 0.2  & 12.3  & 32.9   \\ 

WASP-5  
& 5806  $\pm$ 17
&  5617  $\pm$ 16
&  5690 $\pm$ 80
& 4.58  $\pm$ 0.01
&  4.28  $\pm$ 0.9
&  4.39  $\pm$ 0.03 $^1$ 
&  0.0  $\pm$ 0.01 
& -0.09  $\pm$ 0.01
&  0.11  $\pm$ 0.1
&  2.62  $\pm$ 0.12
&  $\leq$0.5  $\pm$ 0.14
&  3.90  $\pm$ 0.2 \\

%WASP-5  & \multicolumn{1}{p{2cm}}{\centering 5806 \\ $\pm$ 17} 
%&
%\multicolumn{1}{p{2cm}}{\centering 5617 \\ $\pm$ 16} 
%& \multicolumn{1}{p{2cm}}{\centering 5690 \\$\pm$ 80} 
%& \multicolumn{1}{p{2cm}}{\centering 4.58 \\ $\pm$ 0.01}  
%& \multicolumn{1}{p{2cm}}{\centering 4.28 \\ $\pm$ 0.9} 
%& \multicolumn{1}{p{2cm}}{\centering 4.39 \\ $\pm$ 0.03 $^1$} 
%& \multicolumn{1}{p{2cm}}{\centering 0.0 \\ $\pm$ 0.01} 
%\multicolumn{1}{p{2cm}}{\centering -0.09 \\ $\pm$ 0.01} 
%& \multicolumn{1}{p{2cm}}{\centering 0.11 \\ $\pm$ 0.1} 
%& \multicolumn{1}{p{2cm}}{\centering 2.62 \\ $\pm$ 0.12} 
%&\multicolumn{1}{p{2cm}}{\centering $\leq$0.5 \\ $\pm$ 0.14} 
%& \multicolumn{1}{p{2cm}}{\centering 3.90 \\ $\pm$ 0.2} \\

%6     & 5347 & 80 & 5375 & 65  & 4.56  & 0.1 & 4.61 & 0.07  & -0.36  & 0.07 %& -0.17 & 0.09 & 3.22  & 0.8 & 2.4  & 0.5  & 11.9  & 59.9  \\ 
%done
WASP-6  
& 5380  $\pm$ 15
& 5427  $\pm$ 19
& 5375 $\pm$ 65
&  4.57  $\pm$ 0.01  
&  4.61  $\pm$ 0.07  
&  4.52  $\pm$ 0.01 $^1$
&  -0.35  $\pm$ 0.01 
& -0.32  $\pm$ 0.01
&  -0.17  $\pm$ 0.09
&  2.95  $\pm$ 0.21
& 3.01   $\pm$ 0.19
&  2.40  $\pm$ 0.5 \\

%WASP-6  & \multicolumn{1}{p{2cm}}{\centering 5380 \\ $\pm$ 15} 
%&\multicolumn{1}{p{2cm}}{\centering 5427 \\ $\pm$ 19} 
%& \multicolumn{1}{p{2cm}}{\centering 5375 \\$\pm$ 65} 
%& \multicolumn{1}{p{2cm}}{\centering 4.57 \\ $\pm$ 0.01}  
%& \multicolumn{1}{p{2cm}}{\centering 4.61 \\ $\pm$ 0.07}  
%& \multicolumn{1}{p{2cm}}{\centering 4.52 \\ $\pm$ 0.01 $^1$} 
%& \multicolumn{1}{p{2cm}}{\centering -0.35 \\ $\pm$ 0.01} 
%&\multicolumn{1}{p{2cm}}{\centering -0.32 \\ $\pm$ 0.01} 
%& \multicolumn{1}{p{2cm}}{\centering -0.17 \\ $\pm$ 0.09} 
%& \multicolumn{1}{p{2cm}}{\centering 2.95 \\ $\pm$ 0.21} 
%&\multicolumn{1}{p{2cm}}{\centering 3.01  \\ $\pm$ 0.19} 
%& \multicolumn{1}{p{2cm}}{\centering 2.40 \\ $\pm$ 0.5} \\

%7     & 6538 & 80 & 6550 & 70  & 4.43  & 0.1 & 4.32 & 0.06  & -0.021 & 0.07 & 0.16  & 0.06 & 17.32 & 0.8 & 18.1 & 0.02 & 9.5   & 144    \\
% done
WASP-7  
&  6532  $\pm$ 10
& 6494  $\pm$ 9
&  6550 $\pm$ 70
&  4.41  $\pm$ 0.2
&  4.32  $\pm$ 0.07
&  4.22  $\pm$ 0.04 $^1$
&  0.0  $\pm$ 0.01
& -0.01  $\pm$ 0.01
&  0.16  $\pm$ 0.06
& 17.14  $\pm$ 0.15
& 17.27  $\pm$ 0.14
&  18.10   $\pm$ 0.02 \\

%WASP-7  & \multicolumn{1}{p{2cm}}{\centering 6532 \\ $\pm$ 10} 
%&\multicolumn{1}{p{2cm}}{\centering 6494 \\ $\pm$ 9} 
%& \multicolumn{1}{p{2cm}}{\centering 6550 \\$\pm$ 70} 
%& \multicolumn{1}{p{2cm}}{\centering 4.41 \\ $\pm$ 0.2}  
%& \multicolumn{1}{p{2cm}}{\centering 4.32 \\ $\pm$ 0.07} 
%& \multicolumn{1}{p{2cm}}{\centering 4.22 \\ $\pm$ 0.04 $^1$}
%& \multicolumn{1}{p{2cm}}{\centering 0.0 \\ $\pm$ 0.01} 
%&\multicolumn{1}{p{2cm}}{\centering -0.01 \\ $\pm$ 0.01} 
%& \multicolumn{1}{p{2cm}}{\centering 0.16 \\ $\pm$ 0.06} 
%& \multicolumn{1}{p{2cm}}{\centering 17.14 \\ $\pm$ 0.15} 
%&\multicolumn{1}{p{2cm}}{\centering 17.27 \\ $\pm$ 0.14} 
%& \multicolumn{1}{p{2cm}}{\centering 18.10  \\ $\pm$ 0.02}\\
 
%8     & 5525 & 80 & 5560 & 90  & 4.47  & 0.1 & 4.4  & 0.09  & 0.012  & 0.07 & 0.18  & 0.11 & 0.217 & 0.8 & 2.7  & 0.5  & 9.79  & 206.7 \\
% done
WASP-8* 
&  5578  $\pm$ 15 
& 5488  $\pm$ 9 
&  5560 $\pm$ 90 
&   4.56  $\pm$ 0.01  
&  4.40  $\pm$ 0.09 
& 4.48  $\pm$ 0.01 $^1$
&  0.09  $\pm$ 0.01
& 0.03  $\pm$ 0.01
&  0.18   $\pm$ 0.11
&  $\leq$ 0.50  $\pm$ 0.17 
& $\leq$ 0.50  $\pm$ 0.17
&  2.70   $\pm$ 0.50  \\

%WASP-8* & \multicolumn{1}{p{2cm}}{\centering 5578 \\ $\pm$ 15} 
%&\multicolumn{1}{p{2cm}}{\centering 5488 \\ $\pm$ 9} 
%& \multicolumn{1}{p{2cm}}{\centering 5560 \\$\pm$ 90} 
%& \multicolumn{1}{p{2cm}}{\centering  4.56 \\ $\pm$ 0.01}  
%& \multicolumn{1}{p{2cm}}{\centering 4.40 \\ $\pm$ 0.09} 
%& \multicolumn{1}{p{2cm}}{\centering 4.48 \\ $\pm$ 0.01 $^1$}
%& \multicolumn{1}{p{2cm}}{\centering 0.09 \\ $\pm$ 0.01}
%&\multicolumn{1}{p{2cm}}{\centering 0.03 \\ $\pm$ 0.01} 
%& \multicolumn{1}{p{2cm}}{\centering 0.18  \\ $\pm$ 0.11} 
%& \multicolumn{1}{p{2cm}}{\centering $\leq$ 0.50 \\ $\pm$ 0.17} 
%&\multicolumn{1}{p{2cm}}{\centering $\leq$ 0.50 \\ $\pm$ 0.17} 
%& \multicolumn{1}{p{2cm}}{\centering 2.70  \\ $\pm$ 0.50}  \\

%15    & 6455 & 80 & 6405 & 80  & 4.469 & 0.1 & 4.4  & 0.11  & -0.129 & 0.07 & 0     & 0.1  & 3.36  & 0.8 & 4.9  & 0.4  & 11    & 110.3  \\ 
% done
%WASP-15*  & \multicolumn{1}{p{2cm}}{\centering 6428 \\ $\pm$ 14} 
%&\multicolumn{1}{p{2cm}}{\centering 6395 \\ $\pm$ 14} 
%& \multicolumn{1}{p{2cm}}{\centering 6405 \\$\pm$ 80} 
%& \multicolumn{1}{p{2cm}}{\centering  4.44 \\ $\pm$ 0.01}  
%& \multicolumn{1}{p{2cm}}{\centering 4.40 \\ $\pm$ 0.11} 
%& \multicolumn{1}{p{2cm}}{\centering 4.22 \\ $\pm$ 0.02 $^1$}
%& \multicolumn{1}{p{2cm}}{\centering -0.16 \\ $\pm$ 0.01} 
%&\multicolumn{1}{p{2cm}}{\centering 0.03 \\ $\pm$ 0.01} 
%& \multicolumn{1}{p{2cm}}{\centering 0.00  \\ $\pm$ 0.10} 
%& \multicolumn{1}{p{2cm}}{\centering 5.47 \\ $\pm$ 0.19} 
%&\multicolumn{1}{p{2cm}}{\centering 5.38 \\ $\pm$ 0.11} 
%& \multicolumn{1}{p{2cm}}{\centering 4.90  \\ $\pm$ 0.40} \\

WASP-15*  
&  6428  $\pm$ 14 
& 6395  $\pm$ 14 
&  6405 $\pm$ 8
&   4.44  $\pm$ 0.01  
&  4.40  $\pm$ 0.11 
&  4.22  $\pm$ 0.02 $^1$
&  -0.16  $\pm$ 0.01
& 0.03  $\pm$ 0.01 
&  0.00   $\pm$ 0.10
&  5.47  $\pm$ 0.19 
& 5.38  $\pm$ 0.11 
&  4.90   $\pm$ 0.40 \\

%16    & 5670 & 80 & 5550 & 60  & 4.43  & 0.1 & 4.21 & 0.011 & -0.08  & 0.07 & 0.07  & 0.1  & 1.68  & 0.8 & 2.5  & 0.4  & 11.3  & 94.22  \\ 
% done
%WASP-16*  & \multicolumn{1}{p{2cm}}{\centering 5735 \\ $\pm$ 14} 
%&\multicolumn{1}{p{2cm}}{\centering 5561 \\ $\pm$ 13} 
%& \multicolumn{1}{p{2cm}}{\centering 5550 \\$\pm$ 60} 
%& \multicolumn{1}{p{2cm}}{\centering  4.48 \\ $\pm$ 0.03}  
%& \multicolumn{1}{p{2cm}}{\centering 4.21 \\ $\pm$ 0.01} 
%& \multicolumn{1}{p{2cm}}{\centering 4.49 \\ $\pm$ 0.02 $^1$}
%& \multicolumn{1}{p{2cm}}{\centering -0.02 \\ $\pm$ 0.01} 
%&\multicolumn{1}{p{2cm}}{\centering -0.14 \\ $\pm$ 0.01} 
%& \multicolumn{1}{p{2cm}}{\centering 0.07  \\ $\pm$ 0.10} 
%& \multicolumn{1}{p{2cm}}{\centering 1.30 \\ $\pm$ 0.21} 
%&\multicolumn{1}{p{2cm}}{\centering 1.40 \\ $\pm$ 0.20} 
%& \multicolumn{1}{p{2cm}}{\centering 2.50  \\ $\pm$ 0.40}  \\

WASP-16*  
& 5735  $\pm$ 14 
& 5561  $\pm$ 13 
&  5550 $\pm$ 60 
&   4.48  $\pm$ 0.03  
&  4.21  $\pm$ 0.01 
&  4.49  $\pm$ 0.02 $^1$
&  -0.02  $\pm$ 0.01 
& -0.14  $\pm$ 0.01 
&  0.07   $\pm$ 0.10 
& 1.30  $\pm$ 0.21 
& 1.40  $\pm$ 0.20 
& 2.50   $\pm$ 0.40  \\

%17     
% done
%WASP-17  & \multicolumn{1}{p{2cm}}{\centering 6699 \\ $\pm$ 15} 
%&\multicolumn{1}{p{2cm}}{\centering 6753 \\ $\pm$ 15} 
%& \multicolumn{1}{p{2cm}}{\centering 6700 \\$\pm$ 105} 
%& \multicolumn{1}{p{2cm}}{\centering  4.27 \\ $\pm$ 0.01} 
% & \multicolumn{1}{p{2cm}}{\centering 4.34 \\ $\pm$ 0.23} 
% & \multicolumn{1}{p{2cm}}{\centering 4.16 \\ $\pm$ 0.02 $^1$}
% & \multicolumn{1}{p{2cm}}{\centering -0.24 \\ $\pm$ 0.01} 
% &\multicolumn{1}{p{2cm}}{\centering -0.22 \\ $\pm$ 0.01} 
% & \multicolumn{1}{p{2cm}}{\centering -0.12  \\ $\pm$ 0.10} 
% & \multicolumn{1}{p{2cm}}{\centering 7.86 \\ $\pm$ 0.22} 
 %&\multicolumn{1}{p{2cm}}{\centering 7.59 \\ $\pm$ 0.22} 
 %& \multicolumn{1}{p{2cm}}{\centering 9.80  \\ $\pm$ 1.10} \\
 
 WASP-17  
 & 6699  $\pm$ 15 
& 6753  $\pm$ 15 
&  6700 $\pm$ 105 
&   4.27  $\pm$ 0.01 
 &  4.34  $\pm$ 0.23 
 &  4.16  $\pm$ 0.02 $^1$
 &  -0.24  $\pm$ 0.01 
 & -0.22  $\pm$ 0.01 
 &  -0.12   $\pm$ 0.10
 &  7.86  $\pm$ 0.22 
 & 7.59  $\pm$ 0.22 
 &  9.80  $\pm$ 1.10 \\

%18    & 6420 & 80 & 6400 & 75  & 4.468 & 0.1 & 4.32 & 0.09  & -0.047 & 0.07 & 0.08  & 0.08 & 10.25 & 0.8 & 10.9 & 0.7  & 9.3   & 164.4  \\
% done
%WASP-18  & \multicolumn{1}{p{2cm}}{\centering 6434 \\ $\pm$ 13} 
%&\multicolumn{1}{p{2cm}}{\centering 6354 \\ $\pm$ 15} 
%& \multicolumn{1}{p{2cm}}{\centering 6400 \\$\pm$ 75} 
%& \multicolumn{1}{p{2cm}}{\centering  4.47 \\ $\pm$ 0.02}  
%& \multicolumn{1}{p{2cm}}{\centering 4.32 \\ $\pm$ 0.09} 
%& \multicolumn{1}{p{2cm}}{\centering 4.32 \\ $\pm$ 0.03 $^1$}
%& \multicolumn{1}{p{2cm}}{\centering -0.04 \\ $\pm$ 0.01} 
%&\multicolumn{1}{p{2cm}}{\centering -0.09 \\ $\pm$ 0.01} 
%& \multicolumn{1}{p{2cm}}{\centering 0.08  \\ $\pm$ 0.08} 
%& \multicolumn{1}{p{2cm}}{\centering 10.11 \\ $\pm$ 0.17} 
%&\multicolumn{1}{p{2cm}}{\centering 9.95 \\ $\pm$ 0.13} 
%& \multicolumn{1}{p{2cm}}{\centering 10.9  \\ $\pm$ 0.7} \\

WASP-18  
&  6434  $\pm$ 13 
& 6354  $\pm$ 15 
&  6400 $\pm$ 75 
&   4.47  $\pm$ 0.02 
&  4.32  $\pm$ 0.09 
& 4.32  $\pm$ 0.03 $^1$
&  -0.04  $\pm$ 0.01 
& -0.09  $\pm$ 0.01
&  0.08   $\pm$ 0.08 
&  10.11  $\pm$ 0.17 
& 9.95  $\pm$ 0.13 
&  10.9   $\pm$ 0.7 \\
 
%19    & 5484 & 80 & 5460 & 90  & 4.50  & 0.1 & 4.37 & 0.14  & -0.02  & 0.07 & 0.14  & 0.11 & 5.04  & 0.8 & 5.1  & 0.3  & 12.59 & 53.6  \\ 
%done
%WASP-19  & \multicolumn{1}{p{2cm}}{\centering 5573 \\ $\pm$ 17} 
%&\multicolumn{1}{p{2cm}}{\centering 5540 \\ $\pm$ 16} 
%& \multicolumn{1}{p{2cm}}{\centering 5460 \\$\pm$ 90} 
%& \multicolumn{1}{p{2cm}}{\centering  4.51 \\ $\pm$ 0.02}  
%& \multicolumn{1}{p{2cm}}{\centering 4.37 \\ $\pm$ 0.14} 
%& \multicolumn{1}{p{2cm}}{\centering 4.44 \\ $\pm$ 0.01 $^1$}
%& \multicolumn{1}{p{2cm}}{\centering 0.02 \\ $\pm$ 0.01} 
%&\multicolumn{1}{p{2cm}}{\centering 0.02 \\ $\pm$ 0.01} 
%& \multicolumn{1}{p{2cm}}{\centering 0.14  \\ $\pm$ 0.11} 
%& \multicolumn{1}{p{2cm}}{\centering 3.75 \\ $\pm$ 0.13} 
%&\multicolumn{1}{p{2cm}}{\centering 3.71 \\ $\pm$ 0.13} 
%& \multicolumn{1}{p{2cm}}{\centering 5.1  \\ $\pm$ 0.3}   \\

WASP-19  
&  5573  $\pm$ 17 
& 5540  $\pm$ 16
&  5460 $\pm$ 90 
&   4.51  $\pm$ 0.02  
&  4.37  $\pm$ 0.14 
&  4.44  $\pm$ 0.01 $^1$
&  0.02  $\pm$ 0.01 
& 0.02  $\pm$ 0.01 
&  0.14   $\pm$ 0.11 
&  3.75  $\pm$ 0.13 
& 3.71  $\pm$ 0.13 
&  5.1   $\pm$ 0.3   \\

%20    & 5983 & 80 & 6030 & 80  & 4.50  & 0.1 & 4.54 & 0.13  & -0.11  & 0.07 & 0.13  & 0.09 & 3.63  & 0.8 & 4.3  & 0.4  & 10.68 & 210.3  \\ 

%WASP-20  & \multicolumn{1}{p{2cm}}{\centering 5983 \\ $\pm$ 21}
%& \multicolumn{1}{p{2cm}}{\centering 6037 \\ $\pm$ 16} 
%& \multicolumn{1}{p{2cm}}{\centering 6030 \\$\pm$ 80} 
%& \multicolumn{1}{p{2cm}}{\centering  4.50 \\ $\pm$ 0.02}  
%& \multicolumn{1}{p{2cm}}{\centering 4.54 \\ $\pm$ 0.13} 
%& \multicolumn{1}{p{2cm}}{\centering 4.23 \\ $\pm$ 0.02 $^2$} 
%& \multicolumn{1}{p{2cm}}{\centering -0.11 \\ $\pm$ 0.01} 
%&\multicolumn{1}{p{2cm}}{\centering -0.09 \\ $\pm$ 0.01} 
%& \multicolumn{1}{p{2cm}}{\centering 0.13  \\ $\pm$ 0.09} 
%& \multicolumn{1}{p{2cm}}{\centering 3.63 \\ $\pm$ 0.13} 
%&\multicolumn{1}{p{2cm}}{\centering 1.96 \\ $\pm$ 0.18} 
%& \multicolumn{1}{p{2cm}}{\centering 4.30  \\ $\pm$ 0.40} \\

WASP-20  
&  5983  $\pm$ 21
&  6037  $\pm$ 16 
&  6030 $\pm$ 80 
&   4.50  $\pm$ 0.02  
&  4.54  $\pm$ 0.13 
&  4.23  $\pm$ 0.02 $^2$ 
& -0.11  $\pm$ 0.01 
& -0.09  $\pm$ 0.01 
&  0.13   $\pm$ 0.09
&  3.63  $\pm$ 0.13 
& 1.96  $\pm$ 0.18
& 4.30   $\pm$ 0.40 \\

%22    & 6032 & 80 & 6020 & 65  & 4.43  & 0.1 & 4.25 & 0.09  & 0.0    & 0.07 & 0.16  & 0.08 & 2.64  & 0.8 & 4.4  & 0.2  & 12.0  & 77     \\ 

%WASP-22*  & \multicolumn{1}{p{2cm}}{\centering 6032 \\ $\pm$ 20} 
%&\multicolumn{1}{p{2cm}}{\centering 5980 \\ $\pm$ 20} 
%& \multicolumn{1}{p{2cm}}{\centering 6020 \\$\pm$ 65} 
%& \multicolumn{1}{p{2cm}}{\centering  4.43 \\ $\pm$ 0.02} 
% & \multicolumn{1}{p{2cm}}{\centering 4.25 \\ $\pm$ 0.09} 
% & \multicolumn{1}{p{2cm}}{\centering 4.32 \\ $\pm$ 0.02 $^1$}
% & \multicolumn{1}{p{2cm}}{\centering 0.00 \\ $\pm$ 0.01} 
% &\multicolumn{1}{p{2cm}}{\centering -0.03 \\ $\pm$ 0.01} 
% & \multicolumn{1}{p{2cm}}{\centering 0.16  \\ $\pm$ 0.08}
%  & \multicolumn{1}{p{2cm}}{\centering 4.77 \\ $\pm$ 0.14} 
%  &\multicolumn{1}{p{2cm}}{\centering 4.72 \\ $\pm$ 0.19} 
%  & \multicolumn{1}{p{2cm}}{\centering 4.40  \\ $\pm$ 0.20}  \\

  WASP-22*  
  &  6032  $\pm$ 20 
& 5980  $\pm$ 20 
&  6020 $\pm$ 65 
&   4.43  $\pm$ 0.02
 & 4.25  $\pm$ 0.09 
 &  4.32  $\pm$ 0.02 $^1$
 &  0.00  $\pm$ 0.01 
 & -0.03  $\pm$ 0.01 
 &  0.16   $\pm$ 0.08
  &  4.77  $\pm$ 0.14
  & 4.72  $\pm$ 0.19
  &  4.40   $\pm$ 0.20  \\

%23    & 4986 & 80 & 5020 & 50  & 4.51  & 0.1 & 4.31 & 0.12  & -0.17  & 0.07 & 0.04  & 0.07 & 0.21  & 0.8 & 2.4  & 0.3  & 12.7  & 55.2   \\ 

%WASP-23*  & \multicolumn{1}{p{2cm}}{\centering 4986 \\ $\pm$ 14}
% &\multicolumn{1}{p{2cm}}{\centering 4936 \\ $\pm$ 8} 
% & \multicolumn{1}{p{2cm}}{\centering 5020 \\$\pm$ 50} 
% & \multicolumn{1}{p{2cm}}{\centering  4.51 \\ $\pm$ 0.03}  
 %& \multicolumn{1}{p{2cm}}{\centering 4.31 \\ $\pm$ 0.12}
% & \multicolumn{1}{p{2cm}}{\centering 4.59 \\ $\pm$ 0.02 $^1$} 
 %& \multicolumn{1}{p{2cm}}{\centering -0.17 \\ $\pm$ 0.01} 
 %&\multicolumn{1}{p{2cm}}{\centering -0.20 \\ $\pm$ 0.01} 
% & \multicolumn{1}{p{2cm}}{\centering 0.04  \\ $\pm$ 0.07} 
% & \multicolumn{1}{p{2cm}}{\centering 1.27 \\ $\pm$ 0.22} 
 %&\multicolumn{1}{p{2cm}}{\centering 1.22 \\ $\pm$ 0.22} 
% & \multicolumn{1}{p{2cm}}{\centering 2.40  \\ $\pm$ 0.30}  \\

WASP-23*  
 &  4986  $\pm$ 14
 &  4936  $\pm$ 8 
 &   5020 $\pm$ 50
 &   4.51  $\pm$ 0.03  
 & 4.31  $\pm$ 0.12
 &  4.59  $\pm$ 0.02 $^1$
 &  -0.17  $\pm$ 0.01 
 & -0.20  $\pm$ 0.01 
 &  0.04   $\pm$ 0.07 
 &  1.27  $\pm$ 0.22 
 & 1.22  $\pm$ 0.22 
 &  2.40   $\pm$ 0.30  \\

%WASP-24  & \multicolumn{1}{p{2cm}}{\centering 6295 \\ $\pm$ 14} 
%&\multicolumn{1}{p{2cm}}{\centering 6143 \\ $\pm$ 15} 
%& \multicolumn{1}{p{2cm}}{\centering 6080 \\$\pm$ 60} 
%& \multicolumn{1}{p{2cm}}{\centering  4.48 \\ $\pm$ 0.01} 
% & \multicolumn{1}{p{2cm}}{\centering 4.20 \\ $\pm$ 0.11} 
% & \multicolumn{1}{p{2cm}}{\centering 4.25 \\ $\pm$ 0.01 $^1$}
% & \multicolumn{1}{p{2cm}}{\centering -0.11 \\ $\pm$ 0.01} 
% &\multicolumn{1}{p{2cm}}{\centering -0.18 \\ $\pm$ 0.01} 
% & \multicolumn{1}{p{2cm}}{\centering 0.02  \\ $\pm$ 0.08} 
% & \multicolumn{1}{p{2cm}}{\centering 3.21 \\ $\pm$ 0.21} 
% &\multicolumn{1}{p{2cm}}{\centering 3.22 \\ $\pm$ 0.21}
%  & \multicolumn{1}{p{2cm}}{\centering 6.40  \\ $\pm$ 0.20}  \\
  
  WASP-24  
  &  6295  $\pm$ 14 
& 6143  $\pm$ 15 
&  6080 $\pm$ 60 
&   4.48  $\pm$ 0.01 
 &  4.20  $\pm$ 0.11 
 & 4.25  $\pm$ 0.01 $^1$
 &   $\pm$ 0.01
 & -0.18  $\pm$ 0.01 
 &  0.02   $\pm$ 0.08
 &  3.21  $\pm$ 0.21
 & 3.22  $\pm$ 0.21
  &  6.40   $\pm$ 0.20  \\

%29    & 4650 & 80 & 4730 & 70  & 4.4   & 0.1 & 4.48 & 0.16  & -0.018 & 0.07 & 0.24  & 0.12 & 2.52  & 0.8 & 2.4  & 0.5  & 11.30 & 63.1  \\

%WASP-29  & \multicolumn{1}{p{2cm}}{\centering 4650 \\ $\pm$ 20} 
%&\multicolumn{1}{p{2cm}}{\centering 4680 \\ $\pm$ 23} 
%& \multicolumn{1}{p{2cm}}{\centering 4730 \\$\pm$ 50} 
%& \multicolumn{1}{p{2cm}}{\centering  4.40 \\ $\pm$ 0.01}  
%& \multicolumn{1}{p{2cm}}{\centering 4.48 \\ $\pm$ 0.16} 
%& \multicolumn{1}{p{2cm}}{\centering 4.55 \\ $\pm$ 0.02 $^1$}
%& \multicolumn{1}{p{2cm}}{\centering -0.02 \\ $\pm$ 0.01} 
%&\multicolumn{1}{p{2cm}}{\centering 0.05 \\ $\pm$ 0.01} 
%& \multicolumn{1}{p{2cm}}{\centering 0.24  \\ $\pm$ 0.12} 
%& \multicolumn{1}{p{2cm}}{\centering 2.52 \\ $\pm$ 0.19} 
%&\multicolumn{1}{p{2cm}}{\centering 2.52 \\ $\pm$ 0.19} 
%& \multicolumn{1}{p{2cm}}{\centering $\leq 05$  \\ $\pm$ 0.5}  \\

WASP-29  
&  4650  $\pm$ 20 
& 4680  $\pm$ 23 
&  4730 $\pm$ 50 
&   4.40  $\pm$ 0.01  
& 4.48  $\pm$ 0.16 
&  4.55  $\pm$ 0.02 $^1$
&  -0.02  $\pm$ 0.01 
& 0.05  $\pm$ 0.01 
&  0.24   $\pm$ 0.12
& 2.52  $\pm$ 0.19 
& 2.52  $\pm$ 0.19 
&  $\leq 05$   $\pm$ 0.5  \\

%WASP-30  & \multicolumn{1}{p{2cm}}{\centering 6732 \\ $\pm$ 14} 
%&\multicolumn{1}{p{2cm}}{\centering 6891 \\ $\pm$ 13} 
%& \multicolumn{1}{p{2cm}}{\centering 6190 \\$\pm$ 50} 
%& \multicolumn{1}{p{2cm}}{\centering  4.74 \\ $\pm$ 0.01}  
%& \multicolumn{1}{p{2cm}}{\centering 4.18 \\ $\pm$ 0.18} 
%& \multicolumn{1}{p{2cm}}{\centering 4.28 \\ $\pm$ 0.01 $^3$} 
%& \multicolumn{1}{p{2cm}}{\centering -0.09 \\ $\pm$ 0.01} 
%&\multicolumn{1}{p{2cm}}{\centering -0.01 \\ $\pm$ 0.01} 
%& \multicolumn{1}{p{2cm}}{\centering 0.09 \\ $\pm$ 0.07} 
%& \multicolumn{1}{p{2cm}}{\centering 13.40 \\ $\pm$ 0.17} 
%&\multicolumn{1}{p{2cm}}{\centering 11.79 \\ $\pm$ 0.13} 
%& \multicolumn{1}{p{2cm}}{\centering 12.10 \\ $\pm$ 0.50}   \\

WASP-30  
&  6732  $\pm$ 14 
& 6891  $\pm$ 13 
&  6190 $\pm$ 50 
&   4.74 $\pm$ 0.01
&  4.18  $\pm$ 0.18 
&  4.28  $\pm$ 0.01 $^3$ 
&  -0.09  $\pm$ 0.01 
& -0.01  $\pm$ 0.01 
&  0.09  $\pm$ 0.07 
&  13.40  $\pm$ 0.17 
& 11.79  $\pm$ 0.13 
&  12.10  $\pm$ 0.50   \\

%WASP-31  & \multicolumn{1}{p{2cm}}{\centering 6435 \\ $\pm$ 18} 
%&\multicolumn{1}{p{2cm}}{\centering 6381 \\ $\pm$ 14} 
%& \multicolumn{1}{p{2cm}}{\centering 6320 \\$\pm$ 75} 
%& \multicolumn{1}{p{2cm}}{\centering  4.46 \\ $\pm$ 0.02} 
% & \multicolumn{1}{p{2cm}}{\centering 4.36 \\ $\pm$ 0.10} 
% & \multicolumn{1}{p{2cm}}{\centering 4.31 \\ $\pm$ 0.02 $^1$}
% & \multicolumn{1}{p{2cm}}{\centering -0.26 \\ $\pm$ 0.01} 
% &\multicolumn{1}{p{2cm}}{\centering -0.27 \\ $\pm$ 0.01}
%  & \multicolumn{1}{p{2cm}}{\centering -0.09  \\ $\pm$ 0.10} 
%  & \multicolumn{1}{p{2cm}}{\centering 7.51 \\ $\pm$ 0.13}
%   &\multicolumn{1}{p{2cm}}{\centering 7.43 \\ $\pm$ 0.13}
%    & \multicolumn{1}{p{2cm}}{\centering 7.90  \\ $\pm$ 0.30}   \\

WASP-31  
& 6435  $\pm$ 18
& 6381  $\pm$ 14 
&  6320 $\pm$ 75 
& 4.46  $\pm$ 0.02 
 &  4.36  $\pm$ 0.10 
 &  4.31  $\pm$ 0.02 $^1$
 &  -0.26  $\pm$ 0.01 
 & -0.27  $\pm$ 0.01
  &  -0.09   $\pm$ 0.10 
  & 7.51  $\pm$ 0.13
   &7.43  $\pm$ 0.13
    &  7.90   $\pm$ 0.30   \\

%\#53  & 4863 & 80 & 4950 & 60  & 4.41  & 0.1 & 4.4  & 0.2   & -0.048 & 0.07 & 0.2   & 0.11 & 0.40  & 0.8 & 2.7  & 0.3  & 0     & 46.9  \\ 

%WASP-53  & \multicolumn{1}{p{2cm}}{\centering 4863 \\ $\pm$ 16} 
%&\multicolumn{1}{p{2cm}}{\centering 4925 \\ $\pm$ 16} 
%& \multicolumn{1}{p{2cm}}{\centering 4950 \\$\pm$ 60} 
%& \multicolumn{1}{p{2cm}}{\centering  4.41 \\ $\pm$ 0.02}  
%& \multicolumn{1}{p{2cm}}{\centering 4.40 \\ $\pm$ 0.20} 
%& \multicolumn{1}{p{2cm}}{\centering 4.55 \\ $\pm$ 0.02 $^4$}
%& \multicolumn{1}{p{2cm}}{\centering -0.05 \\ $\pm$ 0.01} 
%&\multicolumn{1}{p{2cm}}{\centering -0.16 \\ $\pm$ 0.01} 
%& \multicolumn{1}{p{2cm}}{\centering 0.11  \\ $\pm$ 0.4} 
%& \multicolumn{1}{p{2cm}}{\centering 0.40 \\ $\pm$ 0.17} 
%&\multicolumn{1}{p{2cm}}{\centering 4.00 \\ $\pm$ 0.21} 
%& \multicolumn{1}{p{2cm}}{\centering 2.70  \\ $\pm$ 0.30} \\

WASP-53  
&  4863  $\pm$ 16 
& 4925 $\pm$ 16
&  4950 $\pm$ 60
&   4.41  $\pm$ 0.02  
&  4.40  $\pm$ 0.20 
&  4.55  $\pm$ 0.02 $^4$
&  -0.05  $\pm$ 0.01
& -0.16  $\pm$ 0.01 
&  0.11   $\pm$ 0.4 
&  0.40  $\pm$ 0.17 
& 4.00  $\pm$ 0.21 
&  2.70   $\pm$ 0.30 \\

%69    & 4782 & 80 & 4750 & 55  & 4.59  & 0.1 & 4.36 & 0.19  & 0.10   & 0.07 & 0.29  & 0.11 & 1.67  & 0.8 & 2.9  & 0.3  & 9.88  & 149.4\\

%WASP-69*  & \multicolumn{1}{p{2cm}}{\centering 4782 \\ $\pm$ 15} 
%&\multicolumn{1}{p{2cm}}{\centering 4687 \\ $\pm$ 14} 
%& \multicolumn{1}{p{2cm}}{\centering 4750 \\$\pm$ 55} 
%& \multicolumn{1}{p{2cm}}{\centering  4.59 \\ $\pm$ 0.02}  
%& \multicolumn{1}{p{2cm}}{\centering 4.36 \\ $\pm$ 0.19} 
%& \multicolumn{1}{p{2cm}}{\centering 4.54 \\ $\pm$ 0.02 $^5$} 
%& \multicolumn{1}{p{2cm}}{\centering 0.10 \\ $\pm$ 0.01} 
%&\multicolumn{1}{p{2cm}}{\centering 0.10 \\ $\pm$ 0.01} 
%& \multicolumn{1}{p{2cm}}{\centering 0.29  \\ $\pm$ 0.11} 
%& \multicolumn{1}{p{2cm}}{\centering 1.27 \\ $\pm$ 0.22} 
%&\multicolumn{1}{p{2cm}}{\centering 1.32 \\ $\pm$ 0.22} 
%& \multicolumn{1}{p{2cm}}{\centering 2.90  \\ $\pm$ 0.30} \\

WASP-69*  
&  4782  $\pm$ 15 
& 4687  $\pm$ 14 
&  4750 $\pm$ 55 
&   4.59  $\pm$ 0.02  
&  4.36  $\pm$ 0.19 
& 4.54  $\pm$ 0.02 $^5$ 
& 0.10  $\pm$ 0.01 
& 0.10  $\pm$ 0.01 
&  0.29   $\pm$ 0.11 
&  1.27  $\pm$ 0.22 
& 1.32  $\pm$ 0.22 
&  2.90   $\pm$ 0.30 \\

%80    & 4066 & 80 & 4145 & 100 & 4.6   & 0.1 & 4.6  & 0.1   & -0.33  & 0.07 & -0.16 & 0.16 & 5.04  & 0.8 & 3.5  & 0.3  & 11.9  & 49.2  \\

%WASP-80  & \multicolumn{1}{p{2cm}}{\centering 4066 \\ $\pm$ 22} 
%&\multicolumn{1}{p{2cm}}{\centering 4050 \\ $\pm$ 23} 
%& \multicolumn{1}{p{2cm}}{\centering 4145 \\$\pm$ 100} 
%& \multicolumn{1}{p{2cm}}{\centering  4.60 \\ $\pm$ 0.02}  
%& \multicolumn{1}{p{2cm}}{\centering 4.60 \\ $\pm$ 0.10} 
%& \multicolumn{1}{p{2cm}}{\centering 4.69 \\ $\pm$ 0.01 $^6$} 
%& \multicolumn{1}{p{2cm}}{\centering -0.33 \\ $\pm$ 0.01} 
%&\multicolumn{1}{p{2cm}}{\centering -0.35 \\ $\pm$ 0.01} 
%& \multicolumn{1}{p{2cm}}{\centering -0.16  \\ $\pm$ 0.16} 
%& \multicolumn{1}{p{2cm}}{\centering 5.04 \\ $\pm$ 0.19} 
%&\multicolumn{1}{p{2cm}}{\centering 2.25 \\ $\pm$ 0.17} 
%& \multicolumn{1}{p{2cm}}{\centering 3.50  \\ $\pm$ 0.30}  \\

WASP-80  
&  4066  $\pm$ 22 
& 4050  $\pm$ 23 
&  4145 $\pm$ 100 
&   4.60  $\pm$ 0.02  
&  4.60  $\pm$ 0.10 
&  4.69  $\pm$ 0.01 $^6$ 
&  -0.33  $\pm$ 0.01 
& -0.35  $\pm$ 0.01 
&  -0.16   $\pm$ 0.16 
&  5.04  $\pm$ 0.19 
& 2.25  $\pm$ 0.17 
&  3.50   $\pm$ 0.30  \\

\hline
\label{waspstars}  
               
\end{tabular}}
\tablefoot{References. (1) \citep{Doyle2015}; (2) \protect\cite{Anderson2015}; (3) \protect\cite{Anderson2011}; (4) \protect\cite{Triaud2016}; (3) \protect\cite{Anderson2011}; (5) \protect\cite{Anderson2014}; (6) \protect\cite{Triaud2013} \\ * Macroturbulence was fixed at 0 km\,s$^{-1}$. \\
Uncertainities quoted here are internel and do not represent the true precision of our measurements. This is further discussed in Sect. \ref{precision_w}.  } 

\end{sidewaystable*}

%%%%%%%%%%%%%%%%%%%
% Appendix
%%%%%%%%%%%%%%%%%%%
\iffalse
\begin{appendix} %First appendix

\section{PLATO WP122 results}\label{PLATO_results}
\begin{figure}[ht!]
\centering
\includegraphics[width=0.5\textwidth]{PLATO/plots/star_1_resolution_3000}
\includegraphics[width=0.5\textwidth]{PLATO/plots/star_1_resolution_20000}
\includegraphics[width=0.5\textwidth]{PLATO/plots/star_1_resolution_65000}

\caption{The results of th WP122 blind test for $\alpha\,  \rm Cen\, \rm A$ for resolving power of 3000 (top), 20,000 (center), and 65,000 (bottom).}
\label{alphacenA20000}
\end{figure}

\begin{figure}[ht!]
\centering
\includegraphics[width=0.5\textwidth]{PLATO/plots/star_2_resolution_3000}

\includegraphics[width=0.5\textwidth]{PLATO/plots/star_2_resolution_20000}
\includegraphics[width=0.5\textwidth]{PLATO/plots/star_2_resolution_65000}

\caption{The results of th WP122 blind test for $\beta\,  \rm Hyi$ for resolving power of 3000 (top), 20,000 (center), and 65,000 (bottom)}
\label{alphacenA20000}
\end{figure}

\end{appendix}
\fi
\end{document}